\documentclass[%
 reprint,
superscriptaddress,
 amsmath,amssymb,
 aps,
floatfix,
]{revtex4-2}

\usepackage{enumitem}
\usepackage{graphicx}% Include figure files
\usepackage{dcolumn}% Align table columns on decimal point
\usepackage{bm}% bold math
\usepackage[mathlines]{lineno}% Enable numbering of text and display math
\usepackage{appendix}
\usepackage{braket}
\usepackage[dvipsnames]{xcolor}
\usepackage{amsmath,amssymb,amsfonts} % Typical maths resource package
\usepackage{makecell}% make cells in tables
\usepackage{float}
\usepackage{bbm}
\usepackage{nicefrac}
\usepackage{wasysym}
\usepackage[normalem]{ulem}

\usepackage{hyperref}
\hypersetup{
	colorlinks=true,
	linkcolor=blue,
	filecolor=blue,      
	urlcolor=blue,
	citecolor=blue,
}

\def\Blue#1{\textcolor{blue}{#1}}
\begin{document}

\preprint{APS/123-QED}

\title{Noise-Enhanced Self-Healing Dynamics in Non-Hermitian Systems} 
%\title{Noise-Enhanced Self-Healing in Non-Hermitian Dynamics}

\author{Wuping Yang}
\affiliation{School of Physics, Peking University, Beijing 100871, China}

\author{H. Huang}%
\email[Corresponding author: ]{huanghq07@pku.edu.cn}
\affiliation{School of Physics, Peking University, Beijing 100871, China}
\affiliation{Collaborative Innovation Center of Quantum Matter, Beijing 100084, China}
\affiliation{Center for High Energy Physics, Peking University, Beijing 100871, China}

\date{\today}% It is always \today, today,

%\begin{abstract}
%Self-healing refers to the ability of a wave packet to reconstruct its spatial profile after scattering. In non-Hermitian systems, this phenomenon can arise from the non-Hermitian skin effect and the resulting non-unitary dynamics. Here we investigate how stochastic noise affects such edge self-healing. Counterintuitively, we find that noise can enhance, rather than destroy, the recovery of the wave-packet profile. Weak noise prolongs the self-healing window by sustaining an exponential suppression of deviations, while strong noise drives the system into a robust asymptotic regime in which profile errors remain frozen at a small value. We explain this behavior using finite-time Lyapunov exponents and a strong-noise effective theory. Our results identify a constructive role of noise in non-Hermitian dynamics and provide guidance for realizing robust self-healing in realistic noisy platforms.
%\end{abstract}

\begin{abstract}
Self-healing is the ability of a wave packet to spontaneously restore its spatial profile after scattering. As an emergent feature of non-unitary dynamics, it has attracted significant interest in non-Hermitian physics. Here, we systematically investigate how stochastic noise influences edge self-healing. Counterintuitively, we find that noise can constructively enhance this dynamical process. Weak noise prolongs the self-healing window by aligning the finite-time Lyapunov exponent of the reference state with the maximum imaginary part of the energy spectrum. Remarkably, strong noise universally stabilizes asymptotic profile recovery across the entire spectrum by inducing an effective non-unitary drift-diffusion dynamics. We analytically elucidate these distinct mechanisms using a general finite-time Lyapunov exponent analysis, complemented by a dedicated perturbation theory for the strong-noise regime. Our results provide concrete guidance for realizing robust non-Hermitian dynamics in realistic noisy environments.
\end{abstract}

\maketitle

%\tableofcontents

\Blue{\textit{Introduction.}}---Non-Hermitian physics provides an essential framework for open systems beyond conventional quantum mechanics \cite{RevModPhys.93.015005, doi:10.1080/00018732.2021.1876991,PhysRevLett.85.2478,PhysRevC.67.054322,Rotter_2009,PhysRevA.85.032111,Rotter_2015}, with diverse experimental realizations spanning photonics \cite{RevModPhys.91.015006,feng2017Non, AosRnsurr2012Parity, doi:10.1126/science.1258479, 2015Spawning,doi:10.1126/science.aap9859, doi:10.1126/science.aar4005, Nasari:23, PhysRevLett.127.270602,cerjan2019experimental}, acoustics \cite{PhysRevX.6.021007, doi:10.1126/science.abd8872, PhysRevLett.121.085702, PhysRevLett.127.034301, PhysRevLett.118.174301, PhysRevApplied.16.014012}, electrical circuits \cite{2020Generalized,0Observation, PhysRevB.107.085426, advs.202301128}, and quantum systems \cite{zhao2025two,doi:10.1126/science.abe9869}. This paradigm has predicted numerous exotic phenomena, most notably the non-Hermitian skin effect (NHSE) \cite{PhysRevLett.121.086803}, whose characteristic macroscopic boundary accumulation of eigenstates has sparked massive theoretical interest \cite{PhysRevLett.121.086803, PhysRevLett.124.086801, PhysRevLett.123.246801, PhysRevLett.125.226402, PhysRevX.14.021011, PhysRevB.109.165127,xiong2024nonhermitianskineffectarbitrary, PhysRevLett.131.116601,2021Universal,PhysRevLett.131.076401,HU202551,cwwd-bclc,PhysRevB.111.155121,cwwd-bclc,vxgf-59xt,3ht3-ty3h,wang2025generaltheorygeometrydependentnonhermitian}.

Recently, wave-packet self-healing has attracted significant attention as a remarkable feature of these non-unitary dynamics \cite{PhysRevLett.128.157601}. While canonical diffraction-free beams (e.g., Bessel and Airy waves \cite{PhysRevX.4.011013,McGloin01012005,BOUCHAL1998207,PhysRevLett.58.1499}) have been realized across various platforms \cite{McGloin01012005,BOUCHAL1998207,zhang2014generation,PhysRevX.4.011013,voloch2013generation}, exact self-healing remains fundamentally forbidden for integrable wave functions under unitary evolution.

Non-Hermitian systems overcome this restriction. Pioneering work by Longhi \cite{PhysRevLett.128.157601} revealed that skin modes defined under semi-infinite boundary conditions can exhibit self-healing capabilities within open-boundary lattices. Later work demonstrated that for short-time non-unitary dynamics, skin-mode eigenstates under open boundary conditions (OBCs) can also display self-healing behavior. This phenomenon is well captured by the short-time Lyapunov exponent \cite{xue2025nonblochedgedynamicsnonhermitian}. Furthermore, related studies have established that the long-time bulk dynamics of non-Hermitian wave packets are controlled by the dominant saddle point, implying an intrinsic asymptotic self-healing tendency for bulk modes \cite{llbb-pcgk}.

While theoretically intriguing, the robustness of these idealized self-healing dynamics against unavoidable environmental noise remains a critical open question. Functioning as time-dependent disorder, stochastic noise fundamentally perturbs the underlying non-unitary amplification, mode competition, and boundary accumulation. Understanding its precise role is thus essential for realizing non-Hermitian self-healing in practical settings.

In this Letter, we address this question by systematically investigating the impact of stochastic noise on edge self-healing driven by the NHSE. Counterintuitively, we demonstrate that environmental noise can play a profoundly constructive role. Specifically, we reveal that weak noise prolongs the self-healing window by elevating the finite-time Lyapunov exponent (FTLE) of the reference state toward the maximum imaginary part of the energy spectrum. Most remarkably, strong noise universally stabilizes asymptotic profile recovery by inducing an effective non-unitary drift-diffusion dynamics. To analytically elucidate these distinct mechanisms, we employ a general FTLE framework, complemented by a dedicated perturbation theory for the strong-noise regime. Ultimately, our results bridge the divide between idealized non-Hermitian models and realistic noisy platforms, suggesting novel strategies for robust dynamical control in open wave systems.

%%%%%%%%%%%%%%%%%%%%%%%%%%%%%%%%%%%%%%%%%%%%%
\Blue{\textit{Profile self-healing metric}}.---We consider the dynamics of a state initially prepared in an eigenstate $|\phi(0)\rangle$ of the unperturbed Hamiltonian. The full Hamiltonian is
\begin{equation}
\hat H=\hat H_0+\hat V_{\text{noise}}+\hat V_{\text{scat}},
\end{equation}
where $\hat H_0$ is time independent, $\hat V_{\text{noise}}(t)$ describes stochastic noise, and $\hat V_{\text{scat}}$ is a boundary-localized scattering potential applied during the interval $t\in[t_0,t_1]$. The reference and scattered states evolve as
\[
|\phi(t)\rangle= \mathcal{T}\exp\!\left[-i\int_0^t \bigl(\hat H_0+\hat V_{\text{noise}}(\tau)\bigr)d\tau\right]|\phi(0)\rangle
\]
and
\[
|\psi(t)\rangle= \mathcal{T}\exp\!\left[-i\int_0^t \hat H(\tau)\,d\tau\right]|\phi(0)\rangle,
\]
respectively. To quantify recovery of the wave-packet profile, we define the self-healing metric
\begin{equation}
\eta(t)\equiv 1-\frac{|\langle\psi(t)|\phi(t)\rangle|^2} {\langle\psi(t)|\psi(t)\rangle\,\langle\phi(t)|\phi(t)\rangle}.
\end{equation}
Theoretically, the state is said to exhibit a self-healing property when $\eta(t)$ approaches zero during its time evolution. However, in realistic experimental scenarios, taking into account factors such as finite instrument resolution, self-healing can be practically defined as $\eta(t)$ falling below a sufficiently small threshold during the evolution.

Previous studies~\cite{PhysRevLett.128.157601} characterized self-healing using the deviation state
\[
|\xi(t)\rangle\equiv|\psi(t)\rangle-|\phi(t)\rangle
\]
and the associated deviation metric
\begin{equation}
\epsilon(t)\equiv
\frac{\langle\xi(t)|\xi(t)\rangle}{\langle\phi(t)|\phi(t)\rangle}.
\end{equation}
The two measures satisfy $\eta(t)\le\epsilon(t)$, as derived in Sec.~I of Supplemental Material (SM)~\footnote{\label{fn}See Supplemental Material at \url{http://link.aps.org/supplemental/xxx} for comprehensive analytical derivations and extended numerical verifications. Specifically, the SM provides: (i) a rigorous proof of the self-healing metric inequality $\eta(t) \le \epsilon(t)$; (ii) detailed derivations of the FTLE dynamics under weak noise using biorthogonal expansion; (iii) the complete strong-noise perturbation framework, including the derivation of the effective non-unitary drift-diffusion equation and its universal $1/t$ convergence; (iv) statistical validation of the FTLE-based estimator comparing single-trajectory dynamics with ensemble averages; (v) analysis of the relationship between eigenstate spatial extension (characterized by skin corner weight) and self-healing robustness; (vi) demonstrations of the mechanism's universality across alternative lattice configurations; and (vii) further discussions on the fragility of coherent non-Hermitian saddle-point dynamics under noise. The Supplemental Material includes Refs.~\cite{llbb-pcgk,xue2025nonblochedgedynamicsnonhermitian,PhysRevB.111.155121}.}. Importantly, $\epsilon(t)$ incorporates contributions from both profile distortion and overall amplitude mismatch, whereas $\eta(t)$ strictly characterizes the normalized wavefunction profile. For example, if $|\psi(t)\rangle=\alpha|\phi(t)\rangle$ with $\alpha\neq 1$, then $\epsilon(t)=|\alpha-1|^2$ remains finite even though the two normalized profiles are identical, while $\eta(t)=0$ correctly signals perfect profile recovery.

\begin{figure*}[!htbp]
    \centering
    \includegraphics[width=1\linewidth]{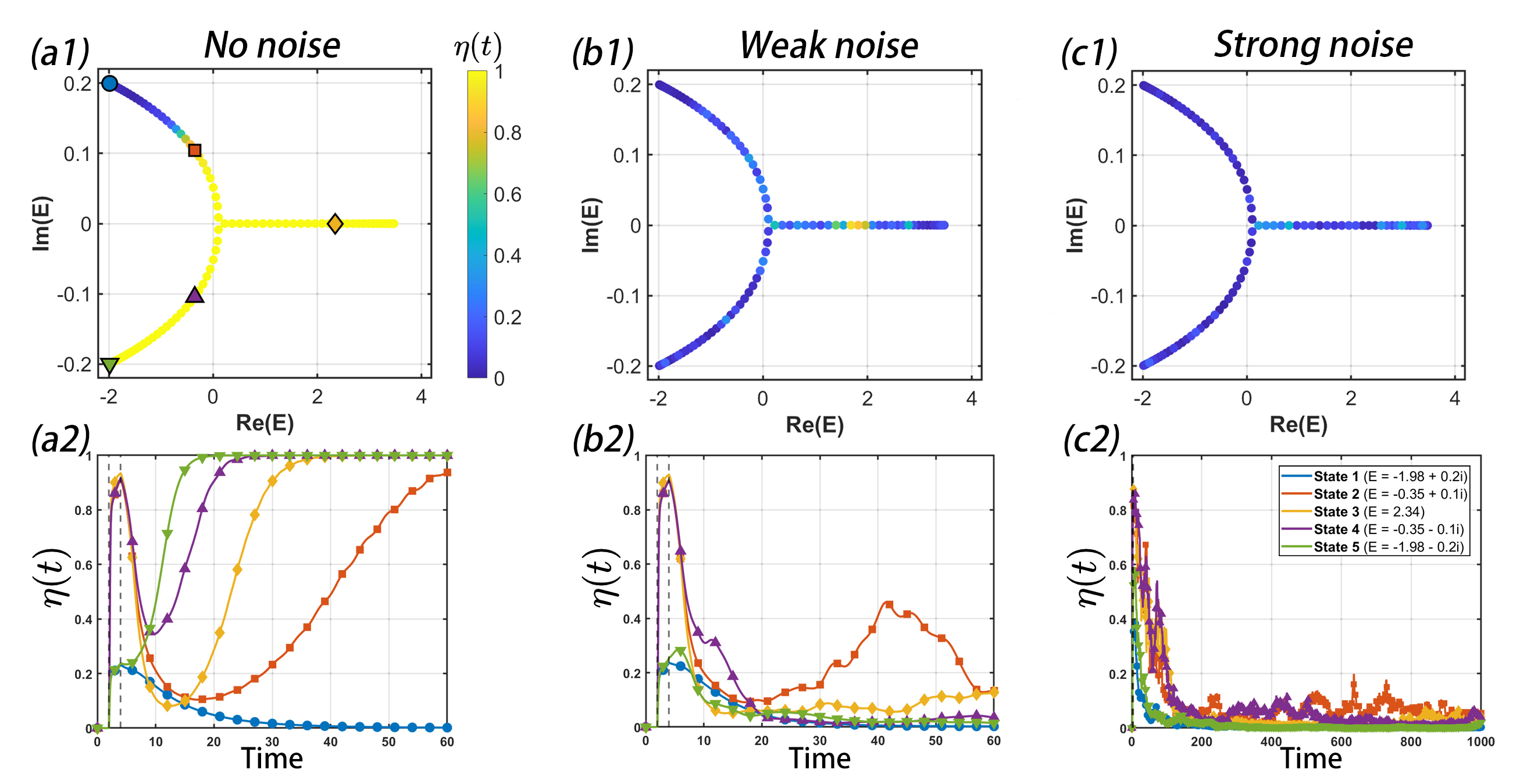}
    \caption{Dynamics of the self-healing metric $\eta(t)$. (a1)--(c1) Snapshots of the energy spectrum, with each eigenstate colored by $\eta(t)$, for no noise, weak noise [$(\theta,\sigma)=(1,0.1)$], and strong noise [$(\theta,\sigma)=(5,10)$]. Snapshots are taken at $t=60$ in (a1) and (b1), and at $t=1000$ in (c1). (a2)--(c2) Time evolution of $\eta(t)$ for representative eigenstates marked by
    the corresponding symbols in (a1), under the same noise conditions. Model parameters are $(t_1,t_{-1},t_2,t_{-2},L)=(0.7,1,0.8,1,100)$. The scattering potential has strength $\gamma=10$ and spatial range $l=10$, and is applied during $t\in[2,4]$.}
    \label{fig1}
\end{figure*}

\Blue{\textit{Noise-enhanced edge self-healing}}.---We consider a general non-Hermitian lattice Hamiltonian
\begin{equation}\label{general Hamiltonian}
\hat H_0=\sum_{i,j} t_{i-j}\hat c_j^\dagger \hat c_i,
\end{equation}
where $t_{i-j}$ denotes the hopping amplitude from site $i$ to site $j$. We assume that the spectrum of $\hat H_0$ under periodic boundary conditions carries a positive winding number\cite{PhysRevLett.125.126402}, so that under OBC the eigenstates exhibit NHSE and accumulate near the left boundary.

To model environmental fluctuations, we introduce a diagonal noise potential
\begin{equation}
\hat V_{\text{noise}}(t)=\sum_j \xi_j(t)\hat c_j^\dagger \hat c_j,
\end{equation}
where $\xi_j(t)$ is the stochastic amplitude at site $j$. We take $\xi_j(t)$ to follow an Ornstein-Uhlenbeck process with the correlation function
\[
\langle \xi_j(t)\xi_{j'}(t+\tau)\rangle
=\frac{\sigma^2}{2\theta}e^{-\theta|\tau|}\delta_{jj'},
\]
where $\sigma$ and $\theta$ are the noise strength and relaxation rate, respectively. $\langle \cdots \rangle$ denotes the ensemble average over noise realizations. The scattering potential is
\begin{equation}
\hat V_{\text{scat}}=-i\gamma\sum_{j=1}^{l} \hat c_j^\dagger \hat c_j,
\end{equation}
where $\gamma$ is real and $l$ is the scattering range.

Fig~\ref{fig1} summarizes the behavior of $\eta(t)$ in the absence of noise, under weak noise, and under strong noise. Panels~(a1)--(c1) show spectral snapshots of $\eta(t)$ for different initial eigenstates, while (a2)--(c2) display the corresponding time traces for representative states. In the noiseless case, self-healing depends strongly on the initial eigenstate: $\eta(t)$ approaches zero only for eigenstates whose imaginary part of the energy exceeds a critical value, consistent with a recent study~\cite{xue2025nonblochedgedynamicsnonhermitian}.

Adding weak noise [$(\theta,\sigma)=(1,0.1)$] significantly enlarges the spectral region that supports self-healing, as evidenced by the reduced high-$\eta$ area in Fig.~\ref{fig1}(b1). Correspondingly, Fig.~\ref{fig1}(b2) shows that eigenstates that do not self-heal in the noiseless case now exhibit a fluctuating but overall decreasing $\eta(t)$.

Most strikingly, in the strong-noise regime [$(\theta,\sigma)=(5,10)$], self-healing becomes nearly universal across the spectrum. As shown in Fig.~\ref{fig1}(c1), $\eta(t)$ remains small for almost all initial eigenstates at long times, and Fig.~\ref{fig1}(c2) further confirms that the profile mismatch saturates at a low value, indicating robust asymptotic recovery. Strong noise, therefore, does not destroy
edge self-healing; instead, it stabilizes and extends it to nearly the entire spectrum.
%%%%%%%%%%%%%%%%%%%%%%%%%%%%

%%%%%%%%%%%%%%%%%%%%%%%%%%%
\Blue{\textit{Finite-time Lyapunov exponent analysis}}.---To analyze the numerical results presented above, we introduce the FTLE as a diagnostic tool for the self-healing dynamics:
\begin{equation}\label{lambda}
\lambda_{\psi}(t) = \frac{1}{2t}
\bigl\langle \ln\langle\psi(t)|\psi(t)\rangle \bigr\rangle,
\end{equation}
The FTLEs of the reference state $|\phi(t)\rangle$ and the deviation state $|\xi(t)\rangle$ are defined analogously and denoted $\lambda_\phi(t)$ and $\lambda_\xi(t)$, respectively.

The deviation metric $\epsilon(t)$ provides an upper bound for $\eta(t)$. For mathematical convenience, we instead consider its geometric mean, $\exp\{ \langle \ln \epsilon(t) \rangle \}$, which can be elegantly expressed via the FTLEs of $|\xi(t)\rangle$ and $|\phi(t)\rangle$ as $e^{2t[\lambda_\xi(t) - \lambda_\phi(t)]}$. Crucially, as explicitly verified in the Sec.~IV of SM~\footnotemark[\value{footnote}], our numerical results closely adhere to the following relation:
\begin{equation}
\langle \eta(t) \rangle \simeq e^{2t[\lambda_\xi(t) - \lambda_\phi(t)]} \le \langle \epsilon(t) \rangle. 
\label{epsilon}
\end{equation}
Consequently, investigating this FTLE-based mean is of profound physical significance, as it provides a robust and excellent proxy for the actual profile self-healing dynamics.

\iffalse
Its ensemble average can be approximated as
\begin{equation}
\langle\epsilon(t)\rangle \simeq e^{2t[\lambda_{\xi}(t)-\lambda_{\phi}(t)]},
\label{epsilon}
\end{equation}
whose derivation and range of validity are discussed in Sec.~II of SM~\footnotemark[\value{footnote}]. Since $\eta(t)\le\langle\epsilon(t)\rangle$, the inequality $\lambda_\xi<\lambda_\phi$ directly implies an exponential suppression of the profile mismatch, and hence self-healing.
\fi

\begin{figure}[!htbp]
    \centering
    \includegraphics[width=1.02\linewidth]{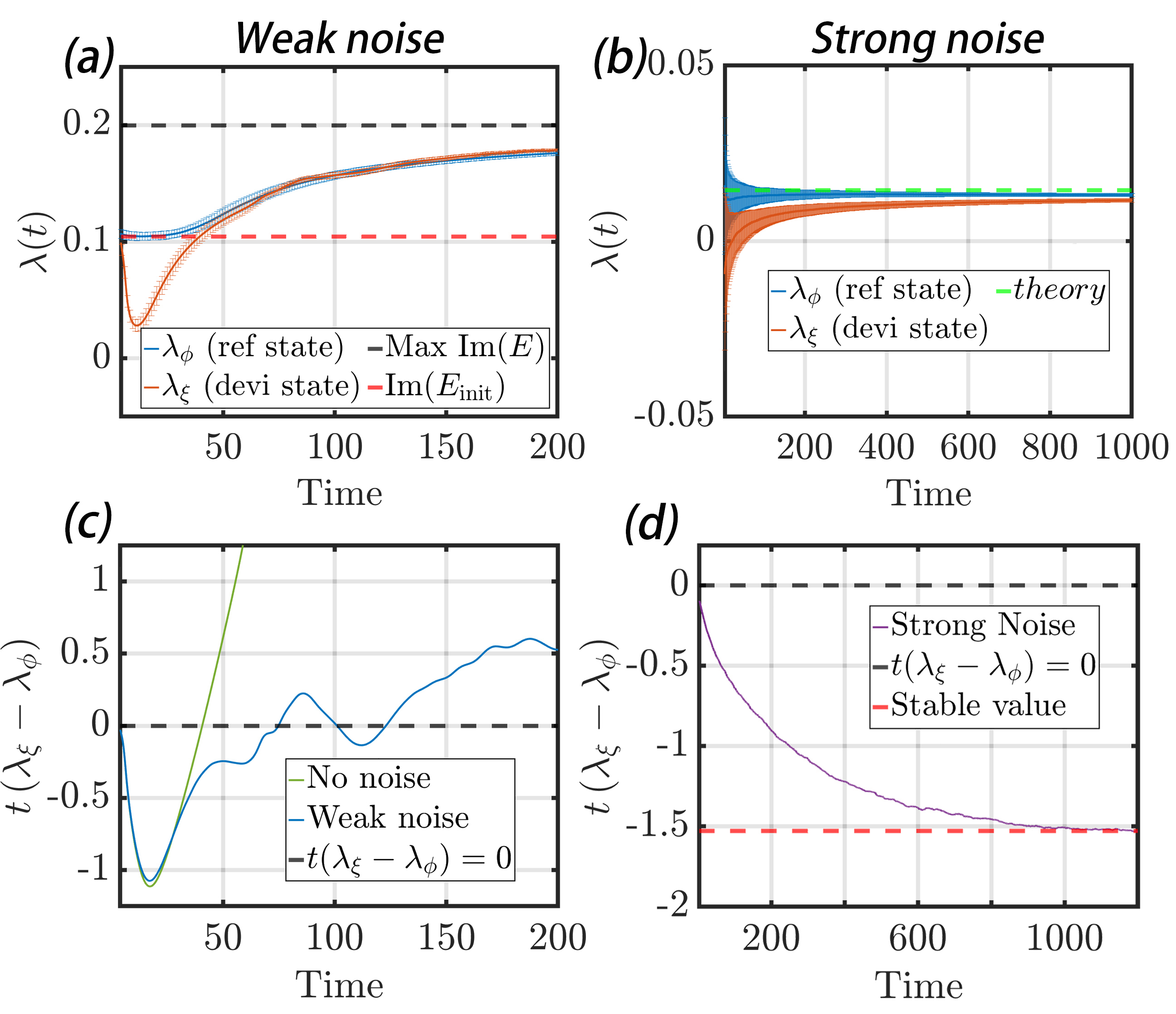}
    \caption{(a),(b) Time evolution of the FTLEs $\lambda_\phi$ (blue) and $\lambda_\xi$ (orange) under weak noise [$(\theta,\sigma)=(1,0.1)$] and strong noise [$(\theta,\sigma)=(5,10)$], respectively. Black dashed lines indicate $\max[\operatorname{Im}(E)]$ of the $\hat H_0$ spectrum; red dashed lines indicate $\operatorname{Im}(E_{\rm init})$ for the representative initial eigenstate $E_{\rm init}=-0.35+0.1i$. The green dashed line in (b) shows the asymptotic FTLE predicted by the strong-noise theory (yielding a value of $0.01444$). All results are averaged over $10^3$ noise realizations. (c),(d) Time evolution of $t(\lambda_\xi-\lambda_\phi)$ for the noiseless, weak-noise, and strong-noise cases. The black dashed line at $t(\lambda_\xi - \lambda_\phi) = 0$ serves as the benchmark for profile self-healing. The red dashed line in (d) indicates the numerically obtained long-time saturation value. System and scattering parameters are the same as in Fig.~\ref{fig1}.}
    \label{fig2}
\end{figure}

Figs~\ref{fig2}(a) and \ref{fig2}(b) show the time evolution of $\lambda_\phi$ and $\lambda_\xi$ for a representative initial eigenstate with $E_{\rm init}=-0.35+0.1i$, under weak and strong noise, respectively. Figs~\ref{fig2}(c) and \ref{fig2}(d) display the corresponding evolution of $t[\lambda_{\xi}(t)-\lambda_{\phi}(t)]$.

In the weak-noise regime [Fig.~\ref{fig2}(a)], $\lambda_{\phi}$ initially tracks $\operatorname{Im}(E_{\rm init})$ before converging to $\max[\operatorname{Im}(E)]$ at long times. By contrast, $\lambda_{\xi}$ exhibits a pronounced transient suppression before approaching the same asymptotic value. Although this initial downward trend is state-dependent, extensive numerical calculations (see Sec.~II in SM~\footnotemark[\value{footnote}]) confirm that the early-time inequality $\lambda_{\xi}<\lambda_{\phi}$ is a robust feature.

These observations can be understood within the biorthogonal expansion framework by treating the short- and long-time regimes separately (see Sec.~II in SM~\footnotemark[\value{footnote}] for a rigorous derivation). In the early post-scattering regime, both $|\phi(t)\rangle$ and $|\xi(t)\rangle$ are primarily governed by the unperturbed Hamiltonian $\hat H_0$. While $\lambda_\phi$ tracks $\operatorname{Im}(E_{\rm init})$, the initial value of $\lambda_\xi$ is determined by the post-scattering norm $\langle \xi(t_1) \mid \xi(t_1) \rangle$: depending on its magnitude, $\lambda_\xi$ either starts from a strongly suppressed value or undergoes a rapid transient dip driven by a large negative instantaneous growth rate. In either case, the condition $\lambda_\xi<\lambda_\phi$ is established early in the evolution. At long times, weak noise induces intermode transitions that drive both $|\phi(t)\rangle$ and $|\xi(t)\rangle$ toward the eigenmode with the largest imaginary part of the energy. As a result, both $\lambda_\phi$ and $\lambda_\xi$ converge to $\max[\operatorname{Im}(E)]$, the condition $\lambda_\xi<\lambda_\phi$ is eventually lost, and self-healing ceases.

The physical mechanism for weak-noise-enhanced self-healing now follows directly. Self-healing persists as long as $\lambda_\phi>\lambda_\xi$, because Eq.~(\ref{epsilon}) then enforces an exponential suppression of the profile mismatch. As shown in Fig.~\ref{fig2}(c), in the noiseless case the initial suppression of $\lambda_\xi$ is rapidly reversed: $t(\lambda_\xi-\lambda_\phi)$ quickly rebounds and crosses zero, closing the self-healing window. Weak noise delays this rebound by elevating $\lambda_\phi$ toward $\max[\operatorname{Im}(E)]$, thereby keeping $t(\lambda_\xi-\lambda_\phi)$ negative for a substantially longer time. The self-healing window is thus extended without altering the ultimate long-time fate.

\Blue{\textit{Strong-noise perturbation theory}}.---The strong-noise regime is qualitatively different. As shown in Fig.~\ref{fig2}(b), both $\lambda_{\phi}$ and $\lambda_{\xi}$ converge to a common long-time limit that differs from $\max[\operatorname{Im}(E)]$, reflecting an intrinsically noise-dominated dynamics. Crucially, $\lambda_{\xi}$ remains below $\lambda_{\phi}$ throughout the evolution. As a consequence, $t[\lambda_{\xi}(t)-\lambda_{\phi}(t)]$ decays initially and then saturates at a finite negative value [Fig.~\ref{fig2}(d)], implying a persistent upper bound on $\eta(t)$ and hence robust long-time self-healing.

To explain how strong noise induces self-healing across nearly all eigenstates, we derive an effective continuum equation for the ensemble-averaged probability density $\rho(x,t)$ in the strong-noise limit. Following Ref.~\cite{PhysRevE.79.050105,yang2026noiseinducedresurrectiondynamicalskin} and carrying out the derivation detailed in Sec.~III of SM~\footnotemark[\value{footnote}], we obtain
\begin{equation}\label{Continuous_Master_Equation_en}
\frac{\partial \rho}{\partial t} = S\rho + v\frac{\partial \rho}{\partial x} + D\frac{\partial^2 \rho}{\partial x^2},
\end{equation}
where the effective coefficients are
\begin{equation}
\begin{split}
S &= 2Q(t)\sum_{n\neq 0}\bigl(|t_n|^2 - t_n t_{-n}\bigr),\\
v &= -2Q(t)\sum_{n\neq 0} na\,|t_n|^2,\\
D &= Q(t)\sum_{n\neq 0}(na)^2|t_n|^2,
\end{split}
\end{equation}
and the noise kernel integral $Q(t)$ is
\begin{equation}
Q(t)=\int_0^t
\Bigl\langle
e^{i[\phi_j(t)-\phi_j(\tau)]}
e^{-i[\phi_l(t)-\phi_l(\tau)]}
\Bigr\rangle d\tau,
\end{equation}
with $\phi_j(t)\equiv\int_0^t\xi_j(\tau)d\tau$. The parameters $S$, $v$, and $D$ are the effective growth rate, drift velocity, and diffusion coefficient, respectively.

Under OBCs, Eq.~(\ref{Continuous_Master_Equation_en}) admits the solution
\begin{equation}
\label{rho_expression}
\rho(x,t) = e^{-\frac{v}{2D}x} e^{\left(S - \frac{v^2}{4D}\right)t} \sum_{n=1}^{\infty} c_n \sin\left(\frac{n\pi}{L} x\right) e^{-D \frac{n^2 \pi^2}{L^2} t},
\end{equation}
where the coefficients $c_n$ are fixed by the initial condition $\rho(x,0)$. At long times, the dynamics are dominated by the fundamental mode $n=1$, yielding the steady-state Lyapunov exponent 
\begin{equation}
\begin{aligned}&\lambda_{\infty} = \frac{S}{2}-\frac{v^{2}}{8 D}-\frac{D \pi^{2}}{2 L^{2}},\\
&\approx Q(\infty)\left[{\sum_{n>0}\left(t_{-n}-t_{n}\right)^{2}}-{\frac{1}{2} \frac{\left[\sum_{n>0} n\left(\left|t_{-n}\right|^{2}-\left|t_{n}\right|^{2}\right)\right]^{2}}{\sum_{n>0} n^{2}\left(\left|t_{-n}\right|^{2}+\left|t_{n}\right|^{2}\right)}}\right].
\end{aligned}
\end{equation}

The predicted value of $\lambda_\infty$ is shown as the green dashed line in Fig.~\ref{fig2}(b), in excellent agreement with the long-time limits of both $\lambda_\xi$ and $\lambda_\phi$, thereby validating the strong-noise theory. Beyond predicting steady-state value, we prove explicitly in Sec.~III.G of SM~\footnotemark[\value{footnote}] that the finite-time convergence obeys a universal $1/t$ scaling law,
\begin{equation}
\lambda_\infty - \lambda(t) \propto 1/t.
\end{equation}
This scaling has a direct consequence: since both $\lambda_\xi(t)$ and $\lambda_\phi(t)$ approach $\lambda_\infty$ with the same $1/t$ correction, their difference $\lambda_\xi(t)-\lambda_\phi(t)$ also decays as $1/t$, so that the quantity $t[\lambda_\xi(t)-\lambda_\phi(t)]$ saturates to a finite constant. Through Eq.~(\ref{epsilon}), this freezes the profile mismatch at a small value and ensures robust long-time self-healing.

To understand why $\lambda_\xi$ remains below $\lambda_\phi$, we introduce the instantaneous growth rate
\begin{equation}
\zeta(t)\equiv\frac{1}{N(t)}\frac{dN(t)}{dt},
\end{equation}
where $N(t)\equiv\int_0^L\rho(x,t)dx$ is the total norm. The FTLE is related to
$\zeta(t)$ by
\begin{equation}
\lambda(t)=\frac{1}{2t}\int_0^t\zeta(\tau)d\tau+\frac{\ln N(0)}{2t}.
\end{equation}
At long times, the initial-memory term $\ln N(0)/2t$ vanishes and the dynamics are governed by the time-averaged accumulation of $\zeta(t)$. Furthermore, since $N_\xi(t_1) \lesssim N_\phi(t_1)$ [Fig.~\ref{fig2}(b)], this initial memory term merely provides a negative shift favoring self-healing. We can thus safely neglect it and focus solely on the non-trivial time integral of $\zeta(t)$.

Integrating Eq.~(\ref{Continuous_Master_Equation_en}) over $[0,L]$ and imposing OBCs (i.e., $\rho(0,t)=\rho(L,t)=0$) yields
\begin{equation}\label{gamma}
\zeta(t)=S+\frac{D}{N}\left(
\left.\frac{\partial\rho}{\partial x}\right|_L
-\left.\frac{\partial\rho}{\partial x}\right|_0
\right).
\end{equation}
As shown in Eq.~(\ref{gamma}), the boundary-gradient term of the instantaneous growth rate $\zeta(t)$ carries an explicit $1/N$ factor, indicating that this specific contribution is effectively determined by the boundary gradients of the \textit{normalized} state. Consequently, a more sharply localized profile exhibits steeper relative gradients, yielding a larger negative contribution that strongly suppresses $\zeta(t)$.

This boundary-gradient mechanism elegantly explains why $\lambda_\xi < \lambda_\phi$ persists under strong noise. In the post-scattering regime, because strong noise severely suppresses wave  transport \cite{PhysRevLett.119.046601,yang2026noiseinducedresurrectiondynamicalskin}, the deviation state is strongly confined within the narrow spatial range $l$ of the scattering potential, with negligible amplitude leaking into the bulk. In contrast, the reference state $|\phi(t_1)\rangle$ exhibits a broader spatial profile, characterized by an exponential tail that extends deep into the lattice.
Consequently, the normalized profile of the spatially truncated deviation state exhibits steeper density gradients at the boundary. By Eq.~(\ref{gamma}), such a gradient yields a larger negative contribution to the instantaneous growth rate, leading to a smaller $\zeta_\xi$ compared to $\zeta_\phi$ and thereby pulling $\lambda_\xi$ below $\lambda_\phi$. Combined with the universal $1/t$ convergence to a common asymptotic FTLE, the cumulative difference $t[\lambda_\xi(t) - \lambda_\phi(t)]$ inevitably saturates to a finite negative constant, as seen in Fig.~\ref{fig2}(d). Through the exponential nature of the proxy relation in Eq.~\eqref{epsilon}, even a finite negative saturation of $t[\lambda_\xi(t) - \lambda_\phi(t)]$ ensures that $\langle\eta(t)\rangle$ is persistently suppressed to a sufficiently small value. This analytically establishes robust asymptotic self-healing in the strong-noise regime.

Building upon this spatial truncation picture, we deduce that more extended skin modes inherently possess superior self-healing capabilities. For a fixed scattering range $l$, an extended mode undergoes more severe spatial truncation, strongly suppressing the initial $\lambda_\xi(t_1)$. This yields a steeper negative initial slope for $t[\lambda_\xi(t) - \lambda_\phi(t)]$ and drives its long-time saturation to a more negative value, ensuring stronger suppression of the profile mismatch. Notably, skin modes become increasingly extended as their eigenenergies approach the spectral edges (see Sec. V of SM~\footnotemark[\value{footnote}]). This explains why eigenstates closer to the spectral edges exhibit significantly better self-healing, as observed in Fig.~\ref{fig1}.

\Blue{\textit{Conclusion}}.---We have demonstrated that stochastic noise can constructively enhance edge self-healing dynamics in non-Hermitian systems. Weak noise extends the self-healing window by elevating the reference state’s FTLE. In the strong-noise limit, the dynamics are governed by an effective drift-diffusion equation, yielding a universal $1/t$ convergence of all FTLEs that freezes the relative deviation at a minimal level and guarantees robust profile recovery.

Crucially, this noise-enhanced edge self-healing property stands in stark contrast to purely coherent non-Hermitian dynamical phenomena. As detailed in the Sec.~VII of SM~\footnotemark[\value{footnote}], previously theorized non-Hermitian coherent dynamical behaviors—such as bulk self-healing—are intrinsically fragile. Due to the dephasing effect, this phenomenon rapidly collapses even under extremely weak noise. By delineating these fragile coherent effects from the robust noise-driven dynamical effects, our study bridges the gap between idealized non-Hermitian theory and realistic noisy environments.

Looking forward, this theoretical framework can be extended to investigate dynamical self-healing phenomena in higher-dimensional systems, as well as to explore the self-healing capabilities associated with various distinct types of NHSE. Experimentally, the robust self-healing mechanisms uncovered here are highly amenable to realization in tunable platforms like topolectrical circuits, active photonic lattices, and acoustic metamaterials, paving the way for robust, defect-immune wave-steering devices.
%%%%%%%%%%%%%%%%%%%%%%%%%%%%%

\begin{acknowledgments}
This work is supported by the National Key R\&D Program of China (Grant No. 2021YFA1401600) and the National Natural Science Foundation of China (Grant No. 12474056). The work was carried out at the National Supercomputer Center in Tianjin, and the calculations were performed on Tianhe new generation supercomputer. The high-performance computing platform of Peking University supported the computational resources.
\end{acknowledgments}

\bibliography{reference}% Produces the bibliography via BibTeX.

\begin{thebibliography}{3}%
\makeatletter
\providecommand \@ifxundefined [1]{%
 \@ifx{#1\undefined}
}%
\providecommand \@ifnum [1]{%
 \ifnum #1\expandafter \@firstoftwo
 \else \expandafter \@secondoftwo
 \fi
}%
\providecommand \@ifx [1]{%
 \ifx #1\expandafter \@firstoftwo
 \else \expandafter \@secondoftwo
 \fi
}%
\providecommand \natexlab [1]{#1}%
\providecommand \enquote  [1]{``#1''}%
\providecommand \bibnamefont  [1]{#1}%
\providecommand \bibfnamefont [1]{#1}%
\providecommand \citenamefont [1]{#1}%
\providecommand \href@noop [0]{\@secondoftwo}%
\providecommand \href [0]{\begingroup \@sanitize@url \@href}%
\providecommand \@href[1]{\@@startlink{#1}\@@href}%
\providecommand \@@href[1]{\endgroup#1\@@endlink}%
\providecommand \@sanitize@url [0]{\catcode `\\12\catcode `\$12\catcode `\&12\catcode `\#12\catcode `\^12\catcode `\_12\catcode `\%12\relax}%
\providecommand \@@startlink[1]{}%
\providecommand \@@endlink[0]{}%
\providecommand \url  [0]{\begingroup\@sanitize@url \@url }%
\providecommand \@url [1]{\endgroup\@href {#1}{\urlprefix }}%
\providecommand \urlprefix  [0]{URL }%
\providecommand \Eprint [0]{\href }%
\providecommand \doibase [0]{https://doi.org/}%
\providecommand \selectlanguage [0]{\@gobble}%
\providecommand \bibinfo  [0]{\@secondoftwo}%
\providecommand \bibfield  [0]{\@secondoftwo}%
\providecommand \translation [1]{[#1]}%
\providecommand \BibitemOpen [0]{}%
\providecommand \bibitemStop [0]{}%
\providecommand \bibitemNoStop [0]{.\EOS\space}%
\providecommand \EOS [0]{\spacefactor3000\relax}%
\providecommand \BibitemShut  [1]{\csname bibitem#1\endcsname}%
\let\auto@bib@innerbib\@empty
%</preamble>
\bibitem [{\citenamefont {Yang}\ and\ \citenamefont {Huang}(2025)}]{PhysRevB.111.155121}%
  \BibitemOpen
  \bibfield  {author} {\bibinfo {author} {\bibfnamefont {W.}~\bibnamefont {Yang}}\ and\ \bibinfo {author} {\bibfnamefont {H.}~\bibnamefont {Huang}},\ }\bibfield  {title} {\bibinfo {title} {Unified multipole bott indices for non-hermitian skin effect in different orders},\ }\href {https://doi.org/10.1103/PhysRevB.111.155121} {\bibfield  {journal} {\bibinfo  {journal} {Phys. Rev. B}\ }\textbf {\bibinfo {volume} {111}},\ \bibinfo {pages} {155121} (\bibinfo {year} {2025})}\BibitemShut {NoStop}%
\bibitem [{\citenamefont {Yang}\ and\ \citenamefont {Fang}(2025)}]{llbb-pcgk}%
  \BibitemOpen
  \bibfield  {author} {\bibinfo {author} {\bibfnamefont {T.-H.}\ \bibnamefont {Yang}}\ and\ \bibinfo {author} {\bibfnamefont {C.}~\bibnamefont {Fang}},\ }\bibfield  {title} {\bibinfo {title} {Real-time edge dynamics of non-hermitian lattices},\ }\href {https://doi.org/10.1103/llbb-pcgk} {\bibfield  {journal} {\bibinfo  {journal} {Phys. Rev. Lett.}\ }\textbf {\bibinfo {volume} {135}},\ \bibinfo {pages} {186401} (\bibinfo {year} {2025})}\BibitemShut {NoStop}%
\bibitem [{\citenamefont {Xue}\ \emph {et~al.}(2025)\citenamefont {Xue}, \citenamefont {Song}, \citenamefont {Hu},\ and\ \citenamefont {Wang}}]{xue2025nonblochedgedynamicsnonhermitian}%
  \BibitemOpen
  \bibfield  {author} {\bibinfo {author} {\bibfnamefont {W.-T.}\ \bibnamefont {Xue}}, \bibinfo {author} {\bibfnamefont {F.}~\bibnamefont {Song}}, \bibinfo {author} {\bibfnamefont {Y.-M.}\ \bibnamefont {Hu}},\ and\ \bibinfo {author} {\bibfnamefont {Z.}~\bibnamefont {Wang}},\ }\href {https://arxiv.org/abs/2503.13671} {\bibinfo {title} {Non-bloch edge dynamics of non-hermitian lattices}} (\bibinfo {year} {2025}),\ \Eprint {https://arxiv.org/abs/2503.13671} {arXiv:2503.13671 [quant-ph]} \BibitemShut {NoStop}%
\end{thebibliography}%


\begin{thebibliography}{59}%
\makeatletter
\providecommand \@ifxundefined [1]{%
 \@ifx{#1\undefined}
}%
\providecommand \@ifnum [1]{%
 \ifnum #1\expandafter \@firstoftwo
 \else \expandafter \@secondoftwo
 \fi
}%
\providecommand \@ifx [1]{%
 \ifx #1\expandafter \@firstoftwo
 \else \expandafter \@secondoftwo
 \fi
}%
\providecommand \natexlab [1]{#1}%
\providecommand \enquote  [1]{``#1''}%
\providecommand \bibnamefont  [1]{#1}%
\providecommand \bibfnamefont [1]{#1}%
\providecommand \citenamefont [1]{#1}%
\providecommand \href@noop [0]{\@secondoftwo}%
\providecommand \href [0]{\begingroup \@sanitize@url \@href}%
\providecommand \@href[1]{\@@startlink{#1}\@@href}%
\providecommand \@@href[1]{\endgroup#1\@@endlink}%
\providecommand \@sanitize@url [0]{\catcode `\\12\catcode `\$12\catcode `\&12\catcode `\#12\catcode `\^12\catcode `\_12\catcode `\%12\relax}%
\providecommand \@@startlink[1]{}%
\providecommand \@@endlink[0]{}%
\providecommand \url  [0]{\begingroup\@sanitize@url \@url }%
\providecommand \@url [1]{\endgroup\@href {#1}{\urlprefix }}%
\providecommand \urlprefix  [0]{URL }%
\providecommand \Eprint [0]{\href }%
\providecommand \doibase [0]{https://doi.org/}%
\providecommand \selectlanguage [0]{\@gobble}%
\providecommand \bibinfo  [0]{\@secondoftwo}%
\providecommand \bibfield  [0]{\@secondoftwo}%
\providecommand \translation [1]{[#1]}%
\providecommand \BibitemOpen [0]{}%
\providecommand \bibitemStop [0]{}%
\providecommand \bibitemNoStop [0]{.\EOS\space}%
\providecommand \EOS [0]{\spacefactor3000\relax}%
\providecommand \BibitemShut  [1]{\csname bibitem#1\endcsname}%
\let\auto@bib@innerbib\@empty
%</preamble>
\bibitem [{\citenamefont {Bergholtz}\ \emph {et~al.}(2021)\citenamefont {Bergholtz}, \citenamefont {Budich},\ and\ \citenamefont {Kunst}}]{RevModPhys.93.015005}%
  \BibitemOpen
  \bibfield  {author} {\bibinfo {author} {\bibfnamefont {E.~J.}\ \bibnamefont {Bergholtz}}, \bibinfo {author} {\bibfnamefont {J.~C.}\ \bibnamefont {Budich}},\ and\ \bibinfo {author} {\bibfnamefont {F.~K.}\ \bibnamefont {Kunst}},\ }\bibfield  {title} {\bibinfo {title} {Exceptional topology of non-hermitian systems},\ }\href {https://doi.org/10.1103/RevModPhys.93.015005} {\bibfield  {journal} {\bibinfo  {journal} {Rev. Mod. Phys.}\ }\textbf {\bibinfo {volume} {93}},\ \bibinfo {pages} {015005} (\bibinfo {year} {2021})}\BibitemShut {NoStop}%
\bibitem [{\citenamefont {Yuto~Ashida}\ and\ \citenamefont {Ueda}(2020)}]{doi:10.1080/00018732.2021.1876991}%
  \BibitemOpen
  \bibfield  {author} {\bibinfo {author} {\bibfnamefont {Z.~G.}\ \bibnamefont {Yuto~Ashida}}\ and\ \bibinfo {author} {\bibfnamefont {M.}~\bibnamefont {Ueda}},\ }\bibfield  {title} {\bibinfo {title} {Non-hermitian physics},\ }\href {https://doi.org/10.1080/00018732.2021.1876991} {\bibfield  {journal} {\bibinfo  {journal} {Adv. Phys.}\ }\textbf {\bibinfo {volume} {69}},\ \bibinfo {pages} {249} (\bibinfo {year} {2020})}\BibitemShut {NoStop}%
\bibitem [{\citenamefont {Persson}\ \emph {et~al.}(2000)\citenamefont {Persson}, \citenamefont {Rotter}, \citenamefont {St\"ockmann},\ and\ \citenamefont {Barth}}]{PhysRevLett.85.2478}%
  \BibitemOpen
  \bibfield  {author} {\bibinfo {author} {\bibfnamefont {E.}~\bibnamefont {Persson}}, \bibinfo {author} {\bibfnamefont {I.}~\bibnamefont {Rotter}}, \bibinfo {author} {\bibfnamefont {H.-J.}\ \bibnamefont {St\"ockmann}},\ and\ \bibinfo {author} {\bibfnamefont {M.}~\bibnamefont {Barth}},\ }\bibfield  {title} {\bibinfo {title} {Observation of resonance trapping in an open microwave cavity},\ }\href {https://doi.org/10.1103/PhysRevLett.85.2478} {\bibfield  {journal} {\bibinfo  {journal} {Phys. Rev. Lett.}\ }\textbf {\bibinfo {volume} {85}},\ \bibinfo {pages} {2478} (\bibinfo {year} {2000})}\BibitemShut {NoStop}%
\bibitem [{\citenamefont {Volya}\ and\ \citenamefont {Zelevinsky}(2003)}]{PhysRevC.67.054322}%
  \BibitemOpen
  \bibfield  {author} {\bibinfo {author} {\bibfnamefont {A.}~\bibnamefont {Volya}}\ and\ \bibinfo {author} {\bibfnamefont {V.}~\bibnamefont {Zelevinsky}},\ }\bibfield  {title} {\bibinfo {title} {Non-hermitian effective hamiltonian and continuum shell model},\ }\href {https://doi.org/10.1103/PhysRevC.67.054322} {\bibfield  {journal} {\bibinfo  {journal} {Phys. Rev. C}\ }\textbf {\bibinfo {volume} {67}},\ \bibinfo {pages} {054322} (\bibinfo {year} {2003})}\BibitemShut {NoStop}%
\bibitem [{\citenamefont {Rotter}(2009)}]{Rotter_2009}%
  \BibitemOpen
  \bibfield  {author} {\bibinfo {author} {\bibfnamefont {I.}~\bibnamefont {Rotter}},\ }\bibfield  {title} {\bibinfo {title} {A non-hermitian hamilton operator and the physics of open quantum systems},\ }\href {https://doi.org/10.1088/1751-8113/42/15/153001} {\bibfield  {journal} {\bibinfo  {journal} {J. Phys. A Math. Theor.}\ }\textbf {\bibinfo {volume} {42}},\ \bibinfo {pages} {153001} (\bibinfo {year} {2009})}\BibitemShut {NoStop}%
\bibitem [{\citenamefont {Reiter}\ and\ \citenamefont {S\o{}rensen}(2012)}]{PhysRevA.85.032111}%
  \BibitemOpen
  \bibfield  {author} {\bibinfo {author} {\bibfnamefont {F.}~\bibnamefont {Reiter}}\ and\ \bibinfo {author} {\bibfnamefont {A.~S.}\ \bibnamefont {S\o{}rensen}},\ }\bibfield  {title} {\bibinfo {title} {Effective operator formalism for open quantum systems},\ }\href {https://doi.org/10.1103/PhysRevA.85.032111} {\bibfield  {journal} {\bibinfo  {journal} {Phys. Rev. A}\ }\textbf {\bibinfo {volume} {85}},\ \bibinfo {pages} {032111} (\bibinfo {year} {2012})}\BibitemShut {NoStop}%
\bibitem [{\citenamefont {Rotter}\ and\ \citenamefont {Bird}(2015)}]{Rotter_2015}%
  \BibitemOpen
  \bibfield  {author} {\bibinfo {author} {\bibfnamefont {I.}~\bibnamefont {Rotter}}\ and\ \bibinfo {author} {\bibfnamefont {J.~P.}\ \bibnamefont {Bird}},\ }\bibfield  {title} {\bibinfo {title} {A review of progress in the physics of open quantum systems: theory and experiment},\ }\href {https://doi.org/10.1088/0034-4885/78/11/114001} {\bibfield  {journal} {\bibinfo  {journal} {Rep. Prog. Phys.}\ }\textbf {\bibinfo {volume} {78}},\ \bibinfo {pages} {114001} (\bibinfo {year} {2015})}\BibitemShut {NoStop}%
\bibitem [{\citenamefont {Ozawa}\ \emph {et~al.}(2019)\citenamefont {Ozawa}, \citenamefont {Price}, \citenamefont {Amo}, \citenamefont {Goldman}, \citenamefont {Hafezi}, \citenamefont {Lu}, \citenamefont {Rechtsman}, \citenamefont {Schuster}, \citenamefont {Simon}, \citenamefont {Zilberberg},\ and\ \citenamefont {Carusotto}}]{RevModPhys.91.015006}%
  \BibitemOpen
  \bibfield  {author} {\bibinfo {author} {\bibfnamefont {T.}~\bibnamefont {Ozawa}}, \bibinfo {author} {\bibfnamefont {H.~M.}\ \bibnamefont {Price}}, \bibinfo {author} {\bibfnamefont {A.}~\bibnamefont {Amo}}, \bibinfo {author} {\bibfnamefont {N.}~\bibnamefont {Goldman}}, \bibinfo {author} {\bibfnamefont {M.}~\bibnamefont {Hafezi}}, \bibinfo {author} {\bibfnamefont {L.}~\bibnamefont {Lu}}, \bibinfo {author} {\bibfnamefont {M.~C.}\ \bibnamefont {Rechtsman}}, \bibinfo {author} {\bibfnamefont {D.}~\bibnamefont {Schuster}}, \bibinfo {author} {\bibfnamefont {J.}~\bibnamefont {Simon}}, \bibinfo {author} {\bibfnamefont {O.}~\bibnamefont {Zilberberg}},\ and\ \bibinfo {author} {\bibfnamefont {I.}~\bibnamefont {Carusotto}},\ }\bibfield  {title} {\bibinfo {title} {Topological photonics},\ }\href {https://doi.org/10.1103/RevModPhys.91.015006} {\bibfield  {journal} {\bibinfo  {journal} {Rev. Mod. Phys.}\ }\textbf {\bibinfo {volume} {91}},\ \bibinfo {pages} {015006} (\bibinfo {year} {2019})}\BibitemShut {NoStop}%
\bibitem [{\citenamefont {Feng}\ \emph {et~al.}(2017)\citenamefont {Feng}, \citenamefont {El-Ganainy},\ and\ \citenamefont {Ge}}]{feng2017Non}%
  \BibitemOpen
  \bibfield  {author} {\bibinfo {author} {\bibfnamefont {L.}~\bibnamefont {Feng}}, \bibinfo {author} {\bibfnamefont {R.}~\bibnamefont {El-Ganainy}},\ and\ \bibinfo {author} {\bibfnamefont {L.}~\bibnamefont {Ge}},\ }\bibfield  {title} {\bibinfo {title} {Non-hermitian photonics based on parity--time symmetry},\ }\href {https://doi.org/https://doi.org/10.1038/s41566-017-0031-1} {\bibfield  {journal} {\bibinfo  {journal} {Nat. Photon.}\ }\textbf {\bibinfo {volume} {11}},\ \bibinfo {pages} {752} (\bibinfo {year} {2017})}\BibitemShut {NoStop}%
\bibitem [{\citenamefont {A.~Regensburger}\ and\ \citenamefont {Peschel}(2012)}]{AosRnsurr2012Parity}%
  \BibitemOpen
  \bibfield  {author} {\bibinfo {author} {\bibfnamefont {M.-A. M. G. O. D. N.~C.}\ \bibnamefont {A.~Regensburger}, \bibfnamefont {C.~Bersch}}\ and\ \bibinfo {author} {\bibfnamefont {U.}~\bibnamefont {Peschel}},\ }\bibfield  {title} {\bibinfo {title} {Parity–time synthetic photonic lattices},\ }\href {https://doi.org/https://doi.org/10.1038/nature11298} {\bibfield  {journal} {\bibinfo  {journal} {Nature}\ }\textbf {\bibinfo {volume} {488}},\ \bibinfo {pages} {167} (\bibinfo {year} {2012})}\BibitemShut {NoStop}%
\bibitem [{\citenamefont {Feng}\ \emph {et~al.}(2014)\citenamefont {Feng}, \citenamefont {Wong}, \citenamefont {Ma}, \citenamefont {Wang},\ and\ \citenamefont {Zhang}}]{doi:10.1126/science.1258479}%
  \BibitemOpen
  \bibfield  {author} {\bibinfo {author} {\bibfnamefont {L.}~\bibnamefont {Feng}}, \bibinfo {author} {\bibfnamefont {Z.~J.}\ \bibnamefont {Wong}}, \bibinfo {author} {\bibfnamefont {R.-M.}\ \bibnamefont {Ma}}, \bibinfo {author} {\bibfnamefont {Y.}~\bibnamefont {Wang}},\ and\ \bibinfo {author} {\bibfnamefont {X.}~\bibnamefont {Zhang}},\ }\bibfield  {title} {\bibinfo {title} {Single-mode laser by parity-time symmetry breaking},\ }\href {https://doi.org/10.1126/science.1258479} {\bibfield  {journal} {\bibinfo  {journal} {Science}\ }\textbf {\bibinfo {volume} {346}},\ \bibinfo {pages} {972} (\bibinfo {year} {2014})}\BibitemShut {NoStop}%
\bibitem [{\citenamefont {Zhen}\ \emph {et~al.}(2015)\citenamefont {Zhen}, \citenamefont {Hsu}, \citenamefont {Igarashi}, \citenamefont {Lu}, \citenamefont {Kaminer}, \citenamefont {Pick}, \citenamefont {Chua}, \citenamefont {Joannopoulos},\ and\ \citenamefont {Soljai}}]{2015Spawning}%
  \BibitemOpen
  \bibfield  {author} {\bibinfo {author} {\bibfnamefont {B.}~\bibnamefont {Zhen}}, \bibinfo {author} {\bibfnamefont {C.~W.}\ \bibnamefont {Hsu}}, \bibinfo {author} {\bibfnamefont {Y.}~\bibnamefont {Igarashi}}, \bibinfo {author} {\bibfnamefont {L.}~\bibnamefont {Lu}}, \bibinfo {author} {\bibfnamefont {I.}~\bibnamefont {Kaminer}}, \bibinfo {author} {\bibfnamefont {A.}~\bibnamefont {Pick}}, \bibinfo {author} {\bibfnamefont {S.~L.}\ \bibnamefont {Chua}}, \bibinfo {author} {\bibfnamefont {J.~D.}\ \bibnamefont {Joannopoulos}},\ and\ \bibinfo {author} {\bibfnamefont {M.}~\bibnamefont {Soljai}},\ }\bibfield  {title} {\bibinfo {title} {Spawning rings of exceptional points out of dirac cones},\ }\href {https://doi.org/https://doi.org/10.1038/nature14889} {\bibfield  {journal} {\bibinfo  {journal} {Nature}\ }\textbf {\bibinfo {volume} {525}},\ \bibinfo {pages} {354} (\bibinfo {year} {2015})}\BibitemShut {NoStop}%
\bibitem [{\citenamefont {Zhou}\ \emph {et~al.}(2018)\citenamefont {Zhou}, \citenamefont {Peng}, \citenamefont {Yoon}, \citenamefont {Hsu}, \citenamefont {Nelson}, \citenamefont {Fu}, \citenamefont {Joannopoulos}, \citenamefont {Soljačić},\ and\ \citenamefont {Zhen}}]{doi:10.1126/science.aap9859}%
  \BibitemOpen
  \bibfield  {author} {\bibinfo {author} {\bibfnamefont {H.}~\bibnamefont {Zhou}}, \bibinfo {author} {\bibfnamefont {C.}~\bibnamefont {Peng}}, \bibinfo {author} {\bibfnamefont {Y.}~\bibnamefont {Yoon}}, \bibinfo {author} {\bibfnamefont {C.~W.}\ \bibnamefont {Hsu}}, \bibinfo {author} {\bibfnamefont {K.~A.}\ \bibnamefont {Nelson}}, \bibinfo {author} {\bibfnamefont {L.}~\bibnamefont {Fu}}, \bibinfo {author} {\bibfnamefont {J.~D.}\ \bibnamefont {Joannopoulos}}, \bibinfo {author} {\bibfnamefont {M.}~\bibnamefont {Soljačić}},\ and\ \bibinfo {author} {\bibfnamefont {B.}~\bibnamefont {Zhen}},\ }\bibfield  {title} {\bibinfo {title} {Observation of bulk fermi arc and polarization half charge from paired exceptional points},\ }\href {https://doi.org/10.1126/science.aap9859} {\bibfield  {journal} {\bibinfo  {journal} {Science}\ }\textbf {\bibinfo {volume} {359}},\ \bibinfo {pages} {1009} (\bibinfo {year} {2018})},\ \Eprint {https://arxiv.org/abs/https://www.science.org/doi/pdf/10.1126/science.aap9859}
  {https://www.science.org/doi/pdf/10.1126/science.aap9859} \BibitemShut {NoStop}%
\bibitem [{\citenamefont {Bandres}\ \emph {et~al.}(2018)\citenamefont {Bandres}, \citenamefont {Wittek}, \citenamefont {Harari}, \citenamefont {Parto}, \citenamefont {Ren}, \citenamefont {Segev}, \citenamefont {Christodoulides},\ and\ \citenamefont {Khajavikhan}}]{doi:10.1126/science.aar4005}%
  \BibitemOpen
  \bibfield  {author} {\bibinfo {author} {\bibfnamefont {M.~A.}\ \bibnamefont {Bandres}}, \bibinfo {author} {\bibfnamefont {S.}~\bibnamefont {Wittek}}, \bibinfo {author} {\bibfnamefont {G.}~\bibnamefont {Harari}}, \bibinfo {author} {\bibfnamefont {M.}~\bibnamefont {Parto}}, \bibinfo {author} {\bibfnamefont {J.}~\bibnamefont {Ren}}, \bibinfo {author} {\bibfnamefont {M.}~\bibnamefont {Segev}}, \bibinfo {author} {\bibfnamefont {D.~N.}\ \bibnamefont {Christodoulides}},\ and\ \bibinfo {author} {\bibfnamefont {M.}~\bibnamefont {Khajavikhan}},\ }\bibfield  {title} {\bibinfo {title} {Topological insulator laser: Experiments},\ }\href {https://doi.org/10.1126/science.aar4005} {\bibfield  {journal} {\bibinfo  {journal} {Science}\ }\textbf {\bibinfo {volume} {359}},\ \bibinfo {pages} {eaar4005} (\bibinfo {year} {2018})}\BibitemShut {NoStop}%
\bibitem [{\citenamefont {Nasari}\ \emph {et~al.}(2023)\citenamefont {Nasari}, \citenamefont {Pyrialakos}, \citenamefont {Christodoulides},\ and\ \citenamefont {Khajavikhan}}]{Nasari:23}%
  \BibitemOpen
  \bibfield  {author} {\bibinfo {author} {\bibfnamefont {H.}~\bibnamefont {Nasari}}, \bibinfo {author} {\bibfnamefont {G.~G.}\ \bibnamefont {Pyrialakos}}, \bibinfo {author} {\bibfnamefont {D.~N.}\ \bibnamefont {Christodoulides}},\ and\ \bibinfo {author} {\bibfnamefont {M.}~\bibnamefont {Khajavikhan}},\ }\bibfield  {title} {\bibinfo {title} {Non-hermitian topological photonics},\ }\href {https://doi.org/10.1364/OME.483361} {\bibfield  {journal} {\bibinfo  {journal} {Opt. Mater. Express}\ }\textbf {\bibinfo {volume} {13}},\ \bibinfo {pages} {870} (\bibinfo {year} {2023})}\BibitemShut {NoStop}%
\bibitem [{\citenamefont {Wang}\ \emph {et~al.}(2021)\citenamefont {Wang}, \citenamefont {Li}, \citenamefont {Xiao}, \citenamefont {Han}, \citenamefont {Yi},\ and\ \citenamefont {Xue}}]{PhysRevLett.127.270602}%
  \BibitemOpen
  \bibfield  {author} {\bibinfo {author} {\bibfnamefont {K.}~\bibnamefont {Wang}}, \bibinfo {author} {\bibfnamefont {T.}~\bibnamefont {Li}}, \bibinfo {author} {\bibfnamefont {L.}~\bibnamefont {Xiao}}, \bibinfo {author} {\bibfnamefont {Y.}~\bibnamefont {Han}}, \bibinfo {author} {\bibfnamefont {W.}~\bibnamefont {Yi}},\ and\ \bibinfo {author} {\bibfnamefont {P.}~\bibnamefont {Xue}},\ }\bibfield  {title} {\bibinfo {title} {Detecting non-bloch topological invariants in quantum dynamics},\ }\href {https://doi.org/10.1103/PhysRevLett.127.270602} {\bibfield  {journal} {\bibinfo  {journal} {Phys. Rev. Lett.}\ }\textbf {\bibinfo {volume} {127}},\ \bibinfo {pages} {270602} (\bibinfo {year} {2021})}\BibitemShut {NoStop}%
\bibitem [{\citenamefont {Cerjan}\ \emph {et~al.}(2019)\citenamefont {Cerjan}, \citenamefont {Huang}, \citenamefont {Chong},\ and\ \citenamefont {Rechtsman}}]{cerjan2019experimental}%
  \BibitemOpen
  \bibfield  {author} {\bibinfo {author} {\bibfnamefont {A.}~\bibnamefont {Cerjan}}, \bibinfo {author} {\bibfnamefont {S.}~\bibnamefont {Huang}}, \bibinfo {author} {\bibfnamefont {Y.}~\bibnamefont {Chong}},\ and\ \bibinfo {author} {\bibfnamefont {M.}~\bibnamefont {Rechtsman}},\ }\bibfield  {title} {\bibinfo {title} {Experimental realization of a weyl exceptional ring},\ }\href {https://doi.org/https://doi.org/10.1038/s41566-019-0453-z} {\bibfield  {journal} {\bibinfo  {journal} {Nat. Photonics}\ }\textbf {\bibinfo {volume} {13}},\ \bibinfo {pages} {623} (\bibinfo {year} {2019})}\BibitemShut {NoStop}%
\bibitem [{\citenamefont {Ding}\ \emph {et~al.}(2016)\citenamefont {Ding}, \citenamefont {Ma}, \citenamefont {Xiao}, \citenamefont {Zhang},\ and\ \citenamefont {Chan}}]{PhysRevX.6.021007}%
  \BibitemOpen
  \bibfield  {author} {\bibinfo {author} {\bibfnamefont {K.}~\bibnamefont {Ding}}, \bibinfo {author} {\bibfnamefont {G.}~\bibnamefont {Ma}}, \bibinfo {author} {\bibfnamefont {M.}~\bibnamefont {Xiao}}, \bibinfo {author} {\bibfnamefont {Z.~Q.}\ \bibnamefont {Zhang}},\ and\ \bibinfo {author} {\bibfnamefont {C.~T.}\ \bibnamefont {Chan}},\ }\bibfield  {title} {\bibinfo {title} {Emergence, coalescence, and topological properties of multiple exceptional points and their experimental realization},\ }\href {https://doi.org/10.1103/PhysRevX.6.021007} {\bibfield  {journal} {\bibinfo  {journal} {Phys. Rev. X}\ }\textbf {\bibinfo {volume} {6}},\ \bibinfo {pages} {021007} (\bibinfo {year} {2016})}\BibitemShut {NoStop}%
\bibitem [{\citenamefont {Tang}\ \emph {et~al.}(2020)\citenamefont {Tang}, \citenamefont {Jiang}, \citenamefont {Ding}, \citenamefont {Xiao}, \citenamefont {Zhang}, \citenamefont {Chan},\ and\ \citenamefont {Ma}}]{doi:10.1126/science.abd8872}%
  \BibitemOpen
  \bibfield  {author} {\bibinfo {author} {\bibfnamefont {W.}~\bibnamefont {Tang}}, \bibinfo {author} {\bibfnamefont {X.}~\bibnamefont {Jiang}}, \bibinfo {author} {\bibfnamefont {K.}~\bibnamefont {Ding}}, \bibinfo {author} {\bibfnamefont {Y.-X.}\ \bibnamefont {Xiao}}, \bibinfo {author} {\bibfnamefont {Z.-Q.}\ \bibnamefont {Zhang}}, \bibinfo {author} {\bibfnamefont {C.~T.}\ \bibnamefont {Chan}},\ and\ \bibinfo {author} {\bibfnamefont {G.}~\bibnamefont {Ma}},\ }\bibfield  {title} {\bibinfo {title} {Exceptional nexus with a hybrid topological invariant},\ }\href {https://doi.org/10.1126/science.abd8872} {\bibfield  {journal} {\bibinfo  {journal} {Science}\ }\textbf {\bibinfo {volume} {370}},\ \bibinfo {pages} {1077} (\bibinfo {year} {2020})}\BibitemShut {NoStop}%
\bibitem [{\citenamefont {Ding}\ \emph {et~al.}(2018)\citenamefont {Ding}, \citenamefont {Ma}, \citenamefont {Zhang},\ and\ \citenamefont {Chan}}]{PhysRevLett.121.085702}%
  \BibitemOpen
  \bibfield  {author} {\bibinfo {author} {\bibfnamefont {K.}~\bibnamefont {Ding}}, \bibinfo {author} {\bibfnamefont {G.}~\bibnamefont {Ma}}, \bibinfo {author} {\bibfnamefont {Z.~Q.}\ \bibnamefont {Zhang}},\ and\ \bibinfo {author} {\bibfnamefont {C.~T.}\ \bibnamefont {Chan}},\ }\bibfield  {title} {\bibinfo {title} {Experimental demonstration of an anisotropic exceptional point},\ }\href {https://doi.org/10.1103/PhysRevLett.121.085702} {\bibfield  {journal} {\bibinfo  {journal} {Phys. Rev. Lett.}\ }\textbf {\bibinfo {volume} {121}},\ \bibinfo {pages} {085702} (\bibinfo {year} {2018})}\BibitemShut {NoStop}%
\bibitem [{\citenamefont {Tang}\ \emph {et~al.}(2021)\citenamefont {Tang}, \citenamefont {Ding},\ and\ \citenamefont {Ma}}]{PhysRevLett.127.034301}%
  \BibitemOpen
  \bibfield  {author} {\bibinfo {author} {\bibfnamefont {W.}~\bibnamefont {Tang}}, \bibinfo {author} {\bibfnamefont {K.}~\bibnamefont {Ding}},\ and\ \bibinfo {author} {\bibfnamefont {G.}~\bibnamefont {Ma}},\ }\bibfield  {title} {\bibinfo {title} {Direct measurement of topological properties of an exceptional parabola},\ }\href {https://doi.org/10.1103/PhysRevLett.127.034301} {\bibfield  {journal} {\bibinfo  {journal} {Phys. Rev. Lett.}\ }\textbf {\bibinfo {volume} {127}},\ \bibinfo {pages} {034301} (\bibinfo {year} {2021})}\BibitemShut {NoStop}%
\bibitem [{\citenamefont {Aur\'egan}\ and\ \citenamefont {Pagneux}(2017)}]{PhysRevLett.118.174301}%
  \BibitemOpen
  \bibfield  {author} {\bibinfo {author} {\bibfnamefont {Y.}~\bibnamefont {Aur\'egan}}\ and\ \bibinfo {author} {\bibfnamefont {V.}~\bibnamefont {Pagneux}},\ }\bibfield  {title} {\bibinfo {title} {$\mathcal{P}\mathcal{T}$-symmetric scattering in flow duct acoustics},\ }\href {https://doi.org/10.1103/PhysRevLett.118.174301} {\bibfield  {journal} {\bibinfo  {journal} {Phys. Rev. Lett.}\ }\textbf {\bibinfo {volume} {118}},\ \bibinfo {pages} {174301} (\bibinfo {year} {2017})}\BibitemShut {NoStop}%
\bibitem [{\citenamefont {Puri}\ \emph {et~al.}(2021)\citenamefont {Puri}, \citenamefont {Ferdous}, \citenamefont {Shakeri}, \citenamefont {Basiri}, \citenamefont {Dubois},\ and\ \citenamefont {Ramezani}}]{PhysRevApplied.16.014012}%
  \BibitemOpen
  \bibfield  {author} {\bibinfo {author} {\bibfnamefont {S.}~\bibnamefont {Puri}}, \bibinfo {author} {\bibfnamefont {J.}~\bibnamefont {Ferdous}}, \bibinfo {author} {\bibfnamefont {A.}~\bibnamefont {Shakeri}}, \bibinfo {author} {\bibfnamefont {A.}~\bibnamefont {Basiri}}, \bibinfo {author} {\bibfnamefont {M.}~\bibnamefont {Dubois}},\ and\ \bibinfo {author} {\bibfnamefont {H.}~\bibnamefont {Ramezani}},\ }\bibfield  {title} {\bibinfo {title} {Tunable non-hermitian acoustic filter},\ }\href {https://doi.org/10.1103/PhysRevApplied.16.014012} {\bibfield  {journal} {\bibinfo  {journal} {Phys. Rev. Appl.}\ }\textbf {\bibinfo {volume} {16}},\ \bibinfo {pages} {014012} (\bibinfo {year} {2021})}\BibitemShut {NoStop}%
\bibitem [{\citenamefont {Helbig}\ \emph {et~al.}(2020)\citenamefont {Helbig}, \citenamefont {Hofmann}, \citenamefont {Imhof}, \citenamefont {Abdelghany},\ and\ \citenamefont {Thomale}}]{2020Generalized}%
  \BibitemOpen
  \bibfield  {author} {\bibinfo {author} {\bibfnamefont {T.}~\bibnamefont {Helbig}}, \bibinfo {author} {\bibfnamefont {T.}~\bibnamefont {Hofmann}}, \bibinfo {author} {\bibfnamefont {S.}~\bibnamefont {Imhof}}, \bibinfo {author} {\bibfnamefont {M.}~\bibnamefont {Abdelghany}},\ and\ \bibinfo {author} {\bibfnamefont {R.}~\bibnamefont {Thomale}},\ }\bibfield  {title} {\bibinfo {title} {Generalized bulk–boundary correspondence in non-hermitian topolectrical circuits},\ }\href {https://doi.org/https://doi.org/10.1038/s41567-020-0922-9} {\bibfield  {journal} {\bibinfo  {journal} {Nat. Phys.}\ }\textbf {\bibinfo {volume} {16}},\ \bibinfo {pages} {086601} (\bibinfo {year} {2020})}\BibitemShut {NoStop}%
\bibitem [{\citenamefont {Zou}\ \emph {et~al.}(2021)\citenamefont {Zou}, \citenamefont {Chen}, \citenamefont {He}, \citenamefont {Bao},\ and\ \citenamefont {Zhang}}]{0Observation}%
  \BibitemOpen
  \bibfield  {author} {\bibinfo {author} {\bibfnamefont {D.}~\bibnamefont {Zou}}, \bibinfo {author} {\bibfnamefont {T.}~\bibnamefont {Chen}}, \bibinfo {author} {\bibfnamefont {W.}~\bibnamefont {He}}, \bibinfo {author} {\bibfnamefont {J.}~\bibnamefont {Bao}},\ and\ \bibinfo {author} {\bibfnamefont {X.}~\bibnamefont {Zhang}},\ }\bibfield  {title} {\bibinfo {title} {Observation of hybrid higher-order skin-topological effect in non-hermitian topolectrical circuits},\ }\href {https://doi.org/https://doi.org/10.1038/s41467-021-26414-5} {\bibfield  {journal} {\bibinfo  {journal} {Nat. Commun.}\ }\textbf {\bibinfo {volume} {12}},\ \bibinfo {pages} {7201} (\bibinfo {year} {2021})}\BibitemShut {NoStop}%
\bibitem [{\citenamefont {Zhang}\ \emph {et~al.}(2023)\citenamefont {Zhang}, \citenamefont {Chen}, \citenamefont {Li}, \citenamefont {Lee},\ and\ \citenamefont {Zhang}}]{PhysRevB.107.085426}%
  \BibitemOpen
  \bibfield  {author} {\bibinfo {author} {\bibfnamefont {H.}~\bibnamefont {Zhang}}, \bibinfo {author} {\bibfnamefont {T.}~\bibnamefont {Chen}}, \bibinfo {author} {\bibfnamefont {L.}~\bibnamefont {Li}}, \bibinfo {author} {\bibfnamefont {C.~H.}\ \bibnamefont {Lee}},\ and\ \bibinfo {author} {\bibfnamefont {X.}~\bibnamefont {Zhang}},\ }\bibfield  {title} {\bibinfo {title} {Electrical circuit realization of topological switching for the non-hermitian skin effect},\ }\href {https://doi.org/10.1103/PhysRevB.107.085426} {\bibfield  {journal} {\bibinfo  {journal} {Phys. Rev. B}\ }\textbf {\bibinfo {volume} {107}},\ \bibinfo {pages} {085426} (\bibinfo {year} {2023})}\BibitemShut {NoStop}%
\bibitem [{\citenamefont {Yuan}\ \emph {et~al.}(2023)\citenamefont {Yuan}, \citenamefont {Zhang}, \citenamefont {Zhou}, \citenamefont {Wang}, \citenamefont {Pan}, \citenamefont {Feng}, \citenamefont {Sun},\ and\ \citenamefont {Zhang}}]{advs.202301128}%
  \BibitemOpen
  \bibfield  {author} {\bibinfo {author} {\bibfnamefont {H.}~\bibnamefont {Yuan}}, \bibinfo {author} {\bibfnamefont {W.}~\bibnamefont {Zhang}}, \bibinfo {author} {\bibfnamefont {Z.}~\bibnamefont {Zhou}}, \bibinfo {author} {\bibfnamefont {W.}~\bibnamefont {Wang}}, \bibinfo {author} {\bibfnamefont {N.}~\bibnamefont {Pan}}, \bibinfo {author} {\bibfnamefont {Y.}~\bibnamefont {Feng}}, \bibinfo {author} {\bibfnamefont {H.}~\bibnamefont {Sun}},\ and\ \bibinfo {author} {\bibfnamefont {X.}~\bibnamefont {Zhang}},\ }\bibfield  {title} {\bibinfo {title} {Non-hermitian topolectrical circuit sensor with high sensitivity},\ }\href {https://doi.org/https://doi.org/10.1002/advs.202301128} {\bibfield  {journal} {\bibinfo  {journal} {Adv. Sci.}\ }\textbf {\bibinfo {volume} {10}},\ \bibinfo {pages} {2301128} (\bibinfo {year} {2023})}\BibitemShut {NoStop}%
\bibitem [{\citenamefont {Zhao}\ \emph {et~al.}(2025)\citenamefont {Zhao}, \citenamefont {Wang}, \citenamefont {He}, \citenamefont {Poon}, \citenamefont {Pak}, \citenamefont {Liu}, \citenamefont {Ren}, \citenamefont {Liu},\ and\ \citenamefont {Jo}}]{zhao2025two}%
  \BibitemOpen
  \bibfield  {author} {\bibinfo {author} {\bibfnamefont {E.}~\bibnamefont {Zhao}}, \bibinfo {author} {\bibfnamefont {Z.}~\bibnamefont {Wang}}, \bibinfo {author} {\bibfnamefont {C.}~\bibnamefont {He}}, \bibinfo {author} {\bibfnamefont {T.~F.~J.}\ \bibnamefont {Poon}}, \bibinfo {author} {\bibfnamefont {K.~K.}\ \bibnamefont {Pak}}, \bibinfo {author} {\bibfnamefont {Y.-J.}\ \bibnamefont {Liu}}, \bibinfo {author} {\bibfnamefont {P.}~\bibnamefont {Ren}}, \bibinfo {author} {\bibfnamefont {X.-J.}\ \bibnamefont {Liu}},\ and\ \bibinfo {author} {\bibfnamefont {G.-B.}\ \bibnamefont {Jo}},\ }\bibfield  {title} {\bibinfo {title} {Two-dimensional non-hermitian skin effect in an ultracold fermi gas},\ }\href {https://doi.org/https://doi.org/10.1038/s41586-024-08347-3} {\bibfield  {journal} {\bibinfo  {journal} {Nature}\ }\textbf {\bibinfo {volume} {637}},\ \bibinfo {pages} {565} (\bibinfo {year} {2025})}\BibitemShut {NoStop}%
\bibitem [{\citenamefont {Öztürk}\ \emph {et~al.}(2021)\citenamefont {Öztürk}, \citenamefont {Lappe}, \citenamefont {Hellmann}, \citenamefont {Schmitt}, \citenamefont {Klaers}, \citenamefont {Vewinger}, \citenamefont {Kroha},\ and\ \citenamefont {Weitz}}]{doi:10.1126/science.abe9869}%
  \BibitemOpen
  \bibfield  {author} {\bibinfo {author} {\bibfnamefont {F.~E.}\ \bibnamefont {Öztürk}}, \bibinfo {author} {\bibfnamefont {T.}~\bibnamefont {Lappe}}, \bibinfo {author} {\bibfnamefont {G.}~\bibnamefont {Hellmann}}, \bibinfo {author} {\bibfnamefont {J.}~\bibnamefont {Schmitt}}, \bibinfo {author} {\bibfnamefont {J.}~\bibnamefont {Klaers}}, \bibinfo {author} {\bibfnamefont {F.}~\bibnamefont {Vewinger}}, \bibinfo {author} {\bibfnamefont {J.}~\bibnamefont {Kroha}},\ and\ \bibinfo {author} {\bibfnamefont {M.}~\bibnamefont {Weitz}},\ }\bibfield  {title} {\bibinfo {title} {Observation of a non-hermitian phase transition in an optical quantum gas},\ }\href {https://doi.org/10.1126/science.abe9869} {\bibfield  {journal} {\bibinfo  {journal} {Science}\ }\textbf {\bibinfo {volume} {372}},\ \bibinfo {pages} {88} (\bibinfo {year} {2021})},\ \Eprint {https://arxiv.org/abs/https://www.science.org/doi/pdf/10.1126/science.abe9869} {https://www.science.org/doi/pdf/10.1126/science.abe9869} \BibitemShut {NoStop}%
\bibitem [{\citenamefont {Yao}\ and\ \citenamefont {Wang}(2018)}]{PhysRevLett.121.086803}%
  \BibitemOpen
  \bibfield  {author} {\bibinfo {author} {\bibfnamefont {S.}~\bibnamefont {Yao}}\ and\ \bibinfo {author} {\bibfnamefont {Z.}~\bibnamefont {Wang}},\ }\bibfield  {title} {\bibinfo {title} {Edge states and topological invariants of non-hermitian systems},\ }\href {https://doi.org/10.1103/PhysRevLett.121.086803} {\bibfield  {journal} {\bibinfo  {journal} {Phys. Rev. Lett.}\ }\textbf {\bibinfo {volume} {121}},\ \bibinfo {pages} {086803} (\bibinfo {year} {2018})}\BibitemShut {NoStop}%
\bibitem [{\citenamefont {Okuma}\ \emph {et~al.}(2020)\citenamefont {Okuma}, \citenamefont {Kawabata}, \citenamefont {Shiozaki},\ and\ \citenamefont {Sato}}]{PhysRevLett.124.086801}%
  \BibitemOpen
  \bibfield  {author} {\bibinfo {author} {\bibfnamefont {N.}~\bibnamefont {Okuma}}, \bibinfo {author} {\bibfnamefont {K.}~\bibnamefont {Kawabata}}, \bibinfo {author} {\bibfnamefont {K.}~\bibnamefont {Shiozaki}},\ and\ \bibinfo {author} {\bibfnamefont {M.}~\bibnamefont {Sato}},\ }\bibfield  {title} {\bibinfo {title} {Topological origin of non-hermitian skin effects},\ }\href {https://doi.org/10.1103/PhysRevLett.124.086801} {\bibfield  {journal} {\bibinfo  {journal} {Phys. Rev. Lett.}\ }\textbf {\bibinfo {volume} {124}},\ \bibinfo {pages} {086801} (\bibinfo {year} {2020})}\BibitemShut {NoStop}%
\bibitem [{\citenamefont {Song}\ \emph {et~al.}(2019)\citenamefont {Song}, \citenamefont {Yao},\ and\ \citenamefont {Wang}}]{PhysRevLett.123.246801}%
  \BibitemOpen
  \bibfield  {author} {\bibinfo {author} {\bibfnamefont {F.}~\bibnamefont {Song}}, \bibinfo {author} {\bibfnamefont {S.}~\bibnamefont {Yao}},\ and\ \bibinfo {author} {\bibfnamefont {Z.}~\bibnamefont {Wang}},\ }\bibfield  {title} {\bibinfo {title} {Non-hermitian topological invariants in real space},\ }\href {https://doi.org/10.1103/PhysRevLett.123.246801} {\bibfield  {journal} {\bibinfo  {journal} {Phys. Rev. Lett.}\ }\textbf {\bibinfo {volume} {123}},\ \bibinfo {pages} {246801} (\bibinfo {year} {2019})}\BibitemShut {NoStop}%
\bibitem [{\citenamefont {Yang}\ \emph {et~al.}(2020)\citenamefont {Yang}, \citenamefont {Zhang}, \citenamefont {Fang},\ and\ \citenamefont {Hu}}]{PhysRevLett.125.226402}%
  \BibitemOpen
  \bibfield  {author} {\bibinfo {author} {\bibfnamefont {Z.}~\bibnamefont {Yang}}, \bibinfo {author} {\bibfnamefont {K.}~\bibnamefont {Zhang}}, \bibinfo {author} {\bibfnamefont {C.}~\bibnamefont {Fang}},\ and\ \bibinfo {author} {\bibfnamefont {J.}~\bibnamefont {Hu}},\ }\bibfield  {title} {\bibinfo {title} {Non-hermitian bulk-boundary correspondence and auxiliary generalized brillouin zone theory},\ }\href {https://doi.org/10.1103/PhysRevLett.125.226402} {\bibfield  {journal} {\bibinfo  {journal} {Phys. Rev. Lett.}\ }\textbf {\bibinfo {volume} {125}},\ \bibinfo {pages} {226402} (\bibinfo {year} {2020})}\BibitemShut {NoStop}%
\bibitem [{\citenamefont {Wang}\ \emph {et~al.}(2024)\citenamefont {Wang}, \citenamefont {Song},\ and\ \citenamefont {Wang}}]{PhysRevX.14.021011}%
  \BibitemOpen
  \bibfield  {author} {\bibinfo {author} {\bibfnamefont {H.-Y.}\ \bibnamefont {Wang}}, \bibinfo {author} {\bibfnamefont {F.}~\bibnamefont {Song}},\ and\ \bibinfo {author} {\bibfnamefont {Z.}~\bibnamefont {Wang}},\ }\bibfield  {title} {\bibinfo {title} {Amoeba formulation of non-bloch band theory in arbitrary dimensions},\ }\href {https://doi.org/10.1103/PhysRevX.14.021011} {\bibfield  {journal} {\bibinfo  {journal} {Phys. Rev. X}\ }\textbf {\bibinfo {volume} {14}},\ \bibinfo {pages} {021011} (\bibinfo {year} {2024})}\BibitemShut {NoStop}%
\bibitem [{\citenamefont {Zhang}\ \emph {et~al.}(2024)\citenamefont {Zhang}, \citenamefont {Yang},\ and\ \citenamefont {Sun}}]{PhysRevB.109.165127}%
  \BibitemOpen
  \bibfield  {author} {\bibinfo {author} {\bibfnamefont {K.}~\bibnamefont {Zhang}}, \bibinfo {author} {\bibfnamefont {Z.}~\bibnamefont {Yang}},\ and\ \bibinfo {author} {\bibfnamefont {K.}~\bibnamefont {Sun}},\ }\bibfield  {title} {\bibinfo {title} {Edge theory of non-hermitian skin modes in higher dimensions},\ }\href {https://doi.org/10.1103/PhysRevB.109.165127} {\bibfield  {journal} {\bibinfo  {journal} {Phys. Rev. B}\ }\textbf {\bibinfo {volume} {109}},\ \bibinfo {pages} {165127} (\bibinfo {year} {2024})}\BibitemShut {NoStop}%
\bibitem [{\citenamefont {Xiong}\ \emph {et~al.}(2024)\citenamefont {Xiong}, \citenamefont {Xing},\ and\ \citenamefont {Hu}}]{xiong2024nonhermitianskineffectarbitrary}%
  \BibitemOpen
  \bibfield  {author} {\bibinfo {author} {\bibfnamefont {Y.}~\bibnamefont {Xiong}}, \bibinfo {author} {\bibfnamefont {Z.-Y.}\ \bibnamefont {Xing}},\ and\ \bibinfo {author} {\bibfnamefont {H.}~\bibnamefont {Hu}},\ }\href {https://arxiv.org/abs/2407.01296} {\bibinfo {title} {Non-hermitian skin effect in arbitrary dimensions: non-bloch band theory and classification}} (\bibinfo {year} {2024}),\ \Eprint {https://arxiv.org/abs/2407.01296} {arXiv:2407.01296 [cond-mat.mes-hall]} \BibitemShut {NoStop}%
\bibitem [{\citenamefont {Li}\ \emph {et~al.}(2023)\citenamefont {Li}, \citenamefont {Trauzettel}, \citenamefont {Neupert},\ and\ \citenamefont {Zhang}}]{PhysRevLett.131.116601}%
  \BibitemOpen
  \bibfield  {author} {\bibinfo {author} {\bibfnamefont {C.-A.}\ \bibnamefont {Li}}, \bibinfo {author} {\bibfnamefont {B.}~\bibnamefont {Trauzettel}}, \bibinfo {author} {\bibfnamefont {T.}~\bibnamefont {Neupert}},\ and\ \bibinfo {author} {\bibfnamefont {S.-B.}\ \bibnamefont {Zhang}},\ }\bibfield  {title} {\bibinfo {title} {Enhancement of second-order non-hermitian skin effect by magnetic fields},\ }\href {https://doi.org/10.1103/PhysRevLett.131.116601} {\bibfield  {journal} {\bibinfo  {journal} {Phys. Rev. Lett.}\ }\textbf {\bibinfo {volume} {131}},\ \bibinfo {pages} {116601} (\bibinfo {year} {2023})}\BibitemShut {NoStop}%
\bibitem [{\citenamefont {Zhang}\ \emph {et~al.}(2022)\citenamefont {Zhang}, \citenamefont {Yang},\ and\ \citenamefont {Fang}}]{2021Universal}%
  \BibitemOpen
  \bibfield  {author} {\bibinfo {author} {\bibfnamefont {K.}~\bibnamefont {Zhang}}, \bibinfo {author} {\bibfnamefont {Z.}~\bibnamefont {Yang}},\ and\ \bibinfo {author} {\bibfnamefont {C.}~\bibnamefont {Fang}},\ }\bibfield  {title} {\bibinfo {title} {Universal non-hermitian skin effect in two and higher dimensions},\ }\href {https://doi.org/10.1038/s41467-022-30161-6} {\bibfield  {journal} {\bibinfo  {journal} {Nat. Commun.}\ }\textbf {\bibinfo {volume} {13}},\ \bibinfo {pages} {2496} (\bibinfo {year} {2022})}\BibitemShut {NoStop}%
\bibitem [{\citenamefont {Jiang}\ and\ \citenamefont {Lee}(2023)}]{PhysRevLett.131.076401}%
  \BibitemOpen
  \bibfield  {author} {\bibinfo {author} {\bibfnamefont {H.}~\bibnamefont {Jiang}}\ and\ \bibinfo {author} {\bibfnamefont {C.~H.}\ \bibnamefont {Lee}},\ }\bibfield  {title} {\bibinfo {title} {Dimensional transmutation from non-hermiticity},\ }\href {https://doi.org/10.1103/PhysRevLett.131.076401} {\bibfield  {journal} {\bibinfo  {journal} {Phys. Rev. Lett.}\ }\textbf {\bibinfo {volume} {131}},\ \bibinfo {pages} {076401} (\bibinfo {year} {2023})}\BibitemShut {NoStop}%
\bibitem [{\citenamefont {Hu}(2025)}]{HU202551}%
  \BibitemOpen
  \bibfield  {author} {\bibinfo {author} {\bibfnamefont {H.}~\bibnamefont {Hu}},\ }\bibfield  {title} {\bibinfo {title} {Topological origin of non-hermitian skin effect in higher dimensions and uniform spectra},\ }\href {https://doi.org/https://doi.org/10.1016/j.scib.2024.07.022} {\bibfield  {journal} {\bibinfo  {journal} {Sci. Bull.}\ }\textbf {\bibinfo {volume} {70}},\ \bibinfo {pages} {51} (\bibinfo {year} {2025})}\BibitemShut {NoStop}%
\bibitem [{\citenamefont {Zhang}\ \emph {et~al.}(2025)\citenamefont {Zhang}, \citenamefont {Shu},\ and\ \citenamefont {Sun}}]{cwwd-bclc}%
  \BibitemOpen
  \bibfield  {author} {\bibinfo {author} {\bibfnamefont {K.}~\bibnamefont {Zhang}}, \bibinfo {author} {\bibfnamefont {C.}~\bibnamefont {Shu}},\ and\ \bibinfo {author} {\bibfnamefont {K.}~\bibnamefont {Sun}},\ }\bibfield  {title} {\bibinfo {title} {Algebraic non-hermitian skin effect and generalized fermi surface formula in arbitrary dimensions},\ }\href {https://doi.org/10.1103/cwwd-bclc} {\bibfield  {journal} {\bibinfo  {journal} {Phys. Rev. X}\ }\textbf {\bibinfo {volume} {15}},\ \bibinfo {pages} {031039} (\bibinfo {year} {2025})}\BibitemShut {NoStop}%
\bibitem [{\citenamefont {Yang}\ and\ \citenamefont {Huang}(2025)}]{PhysRevB.111.155121}%
  \BibitemOpen
  \bibfield  {author} {\bibinfo {author} {\bibfnamefont {W.}~\bibnamefont {Yang}}\ and\ \bibinfo {author} {\bibfnamefont {H.}~\bibnamefont {Huang}},\ }\bibfield  {title} {\bibinfo {title} {Unified multipole bott indices for non-hermitian skin effect in different orders},\ }\href {https://doi.org/10.1103/PhysRevB.111.155121} {\bibfield  {journal} {\bibinfo  {journal} {Phys. Rev. B}\ }\textbf {\bibinfo {volume} {111}},\ \bibinfo {pages} {155121} (\bibinfo {year} {2025})}\BibitemShut {NoStop}%
\bibitem [{\citenamefont {Kawabata}\ and\ \citenamefont {Nakamura}(2025)}]{vxgf-59xt}%
  \BibitemOpen
  \bibfield  {author} {\bibinfo {author} {\bibfnamefont {K.}~\bibnamefont {Kawabata}}\ and\ \bibinfo {author} {\bibfnamefont {D.}~\bibnamefont {Nakamura}},\ }\bibfield  {title} {\bibinfo {title} {Hopf bifurcation of nonlinear non-hermitian skin effect},\ }\href {https://doi.org/10.1103/vxgf-59xt} {\bibfield  {journal} {\bibinfo  {journal} {Phys. Rev. Lett.}\ }\textbf {\bibinfo {volume} {135}},\ \bibinfo {pages} {126610} (\bibinfo {year} {2025})}\BibitemShut {NoStop}%
\bibitem [{\citenamefont {Pi}\ \emph {et~al.}(2026)\citenamefont {Pi}, \citenamefont {Li},\ and\ \citenamefont {Yan}}]{3ht3-ty3h}%
  \BibitemOpen
  \bibfield  {author} {\bibinfo {author} {\bibfnamefont {J.}~\bibnamefont {Pi}}, \bibinfo {author} {\bibfnamefont {X.}~\bibnamefont {Li}},\ and\ \bibinfo {author} {\bibfnamefont {Y.}~\bibnamefont {Yan}},\ }\bibfield  {title} {\bibinfo {title} {Scaling behavior of dissipative systems with imaginary gap closing},\ }\href {https://doi.org/10.1103/3ht3-ty3h} {\bibfield  {journal} {\bibinfo  {journal} {Phys. Rev. B}\ }\textbf {\bibinfo {volume} {113}},\ \bibinfo {pages} {064302} (\bibinfo {year} {2026})}\BibitemShut {NoStop}%
\bibitem [{\citenamefont {Wang}\ \emph {et~al.}(2025)\citenamefont {Wang}, \citenamefont {Pi}, \citenamefont {Liu}, \citenamefont {Li},\ and\ \citenamefont {Liu}}]{wang2025generaltheorygeometrydependentnonhermitian}%
  \BibitemOpen
  \bibfield  {author} {\bibinfo {author} {\bibfnamefont {C.}~\bibnamefont {Wang}}, \bibinfo {author} {\bibfnamefont {J.}~\bibnamefont {Pi}}, \bibinfo {author} {\bibfnamefont {Q.}~\bibnamefont {Liu}}, \bibinfo {author} {\bibfnamefont {Y.}~\bibnamefont {Li}},\ and\ \bibinfo {author} {\bibfnamefont {Y.-C.}\ \bibnamefont {Liu}},\ }\href {https://arxiv.org/abs/2506.22743} {\bibinfo {title} {General theory for geometry-dependent non-hermitian bands}} (\bibinfo {year} {2025}),\ \Eprint {https://arxiv.org/abs/2506.22743} {arXiv:2506.22743 [cond-mat.mes-hall]} \BibitemShut {NoStop}%
\bibitem [{\citenamefont {Longhi}(2022)}]{PhysRevLett.128.157601}%
  \BibitemOpen
  \bibfield  {author} {\bibinfo {author} {\bibfnamefont {S.}~\bibnamefont {Longhi}},\ }\bibfield  {title} {\bibinfo {title} {Self-healing of non-hermitian topological skin modes},\ }\href {https://doi.org/10.1103/PhysRevLett.128.157601} {\bibfield  {journal} {\bibinfo  {journal} {Phys. Rev. Lett.}\ }\textbf {\bibinfo {volume} {128}},\ \bibinfo {pages} {157601} (\bibinfo {year} {2022})}\BibitemShut {NoStop}%
\bibitem [{\citenamefont {Grillo}\ \emph {et~al.}(2014)\citenamefont {Grillo}, \citenamefont {Karimi}, \citenamefont {Gazzadi}, \citenamefont {Frabboni}, \citenamefont {Dennis},\ and\ \citenamefont {Boyd}}]{PhysRevX.4.011013}%
  \BibitemOpen
  \bibfield  {author} {\bibinfo {author} {\bibfnamefont {V.}~\bibnamefont {Grillo}}, \bibinfo {author} {\bibfnamefont {E.}~\bibnamefont {Karimi}}, \bibinfo {author} {\bibfnamefont {G.~C.}\ \bibnamefont {Gazzadi}}, \bibinfo {author} {\bibfnamefont {S.}~\bibnamefont {Frabboni}}, \bibinfo {author} {\bibfnamefont {M.~R.}\ \bibnamefont {Dennis}},\ and\ \bibinfo {author} {\bibfnamefont {R.~W.}\ \bibnamefont {Boyd}},\ }\bibfield  {title} {\bibinfo {title} {Generation of nondiffracting electron bessel beams},\ }\href {https://doi.org/10.1103/PhysRevX.4.011013} {\bibfield  {journal} {\bibinfo  {journal} {Phys. Rev. X}\ }\textbf {\bibinfo {volume} {4}},\ \bibinfo {pages} {011013} (\bibinfo {year} {2014})}\BibitemShut {NoStop}%
\bibitem [{\citenamefont {McGloin}\ and\ \citenamefont {Dholakia}(2005)}]{McGloin01012005}%
  \BibitemOpen
  \bibfield  {author} {\bibinfo {author} {\bibfnamefont {D.}~\bibnamefont {McGloin}}\ and\ \bibinfo {author} {\bibfnamefont {K.}~\bibnamefont {Dholakia}},\ }\bibfield  {title} {\bibinfo {title} {Bessel beams: Diffraction in a new light},\ }\href {https://doi.org/10.1080/0010751042000275259} {\bibfield  {journal} {\bibinfo  {journal} {Contemporary Physics}\ }\textbf {\bibinfo {volume} {46}},\ \bibinfo {pages} {15} (\bibinfo {year} {2005})}\BibitemShut {NoStop}%
\bibitem [{\citenamefont {Bouchal}\ \emph {et~al.}(1998)\citenamefont {Bouchal}, \citenamefont {Wagner},\ and\ \citenamefont {Chlup}}]{BOUCHAL1998207}%
  \BibitemOpen
  \bibfield  {author} {\bibinfo {author} {\bibfnamefont {Z.}~\bibnamefont {Bouchal}}, \bibinfo {author} {\bibfnamefont {J.}~\bibnamefont {Wagner}},\ and\ \bibinfo {author} {\bibfnamefont {M.}~\bibnamefont {Chlup}},\ }\bibfield  {title} {\bibinfo {title} {Self-reconstruction of a distorted nondiffracting beam},\ }\href {https://doi.org/https://doi.org/10.1016/S0030-4018(98)00085-6} {\bibfield  {journal} {\bibinfo  {journal} {Optics Communications}\ }\textbf {\bibinfo {volume} {151}},\ \bibinfo {pages} {207} (\bibinfo {year} {1998})}\BibitemShut {NoStop}%
\bibitem [{\citenamefont {Durnin}\ \emph {et~al.}(1987)\citenamefont {Durnin}, \citenamefont {Miceli},\ and\ \citenamefont {Eberly}}]{PhysRevLett.58.1499}%
  \BibitemOpen
  \bibfield  {author} {\bibinfo {author} {\bibfnamefont {J.}~\bibnamefont {Durnin}}, \bibinfo {author} {\bibfnamefont {J.~J.}\ \bibnamefont {Miceli}},\ and\ \bibinfo {author} {\bibfnamefont {J.~H.}\ \bibnamefont {Eberly}},\ }\bibfield  {title} {\bibinfo {title} {Diffraction-free beams},\ }\href {https://doi.org/10.1103/PhysRevLett.58.1499} {\bibfield  {journal} {\bibinfo  {journal} {Phys. Rev. Lett.}\ }\textbf {\bibinfo {volume} {58}},\ \bibinfo {pages} {1499} (\bibinfo {year} {1987})}\BibitemShut {NoStop}%
\bibitem [{\citenamefont {Zhang}\ \emph {et~al.}(2014)\citenamefont {Zhang}, \citenamefont {Li}, \citenamefont {Zhu}, \citenamefont {Zhu}, \citenamefont {Yang}, \citenamefont {Wang}, \citenamefont {Yin},\ and\ \citenamefont {Zhang}}]{zhang2014generation}%
  \BibitemOpen
  \bibfield  {author} {\bibinfo {author} {\bibfnamefont {P.}~\bibnamefont {Zhang}}, \bibinfo {author} {\bibfnamefont {T.}~\bibnamefont {Li}}, \bibinfo {author} {\bibfnamefont {J.}~\bibnamefont {Zhu}}, \bibinfo {author} {\bibfnamefont {X.}~\bibnamefont {Zhu}}, \bibinfo {author} {\bibfnamefont {S.}~\bibnamefont {Yang}}, \bibinfo {author} {\bibfnamefont {Y.}~\bibnamefont {Wang}}, \bibinfo {author} {\bibfnamefont {X.}~\bibnamefont {Yin}},\ and\ \bibinfo {author} {\bibfnamefont {X.}~\bibnamefont {Zhang}},\ }\bibfield  {title} {\bibinfo {title} {Generation of acoustic self-bending and bottle beams by phase engineering},\ }\href {https://doi.org/https://doi.org/10.1038/ncomms5316} {\bibfield  {journal} {\bibinfo  {journal} {Nature communications}\ }\textbf {\bibinfo {volume} {5}},\ \bibinfo {pages} {4316} (\bibinfo {year} {2014})}\BibitemShut {NoStop}%
\bibitem [{\citenamefont {Voloch-Bloch}\ \emph {et~al.}(2013)\citenamefont {Voloch-Bloch}, \citenamefont {Lereah}, \citenamefont {Lilach}, \citenamefont {Gover},\ and\ \citenamefont {Arie}}]{voloch2013generation}%
  \BibitemOpen
  \bibfield  {author} {\bibinfo {author} {\bibfnamefont {N.}~\bibnamefont {Voloch-Bloch}}, \bibinfo {author} {\bibfnamefont {Y.}~\bibnamefont {Lereah}}, \bibinfo {author} {\bibfnamefont {Y.}~\bibnamefont {Lilach}}, \bibinfo {author} {\bibfnamefont {A.}~\bibnamefont {Gover}},\ and\ \bibinfo {author} {\bibfnamefont {A.}~\bibnamefont {Arie}},\ }\bibfield  {title} {\bibinfo {title} {Generation of electron airy beams},\ }\href {https://doi.org/https://doi.org/10.1038/nature11840} {\bibfield  {journal} {\bibinfo  {journal} {Nature}\ }\textbf {\bibinfo {volume} {494}},\ \bibinfo {pages} {331} (\bibinfo {year} {2013})}\BibitemShut {NoStop}%
\bibitem [{\citenamefont {Xue}\ \emph {et~al.}(2025)\citenamefont {Xue}, \citenamefont {Song}, \citenamefont {Hu},\ and\ \citenamefont {Wang}}]{xue2025nonblochedgedynamicsnonhermitian}%
  \BibitemOpen
  \bibfield  {author} {\bibinfo {author} {\bibfnamefont {W.-T.}\ \bibnamefont {Xue}}, \bibinfo {author} {\bibfnamefont {F.}~\bibnamefont {Song}}, \bibinfo {author} {\bibfnamefont {Y.-M.}\ \bibnamefont {Hu}},\ and\ \bibinfo {author} {\bibfnamefont {Z.}~\bibnamefont {Wang}},\ }\href {https://arxiv.org/abs/2503.13671} {\bibinfo {title} {Non-bloch edge dynamics of non-hermitian lattices}} (\bibinfo {year} {2025}),\ \Eprint {https://arxiv.org/abs/2503.13671} {arXiv:2503.13671 [quant-ph]} \BibitemShut {NoStop}%
\bibitem [{\citenamefont {Yang}\ and\ \citenamefont {Fang}(2025)}]{llbb-pcgk}%
  \BibitemOpen
  \bibfield  {author} {\bibinfo {author} {\bibfnamefont {T.-H.}\ \bibnamefont {Yang}}\ and\ \bibinfo {author} {\bibfnamefont {C.}~\bibnamefont {Fang}},\ }\bibfield  {title} {\bibinfo {title} {Real-time edge dynamics of non-hermitian lattices},\ }\href {https://doi.org/10.1103/llbb-pcgk} {\bibfield  {journal} {\bibinfo  {journal} {Phys. Rev. Lett.}\ }\textbf {\bibinfo {volume} {135}},\ \bibinfo {pages} {186401} (\bibinfo {year} {2025})}\BibitemShut {NoStop}%
\bibitem [{Note1()}]{Note1}%
  \BibitemOpen
  \bibinfo {note} {\label {fn}See Supplemental Material at \protect \url {http://link.aps.org/supplemental/xxx} for comprehensive analytical derivations and extended numerical verifications. Specifically, the SM provides: (i) a rigorous proof of the self-healing metric inequality $\eta (t) \le \epsilon (t)$; (ii) detailed derivations of the FTLE dynamics under weak noise using biorthogonal expansion; (iii) the complete strong-noise perturbation framework, including the derivation of the effective non-unitary drift-diffusion equation and its universal $1/t$ convergence; (iv) statistical validation of the FTLE-based estimator comparing single-trajectory dynamics with ensemble averages; (v) analysis of the relationship between eigenstate spatial extension (characterized by skin corner weight) and self-healing robustness; (vi) demonstrations of the mechanism's universality across alternative lattice configurations; and (vii) further discussions on the fragility of coherent non-Hermitian saddle-point dynamics
  under noise. The Supplemental Material includes Refs.~\cite {llbb-pcgk,xue2025nonblochedgedynamicsnonhermitian,PhysRevB.111.155121}.}\BibitemShut {Stop}%
\bibitem [{\citenamefont {Zhang}\ \emph {et~al.}(2020)\citenamefont {Zhang}, \citenamefont {Yang},\ and\ \citenamefont {Fang}}]{PhysRevLett.125.126402}%
  \BibitemOpen
  \bibfield  {author} {\bibinfo {author} {\bibfnamefont {K.}~\bibnamefont {Zhang}}, \bibinfo {author} {\bibfnamefont {Z.}~\bibnamefont {Yang}},\ and\ \bibinfo {author} {\bibfnamefont {C.}~\bibnamefont {Fang}},\ }\bibfield  {title} {\bibinfo {title} {Correspondence between winding numbers and skin modes in non-hermitian systems},\ }\href {https://doi.org/10.1103/PhysRevLett.125.126402} {\bibfield  {journal} {\bibinfo  {journal} {Phys. Rev. Lett.}\ }\textbf {\bibinfo {volume} {125}},\ \bibinfo {pages} {126402} (\bibinfo {year} {2020})}\BibitemShut {NoStop}%
\bibitem [{\citenamefont {Amir}\ \emph {et~al.}(2009)\citenamefont {Amir}, \citenamefont {Lahini},\ and\ \citenamefont {Perets}}]{PhysRevE.79.050105}%
  \BibitemOpen
  \bibfield  {author} {\bibinfo {author} {\bibfnamefont {A.}~\bibnamefont {Amir}}, \bibinfo {author} {\bibfnamefont {Y.}~\bibnamefont {Lahini}},\ and\ \bibinfo {author} {\bibfnamefont {H.~B.}\ \bibnamefont {Perets}},\ }\bibfield  {title} {\bibinfo {title} {Classical diffusion of a quantum particle in a noisy environment},\ }\href {https://doi.org/10.1103/PhysRevE.79.050105} {\bibfield  {journal} {\bibinfo  {journal} {Phys. Rev. E}\ }\textbf {\bibinfo {volume} {79}},\ \bibinfo {pages} {050105} (\bibinfo {year} {2009})}\BibitemShut {NoStop}%
\bibitem [{\citenamefont {Yang}\ and\ \citenamefont {Huang}(2026)}]{yang2026noiseinducedresurrectiondynamicalskin}%
  \BibitemOpen
  \bibfield  {author} {\bibinfo {author} {\bibfnamefont {W.}~\bibnamefont {Yang}}\ and\ \bibinfo {author} {\bibfnamefont {H.}~\bibnamefont {Huang}},\ }\href {https://arxiv.org/abs/2604.11455} {\bibinfo {title} {Noise-induced resurrection of dynamical skin effects in quasiperiodic non-hermitian systems}} (\bibinfo {year} {2026}),\ \Eprint {https://arxiv.org/abs/2604.11455} {arXiv:2604.11455 [quant-ph]} \BibitemShut {NoStop}%
\bibitem [{\citenamefont {Gopalakrishnan}\ \emph {et~al.}(2017)\citenamefont {Gopalakrishnan}, \citenamefont {Islam},\ and\ \citenamefont {Knap}}]{PhysRevLett.119.046601}%
  \BibitemOpen
  \bibfield  {author} {\bibinfo {author} {\bibfnamefont {S.}~\bibnamefont {Gopalakrishnan}}, \bibinfo {author} {\bibfnamefont {K.~R.}\ \bibnamefont {Islam}},\ and\ \bibinfo {author} {\bibfnamefont {M.}~\bibnamefont {Knap}},\ }\bibfield  {title} {\bibinfo {title} {Noise-induced subdiffusion in strongly localized quantum systems},\ }\href {https://doi.org/10.1103/PhysRevLett.119.046601} {\bibfield  {journal} {\bibinfo  {journal} {Phys. Rev. Lett.}\ }\textbf {\bibinfo {volume} {119}},\ \bibinfo {pages} {046601} (\bibinfo {year} {2017})}\BibitemShut {NoStop}%
\end{thebibliography}%

\end{document}

\beginsupplement

\title{Supplementary Material for "Noise-Enhanced Self-Healing Dynamics in Non-Hermitian Systems"}

\author{Wuping Yang}
\affiliation{School of Physics, Peking University, Beijing 100871, China}

\author{H. Huang}%
 \email[Corresponding author: ]{huanghq07@pku.edu.cn}
\affiliation{School of Physics, Peking University, Beijing 100871, China}
\affiliation{Collaborative Innovation Center of Quantum Matter, Beijing 100084, China}
\affiliation{Center for High Energy Physics, Peking University, Beijing 100871, China}

\date{\today}

\maketitle
\tableofcontents
\clearpage

\section{Proof of the inequality $\eta(t) \le \epsilon(t)$}

In the main text, we introduced the inequality $\eta(t) \le \epsilon(t)$. In this section, we provide a rigorous proof for this relation. First, we decompose the deviation wavefunction $|\xi(t)\rangle$ as:
\begin{equation}
|\xi(t)\rangle = c(t) |\phi(t)\rangle + |\xi_{\perp}(t)\rangle,
\label{eq:decomposition}
\end{equation}
where $c(t)$ is a complex scalar representing the projection coefficient of the perturbation onto the reference state, and $|\xi_{\perp}(t)\rangle$ denotes the orthogonal component satisfying $\langle \phi(t) \mid \xi_{\perp}(t) \rangle = 0$. For brevity, the time dependence $t$ will be omitted in the subsequent derivation. Based on Eq.~\eqref{eq:decomposition}, the norm of the deviation is given by:
\begin{equation}
\begin{aligned}
\langle \xi \mid \xi \rangle &= (\langle \phi | c^* + \langle \xi_{\perp} |)(c |\phi\rangle + |\xi_{\perp}\rangle) \\
&= |c|^2 \langle \phi \mid \phi \rangle + c^* \underbrace{\langle \phi \mid \xi_{\perp} \rangle}_{0} + c \underbrace{\langle \xi_{\perp} \mid \phi \rangle}_{0} + \langle \xi_{\perp} \mid \xi_{\perp} \rangle \\
&= |c|^2 N_{\phi} + x,
\end{aligned}
\end{equation}
where we have defined $x\equiv\langle \xi_{\perp} \mid \xi_{\perp} \rangle$ and $N_{\phi}\equiv\langle \phi \mid \phi \rangle$. Substituting this into the definition of $\epsilon = \frac{\langle \xi \mid \xi \rangle}{\langle \phi \mid \phi \rangle}$, we obtain:
\begin{equation}
\epsilon = \frac{|c|^2 N_{\phi} + x}{N_{\phi}} = |c|^2 + \tilde{x},
\end{equation}
where $\tilde{x} \equiv x/N_{\phi}$ represents the normalized orthogonal deviation.

Next, we consider the profile self-healing factor defined by $\eta = 1 - \frac{|\langle \psi \mid \phi \rangle|^2}{\langle \psi \mid \psi \rangle \langle \phi \mid \phi \rangle}$. The inner product in the numerator is calculated as:
\begin{equation}
\langle \psi \mid \phi \rangle = ((1+c)^* \langle \phi | + \langle \xi_{\perp} |) |\phi\rangle = (1+c)^* N_{\phi},
\end{equation}
which yields $|\langle \psi \mid \phi \rangle|^2 = |1+c|^2 N_{\phi}^2$. Similarly, the denominator term is:
\begin{equation}
\langle \psi \mid \psi \rangle = |1+c|^2 N_{\phi} + x.
\end{equation}
Substituting these results into the definition of $\eta$, we derive:
\begin{equation}
\eta = 1 - \frac{|1+c|^2 N_{\phi}}{|1+c|^2 N_{\phi} + x} = \frac{x/N_{\phi}}{|1+c|^2 + x/N_{\phi}} = \frac{\tilde{x}}{|1+c|^2 + \tilde{x}}.
\end{equation}

Therefore, proving $\epsilon \ge \eta$ is equivalent to proving:
\begin{equation}
|c|^2 + \tilde{x} \ge \frac{\tilde{x}}{|1+c|^2 + \tilde{x}},
\end{equation}
where $\tilde{x} \ge 0$ and $c$ is an arbitrary complex number.

If $\tilde{x}=0$, the inequality holds trivially since $|c|^2 \ge 0$. For the case $\tilde{x} > 0$, since the denominator $|1+c|^2 + \tilde{x}$ is positive, the inequality is equivalent to:
\begin{equation}
(|c|^2 + \tilde{x}) (|1+c|^2 + \tilde{x}) -\tilde{x} \ge 0.
\end{equation}
Expanding the left-hand side and rearranging terms, we have:
\begin{align}
& (|c|^2 + \tilde{x}) (|1+c|^2 + \tilde{x}) - \tilde{x} \notag \\
=& (|c|^2 + \tilde{x}) (1 + 2\mathrm{Re}(c) + |c|^2 + \tilde{x}) - \tilde{x} \notag \\
=& \tilde{x}^2 + 2\tilde{x}(|c|^2 + \mathrm{Re}(c)) + \left( |c|^4 + 2|c|^2\mathrm{Re}(c) + |c|^2 \right) \notag \\
=& \tilde{x}^2 + 2\tilde{x}(|c|^2 + \mathrm{Re}(c)) + \left( |c|^4 + 2|c|^2\mathrm{Re}(c) + \mathrm{Re}(c)^2 \right) + \left( |c|^2 - \mathrm{Re}(c)^2 \right) \notag \\
=& \left( \tilde{x} + |c|^2 + \mathrm{Re}(c) \right)^2 + \mathrm{Im}(c)^2 \ge 0.
\end{align}
Consequently, the inequality $\eta(t) \le \epsilon(t)$ is proven.

\section{Derivation of the Lyapunov Exponent Evolution under Weak Noise}
In the main text, we analyzed how weak stochastic noise prolongs the self-healing window by modulating the relative dynamics of the finite-time Lyapunov exponents (FTLEs) for the reference state, $\lambda_{\phi}(t)$, and the deviation state, $\lambda_{\xi}(t)$. To establish a rigorous mathematical foundation for this physical picture, this section provides detailed analytical derivations of the FTLE dynamics. By employing a biorthogonal expansion framework within the weak-noise limit, we systematically evaluate both the short-time transient behaviors and the long-time asymptotic limits of $\lambda_{\phi}(t)$ and $\lambda_{\xi}(t)$, fully supporting the theoretical conclusions discussed in the main text.

\subsection{Derivation of the Long-Time Evolution Behavior of $\lambda_{\phi}(t)$}
As mentioned in the main text, the time evolution of the reference state $|\phi(t)\rangle$ is driven by the total Hamiltonian $\hat{H}(t)=\hat{H}_{0}+\hat{V}_{noise}(t)$ and obeys the Schrödinger equation:
\begin{equation}\label{SE1}
i \partial_t |\phi(t)\rangle = \left( \hat{H}_0 + \hat{V}_{\text{noise}}(t) \right) |\phi(t)\rangle.
\end{equation}
Decomposing the state as $|\phi(t)\rangle = e^{-i E_{\text{init}} t}|\phi(0)\rangle + |\delta \phi(t)\rangle$ and substituting it back into Eq.~(\ref{SE1}), while utilizing the eigenvalue relation $\hat{H}_0 |\phi(0)\rangle = E_{\text{init}} |\phi(0)\rangle$, we obtain:
\begin{equation}\label{SE2}
i \partial_t |\delta \phi(t)\rangle = \hat{H}_0 |\delta \phi(t)\rangle + \hat{V}_{\text{noise}}(t) e^{-i E_{\text{init}} t}|\phi(0)\rangle + \hat{V}_{\text{noise}}(t) |\delta \phi(t)\rangle.
\end{equation}
In the weak noise limit, we can approximate Eq.~(\ref{SE2}) by neglecting the higher-order infinitesimal term:
\begin{equation}\label{SE3}
i \partial_t |\delta \phi(t)\rangle \approx \hat{H}_0 |\delta \phi(t)\rangle + \hat{V}_{\text{noise}}(t) e^{-i E_{\text{init}} t}|\phi(0)\rangle.
\end{equation}
Rearranging Eq.~(\ref{SE3}) and multiplying from the left by the integrating factor $e^{i \hat{H}_0 t}$ yields:
\begin{equation}
\begin{aligned}
e^{i \hat{H}_0 t} \left( i \partial_t |\delta \phi(t)\rangle - \hat{H}_0 |\delta \phi(t)\rangle \right) = e^{i \hat{H}_0 t} \hat{V}_{\text{noise}}(t) e^{-i E_{\text{init}} t}|\phi(0)\rangle,\\
i \partial_t \left( e^{i \hat{H}_0 t} |\delta \phi(t)\rangle \right) = e^{i \hat{H}_0 t} \hat{V}_{\text{noise}}(t) e^{-i E_{\text{init}} t}|\phi(0)\rangle.
\end{aligned}
\end{equation}
Integrating both sides from $0$ to $t$ and applying the initial condition $|\delta \phi(0)\rangle = 0$, we have:
\begin{equation}
|\delta \phi(t)\rangle = -i \int_0^t e^{-i \hat{H}_0 (t-\tau)} \hat{V}_{\text{noise}}(\tau) e^{-i E_{\text{init}} \tau}|\phi(0)\rangle d\tau.
\end{equation}
Inserting the non-Hermitian biorthogonal completeness relation $\hat{I} = \sum_n |R_n\rangle \langle L_n|$ and simplifying, we get:
\begin{equation}
\begin{aligned}
|\delta \phi(t)\rangle =& -i \int_0^t \left( \sum_n e^{-i E_n (t-\tau)} |R_n\rangle \langle L_n| \right) \hat{V}_{\text{noise}}(\tau) e^{-i E_{\text{init}} \tau}|\phi(0)\rangle d\tau\\
=& -i \sum_n e^{-i E_n t} |R_n\rangle \underbrace{\int_0^t e^{i (E_n - E_{\text{init}}) \tau} \langle L_n | \hat{V}_{\text{noise}}(\tau) | \phi(0) \rangle d\tau}_{C^{\phi}_n(t)}\\
=& -i \sum_n C^{\phi}_n(t) e^{-i E_n t} |R_n\rangle.
\label{C_phi}
\end{aligned}
\end{equation}

To formulate the temporal evolution of $\lambda_{\phi}(t)$, we evaluate the squared norm, which is defined as $\|\phi(t)\|^2 \equiv \langle \phi(t) | \phi(t) \rangle$, by expanding the state as $|\phi(t)\rangle = e^{-i E_{\text{init}} t}|\phi(0)\rangle + |\delta \phi(t)\rangle$. Substituting the biorthogonal expansion of $|\delta \phi(t)\rangle$, the total norm can be compactly expressed as the sum of the unperturbed evolution, the pure noise-induced deviation, and their cross-interference:
\begin{equation}\label{phi_norm}
\begin{aligned} \|\phi(t)\|^2 &= e^{i (E_{\text{init}}^* - E_{\text{init}}) t} \langle \phi(0) | \phi(0) \rangle + \sum_{m,n} C_m^{\phi*}(t) C^{\phi}_n(t) e^{i (E_m^* - E_n) t} \langle R_m | R_n \rangle \\ &\quad - \left( i \sum_n C^{\phi}_n(t) e^{i (E_{\text{init}}^* - E_n) t} \langle \phi(0) | R_n \rangle + \text{h.c.} \right). 
\end{aligned}
\end{equation}
When evaluating the macroscopic dynamics via the ensemble average $\langle \|\phi(t)\|^2 \rangle$, the long-time limit is ultimately dictated by the term that prevails in the exponential growth competition. The unperturbed baseline and linear cross-interference terms are intrinsically bottlenecked by the amplification rate of the initial state, $\operatorname{Im}(E_{\text{init}})$. In contrast, the deviation norm (the double summation) contains strictly positive diagonal elements ($m=n$) with expectation values $\langle|C_m^\phi(t)|^2\rangle e^{2\operatorname{Im}(E_m)t}$. Crucially, independent of the coefficient's exact magnitude, this term's maximum contribution to the Lyapunov exponent can reach the largest imaginary part of the entire energy spectrum, entirely unconstrained by the initial state's eigenvalue. Therefore, identifying the dynamically dominant mode directly hinges on analyzing the time evolution of the squared projection coefficient $\langle|C_m^\phi(t)|^2\rangle$. To do this rigorously, we must separately evaluate the initial eigenmode component ($m = \text{init}$) and the remaining arbitrary eigenmodes ($m \neq \text{init}$).

For any arbitrary eigenstate $|R_m\rangle$ distinct from the initial state ($m \neq \text{init}$), the variance of its coefficient $\langle |C_m^\phi(t)|^2 \rangle$ can be evaluated via its double-integral representation. Using Eq. (S15), we obtain:
\begin{equation}
\langle |C_m^\phi(t)|^2 \rangle = \int_0^t \int_0^t e^{i (E_m - E_{\text{init}}) \tau_1} e^{-i (E_m^* - E_{\text{init}}^*) \tau_2} \langle \langle L_m | \hat{V}_{\text{noise}}(\tau_1) | \phi(0) \rangle \langle \phi(0) | \hat{V}_{\text{noise}}(\tau_2) | L_m \rangle \rangle d\tau_1 d\tau_2
\end{equation}

For Ornstein-Uhlenbeck noise, the temporal correlation function is given by $\langle \xi_j(\tau_1) \xi_k(\tau_2) \rangle = \frac{\sigma^2}{2\theta} e^{-\theta |\tau_1 - \tau_2|} \delta_{jk}$. Utilizing this property alongside the definition of the noise operator $\hat{V}_{\text{noise}}(\tau) = \sum_j \xi_j(\tau) |j\rangle \langle j|$, the expectation value in the integrand simplifies to $K_m e^{-\theta |\tau_1 - \tau_2|}$, where $K_m \equiv \frac{\sigma^2}{2\theta} \sum_j |\langle L_m | j \rangle \langle j | \phi(0) \rangle|^2$. Defining the complex energy difference as $\Delta E_m \equiv E_m - E_{\text{init}}$, we evaluate the asymptotic limit as $t \to \infty$:
\begin{equation}
\begin{aligned}
\langle | C_m^{\phi}(\infty)|^{2}\rangle &= K_m \int_{0}^{\infty} d\tau_{1} \int_{0}^{\infty} d\tau_{2} \, e^{i \Delta E_m \tau_{1}} e^{-i \Delta E_m^{*} \tau_{2}} e^{-\theta\left|\tau_{1}-\tau_{2}\right|}\\
&= 2K_m \operatorname{Re} \left( \int_{0}^{\infty} d\tau_{1} \, e^{(i \Delta E_m - \theta)\tau_{1}} \left[ \frac{e^{(\theta - i \Delta E_m^{*})\tau_{1}} - 1}{\theta - i \Delta E_m^{*}} \right] \right)\\
&= 2K_m \operatorname{Re} \left( \frac{1}{\theta - i \Delta E_m^{*}} \left[ \frac{1}{2\operatorname{Im}(\Delta E_m)} - \frac{1}{\theta - i \Delta E_m} \right] \right).
\end{aligned}
\end{equation}

Because $\langle |C_m^\phi(\infty)|^2 \rangle$ converges to a finite value for any $m \neq \text{init}$, we can establish a strict upper bound for the ensemble-averaged logarithmic term by invoking Jensen's inequality. Consequently, we obtain the inequality $\langle \ln |C_m^\phi(t)|^2 \rangle \le \ln ( \langle |C_m^\phi(t)|^2 \rangle )$, which dictates that:
\begin{equation}
\lim_{t \to \infty} \frac{1}{2t} \langle \ln |C_m^\phi(t)|^2 \rangle = 0.
\end{equation}
Subsequently, we address the projection onto the initial eigenmode, where $m = \text{init}$. In this specific case, the complex energy difference strictly vanishes ($\Delta E_{\text{init}} = 0$). By splitting the integration domain to account for the absolute value in the exponent, the double integral for its variance is rigorously derived as follows:
\begin{equation}
\begin{aligned}
\langle |C_{\text{init}}^\phi(t)|^2 \rangle &= K_{\text{init}} \int_0^t \int_0^t e^{-\theta |\tau_1 - \tau_2|} d\tau_1 d\tau_2\\
&= 2 K_{\text{init}} \int_0^t d\tau_1 \int_0^{\tau_1} d\tau_2 \, e^{-\theta (\tau_1 - \tau_2)}\\
&= 2 K_{\text{init}} \int_0^t d\tau_1 \, e^{-\theta \tau_1} \left[ \frac{e^{\theta \tau_2}}{\theta} \right]_0^{\tau_1}\\
&= \frac{2 K_{\text{init}}}{\theta} \int_0^t (1 - e^{-\theta \tau_1}) d\tau_1\\
&= \frac{2 K_{\text{init}}}{\theta} \left( t - \frac{1 - e^{-\theta t}}{\theta} \right).
\end{aligned}
\end{equation}
In the long-time limit, this specific coefficient diverges algebraically:
\begin{equation}
\langle|C_{\text{init}}^\phi(t)|^2\rangle \approx \frac{2K_{\text{init}}}{\theta} t.
\end{equation}
Although the initial mode component contributes an algebraically divergent coefficient over time, the overall squared norm $||\phi(t)||^2$ is overwhelmingly dominated by the exponential growth inherent to the non-unitary dynamics. Let $E_{\text{max}}$ denote the eigenvalue of $\hat{H}_0$ possessing the maximum imaginary part and $C_{\text{max}}^\phi(t)$ denotes its associated coefficient. Assuming the initial state differs from this maximally amplifying state ($E_{\text{init}} \neq E_{\text{max}}$), the ratio of their respective contributions to the global norm evolves as:
\begin{equation}
\frac{\langle|C_{\text{init}}^\phi(t)|^2\rangle e^{2\operatorname{Im}(E_{\text{init}})t}}{\langle|C_{\text{max}}^\phi(t)|^2\rangle e^{2\operatorname{Im}(E_{\text{max}})t}} \propto t e^{-2(\operatorname{Im}(E_{\text{max}}) - \operatorname{Im}(E_{\text{init}}))t}.
\end{equation}
Taking the long-time limit ($t \to \infty$), this ratio strictly approaches zero. This analysis demonstrates that the algebraic growth contributed by the initial mode cannot compete with the exponential divergence of the mode featuring the maximum imaginary eigenvalue.

Consequently, the long-time asymptotic form of the reference state norm is strictly dictated by the maximum imaginary mode:
\begin{equation}
||\phi(t)||^2 \sim |C^{\phi}_{\text{max}}(t)|^2 e^{i(E_{\text{max}}^* - E_{\text{max}}) t}.
\end{equation}

Since $E_{\text{max}}$ falls into the category of non-initial modes ($m \neq \text{init}$), its corresponding logarithmic term vanishes when divided by $t$, yielding $\lim_{t \to \infty} \frac{1}{2t} \langle \ln |C_{\text{max}}^\phi(t)|^2 \rangle = 0$. Ultimately, this confirms that the Lyapunov exponent, formulated as $\lambda_{\phi}(t) = \operatorname{Im}(E_{\text{max}}) + \frac{1}{2t} \langle \ln |C^{\phi}_{\text{max}}(t)|^2 \rangle$, strictly converges to the maximum imaginary part of the spectrum:
\begin{equation}
\lim_{t \to \infty} \lambda_{\phi}(t) = \operatorname{Im}(E_{\text{max}}).
\end{equation}

\subsection{Derivation of the Long-Time Evolution Behavior of $\lambda_{\xi}(t)$}

During the evolution in the post-scattering regime ($t > t_1$), the wave function can be expanded in the biorthogonal basis of the unperturbed non-Hermitian Hamiltonian $\hat{H}_{0}$ as follows:
\begin{equation}
|\xi(t)\rangle = \sum_{n} C_{n}^{\xi}(t) e^{-iE_{n}(t - t_1)} |R_{n}\rangle.
\label{xi_expand}
\end{equation}
Substituting Eq.~(\ref{xi_expand}) into the Schrödinger equation,
\begin{equation}
i\frac{\partial}{\partial t}|\xi(t)\rangle = [\hat{H}_{0} + \hat{V}_{\text{noise}}(t)] |\xi(t)\rangle,
\label{xi_SE}
\end{equation}
we obtain:
\begin{equation}
\begin{split}
{ \sum_{n} i \dot{C}_{n}^{\xi}(t) e^{-iE_n t} |R_n\rangle + \sum_{n} C_{n}^{\xi}(t) E_n e^{-iE_n t} |R_n\rangle } &= { \sum_{n} C_{n}^{\xi}(t) E_n e^{-iE_n t} |R_n\rangle + \sum_{n} C_{n}^{\xi}(t) e^{-iE_n t} \hat{V}_{\text{noise}}(t) |R_n\rangle },\\
\sum_{n} i \dot{C}_{n}^{\xi}(t) e^{-iE_n t} |R_n\rangle &=  \sum_{n} C_{n}^{\xi}(t) e^{-iE_n t} \hat{V}_{\text{noise}}(t) |R_n\rangle.
\end{split}
\end{equation}
By multiplying both sides from the left by a specific left eigenvector $\langle L_m|$ and utilizing the biorthogonal orthonormality relation $\langle L_m | R_n \rangle = \delta_{mn}$, we have:
\begin{equation}
\begin{split}
\langle L_m | \left( \sum_{n} i \dot{C}_{n}^{\xi}(t) e^{-iE_n t} |R_n\rangle \right) &= \langle L_m | \left( \sum_{n} C_{n}^{\xi}(t) e^{-iE_n t} \hat{V}_{\text{noise}}(t) |R_n\rangle \right),\\
i \dot{C}_{m}^{\xi}(t) e^{-iE_m t} &= \sum_{n} C_{n}^{\xi}(t) e^{-iE_n t} \langle L_m | \hat{V}_{\text{noise}}(t) |R_n\rangle.
\end{split}
\end{equation}
Multiplying both sides by $-i e^{iE_m (t-t_1)}$ yields the differential equation:
\begin{equation}
\dot{C}_{m}^{\xi}(t) = -i \sum_{n} C_{n}^{\xi}(t) e^{i(E_m - E_n)(t-t_1)} \langle L_m | \hat{V}_{\text{noise}}(t) |R_n\rangle.
\label{C_xi1}
\end{equation}
Integrating Eq.~(\ref{C_xi1}) from the initial time $t_1$ to $t$, we obtain:
\begin{equation}
\begin{split}
C_{m}^{\xi}(t) &= C_{m}^{\xi}(t_1) - i \int_{t_1}^{t} \sum_{n} C_{n}^{\xi}(\tau) e^{i(E_m - E_n)(\tau-t_1)} \langle L_m | \hat{V}_{\text{noise}}(\tau) |R_n\rangle d\tau, \\
&\approx C_{m}^{\xi}(t_1)  - i \sum_{n}C_{m}^{\xi}(t_1)  \int_{t_1}^{t} e^{i(E_m - E_n)(\tau-t_1)} \langle L_m | \hat{V}_{\text{noise}}(\tau) |R_n\rangle d\tau.
\label{C_xi2}
\end{split}
\end{equation}
It is evident that the first term in Eq.~(\ref{C_xi2}) is a constant, while the integral in the second term is nearly identical to that in Eq.~(\ref{C_phi}). Therefore, by employing a similar approach to the previous section, we can establish that:
\begin{equation}
\lim_{t \to \infty} \frac{1}{2t} \langle \ln |C^{\xi}_{\text{max}}(t)|^2 \rangle = 0.
\end{equation}

The squared global norm of $|\xi(t)\rangle$ is given by:
\begin{equation}
||\xi(t)||^{2} = \langle \xi(t) | \xi(t) \rangle = \sum_{m,n} (C_{m}^{\xi}(t))^{*} C_{n}^{\xi}(t) e^{iE_{m}^{*}(t - t_1)} e^{-iE_{n}(t - t_1)} \langle R_{m} | R_{n} \rangle.
\end{equation}

In the long-time limit $t \gg t_1$, this asymptotically reduces to the diagonal term corresponding to the mode with the maximum imaginary part:
\begin{equation}
\begin{aligned}
\|\xi(t)\|^2 
& \sim |C^{\xi}_{\text{max}}(t)|^2 e^{i(E_{\text{max}}^* - E_{\text{max}}) t}.
\end{aligned}
\end{equation}

Consequently, $\lambda_{\xi}(t)$ also approaches $\operatorname{Im}(E_{\text{max}})$ in the long-time limit.

It is worth emphasizing that for relatively large $t$, taking the difference between $\lambda_{\xi}(t)$ and $\lambda_{\phi}(t)$ yields the approximation:
\begin{equation}
\lambda_{\xi}(t) - \lambda_{\phi}(t) \approx \frac{1}{2t} \left\langle \ln \left( \frac{|C_{max}^{\xi}(t)|^{2}}{|C_{max}^{\phi}(t)|^{2}} \right) \right\rangle.
\end{equation}
Since $|C_{max}^{\phi}(t)|^{2}$ is proportional to the noise intensity $\sigma^{2}$, whereas $|C_{max}^{\xi}(t)|^{2}$ contains a nonzero constant term, the difference $\lambda_{\xi}(t) - \lambda_{\phi}(t)$ is always strictly positive. This implies that although weak noise can enhance the self-healing property of the system, this self-healing capability will inevitably vanish in the long-time limit.

\subsection{Derivation of the Short-Time Evolution Behavior of $\lambda_{\xi}(t)$}

Consider the evolution in the post-scattering regime ($t > t_1$). To evaluate the initial dynamical behavior of the squared wave-function norm $||\xi(t)||^{2} = \langle \xi(t) | \xi(t) \rangle$, we calculate its instantaneous rate of change applying the product rule for differentiation:
\begin{equation}
\begin{aligned}
\frac{d}{dt}||\xi(t)||^{2} &= \left( \frac{\partial}{\partial t} \langle \xi(t) | \right) |\xi(t)\rangle + \langle \xi(t) | \left( \frac{\partial}{\partial t} |\xi(t)\rangle \right), \\
&= i\langle \xi(t) | [\hat{H}_{0}^{\dagger} + \hat{V}_{\text{noise}}^{\dagger}(t)] |\xi(t)\rangle - i\langle \xi(t) | [\hat{H}_{0} + \hat{V}_{\text{noise}}(t)] |\xi(t)\rangle, \\
&= \langle \xi(t) | i(\hat{H}_{0}^{\dagger} - \hat{H}_{0}) | \xi(t) \rangle.
\end{aligned}
\end{equation}
Notably, the explicit noise term $\hat{V}_{\text{noise}}(t)$ vanishes from this expression. Although this explicit cancellation makes the instantaneous variation of $\|\xi(t)\|^2$ appear independent of the noise, its profound influence is actually hidden within the wave function $|\xi(t)\rangle$ itself.

Let us define the evolution time as $t = t_1 + \Delta t$ (with $\Delta t \rightarrow 0$). Furthermore, we define $\Gamma$ as:
\begin{equation}
\Gamma \equiv \frac{d}{dt}||\xi(t)||^{2} \Big|_{t=t_1} = \langle \xi(t_1) | i(\hat{H}_{0}^{\dagger} - \hat{H}_{0}) | \xi(t_1) \rangle.
\end{equation}
Consequently, the squared norm of $| \xi(t) \rangle$ at $t_1 + \Delta t$ can be approximated as:
\begin{equation}
||\xi(t_1 + \Delta t)||^{2} \approx ||\xi(t_1)||^{2} + \Gamma \Delta t =||\xi(t_1)||^{2} \left( 1 + \frac{\Gamma}{||\xi(t_1)||^{2}} \Delta t \right).
\end{equation}
According to the definition of FTLE in the main text, we obtain:
\begin{equation}
\begin{aligned}
\lambda_{\xi}(t_1 + \Delta t) &= \frac{1}{2(t_1 + \Delta t)} \ln ||\xi(t_1 + \Delta t)||^{2}, \\
&\approx \frac{1}{2(t_1 + \Delta t)} \ln \left[ ||\xi(t_1)||^{2} \left( 1 + \frac{\Gamma}{||\xi(t_1)||^{2}} \Delta t \right) \right], \\
&= \frac{1}{2(t_1 + \Delta t)} \left[ \ln ||\xi(t_1)||^{2} + \ln \left( 1 + \frac{\Gamma}{||\xi(t_1)||^{2}} \Delta t \right) \right].
\end{aligned}\label{xi_lamuda_short_Time1}
\end{equation}
Expanding Eq.~(\ref{xi_lamuda_short_Time1}) to the first order in $\Delta t$, we arrive at the analytical expression for the initial slope:
\begin{equation}
\lambda_{\xi}(t_1 + \Delta t) \approx { \frac{\ln ||\xi(t_1)||^{2}}{2t_1} } + { \left[ \frac{\Gamma}{2 t_1 ||\xi(t_1)||^{2}} - \frac{\ln ||\xi(t_1)||^{2}}{2 t_1^{2}} \right] } \Delta t.
\label{xi_lamuda_short_Time2}
\end{equation}
Eq.~(\ref{xi_lamuda_short_Time2}) captures the two distinct regimes of the initial dynamics:
\begin{enumerate}
    \item \textbf{Low Initial Norm Regime}: If $||\xi(t_1)||^{2}$ is relatively small, the constant term $\frac{\ln ||\xi(t_1)||^{2}}{2t_1}$ is a large negative value, while the slope term becomes a significant positive value. In this case, $\lambda_{\xi}$ starts from a deep negative value and monotonically increases toward $\operatorname{Im}(E_{\text{max}})$.
    
    \item \textbf{High Initial Norm Regime}: If $||\xi(t_1)||^{2}$ is relatively large, the initial value $\frac{\ln ||\xi(t_1)||^{2}}{2t_1}$ is positive. However, the instantaneous growth rate $\Gamma$ is typically negative and large, leading to a negative slope. Consequently, $\lambda_{\xi}$ exhibits a transient dip, initially decreasing before eventually rebounding to approach $\operatorname{Im}(E_{\text{max}})$.
\end{enumerate}

In summary, $\lambda_{\xi}$ either originates from a significantly suppressed value or undergoes an initial decrease, ensuring that $\lambda_{\xi} < \lambda_{\phi}$ persists throughout the early stages of evolution.
\section{Derivation of Strong-Noise Perturbation Theory}
In this section, we develop a comprehensive strong-noise perturbation theory. We systematically derive a discrete master equation, which is subsequently expanded into a continuum non-unitary drift-diffusion equation. This theoretical framework allows us not only to obtain analytical solutions for the probability density evolution under both OBC and PBC, but also to rigorously determine the steady-state Lyapunov exponent ($\lambda_{\infty}$). Furthermore, this continuous formulation provides a powerful analytical tool to dissect the transient dynamics: it enables us to mathematically elucidate how boundary-induced density gradients inherently suppress the instantaneous growth rate, and to strictly prove the universal $1/t$ scaling law governing the FTLE convergence. Collectively, these derivations establish a rigorous mathematical foundation for the strong-noise self-healing mechanisms discussed in the main text.
\subsection{Wave Function Correction under Strong Noise}
Consider a general non-Hermitian tight-binding Hamiltonian:
\begin{equation}
\hat{H}_{0} = \sum_{i,j} t_{i-j} \hat{c}_{j}^\dagger \hat{c}_{i},
\end{equation}
where $t_{i-j}$ denotes the hopping amplitude from site $i$ to site $j$. Upon introducing site-independent noise $\xi_j(t)$, the system dynamics follow the stochastic Schrödinger equation:
\begin{equation}
i\partial_t \psi_j(t) = \sum_{l \neq j} t_{j-l} \psi_l(t) + \xi_j(t) \psi_j(t).
\end{equation}
In the strong-noise limit, we perform a perturbative expansion of the wave function in terms of the hopping order: $\psi_j(t) = \psi_j^{(0)}(t) + \psi_j^{(1)}(t) + \dots$. The zeroth-order solution, which neglects the hopping terms, is given by:
\begin{equation}
\psi_j^{(0)}(t) = \psi_j(0) \exp\left( -i \int_0^t \xi_j(\tau) d\tau \right) = \psi_j(0) e^{-i\phi_j(t)},
\end{equation}
where $\phi_j(t) \equiv \int_0^t \xi_j(\tau) d\tau$ defines the accumulated phase factor. To capture how the inter-site hopping fundamentally perturbs the system, we introduce the first-order correction, which is governed by the following differential equation:
\begin{equation}
\partial_t \psi_j^{(1)}(t) + i \xi_j(t) \psi_j^{(1)}(t) = -i \sum_{l \neq j} t_{j-l} \psi_l^{(0)}(t).
\end{equation}
By introducing the integrating factor $e^{i\phi_j(t)}$, we can rewrite the left-hand side of this equation as a total derivative:
\begin{equation}
\partial_t \left( e^{i\phi_j(t)} \psi_j^{(1)}(t) \right) = -i e^{i\phi_j(t)} \sum_{l \neq j} t_{j-l} \psi_l^{(0)}(t).
\end{equation}
Substituting the explicit form of the zeroth-order solution, $\psi_l^{(0)}(t) = \psi_l(0) e^{-i\phi_l(t)}$, into the right-hand side yields:
\begin{equation}
\partial_t \left( e^{i\phi_j(t)} \psi_j^{(1)}(t) \right) = -i \sum_{l \neq j} t_{j-l} \psi_l(0) e^{i[\phi_j(t) - \phi_l(t)]}.
\end{equation}
Integrating both sides over time from $0$ to $t$ and multiplying by the inverse phase factor $e^{-i\phi_j(t)}$, we obtain the formal solution for the first-order correction:
\begin{equation}
\psi_j^{(1)}(t) = -i e^{-i\phi_j(t)} \int_0^t \sum_{l \neq j} t_{j-l} \psi_l(0) e^{i[\phi_j(\tau) - \phi_l(\tau)]} d\tau.
\end{equation}
\subsection{Derivation of the Discrete Master Equation}
We define the probability density at site $j$ as $P_j(t) \equiv |\psi_j(t)|^2$. The evolution equation it obeys satisfies an exact relation derived directly from the stochastic Schrödinger equation. By applying the product rule for differentiation, we obtain:
\begin{equation}
\frac{d}{dt} P_j = (\partial_t \psi_j^*) \psi_j + \psi_j^* (\partial_t \psi_j).
\end{equation}
Substituting $\partial_t \psi_j = -i \sum_{l \neq j} t_{j-l} \psi_l - i \xi_j(t) \psi_j$ and its complex conjugate into the expression, the real noise terms cancel perfectly ($i \xi_j \psi_j^* \psi_j - i \xi_j \psi_j^* \psi_j = 0$), yielding:
\begin{equation}
\frac{d}{dt} P_j = i \sum_{l \neq j} \left( t_{j-l}^* \psi_l^* \psi_j - t_{j-l} \psi_j^* \psi_l \right) = 2 \text{Im} \left( \psi_j^* \sum_{l \neq j} t_{j-l} \psi_l \right).
\end{equation}
To evaluate this under the strong-noise condition, we substitute the first-order perturbative expansion $\psi_j(t) \approx \psi_j^{(0)}(t) + \psi_j^{(1)}(t)$ into the equation. The leading term involves $\psi_j^{(0)*} \psi_l^{(0)}$, but upon performing an ensemble average $\langle \dots \rangle$ over the noise, this zeroth-order cross term vanishes for $j \neq l$ due to the independence of the on-site noise fluctuations (i.e., $\langle e^{i[\phi_j(t) - \phi_l(t)]} \rangle = 0$). Consequently, the evolution of the expected probability density is governed by the first-order cross terms:
\begin{equation}
\frac{d}{dt} \langle P_j \rangle \approx 2 \text{Im} \sum_{l \neq j} t_{j-l} \left( \langle \psi_j^{(0)*} \psi_l^{(1)} \rangle + \langle \psi_j^{(1)*} \psi_l^{(0)} \rangle \right).
\end{equation}
Using the formal solution for the first-order correction derived previously, the first correlation term can be evaluated. Given the statistical independence of noise across different sites, the cross-correlations $\langle e^{i\phi_j} e^{-i\phi_m} \rangle$ vanish for $j \neq m$. The non-vanishing contribution simplifies to:
\begin{equation}
\langle \psi_j^{(0)*}(t) \psi_l^{(1)}(t) \rangle \approx -i t_{l-j} \langle P_j \rangle \int_0^t \langle e^{i[\phi_j(t) - \phi_j(\tau)]} e^{-i[\phi_l(t) - \phi_l(\tau)]} \rangle d\tau.
\end{equation}
Substituting this back into the evolution rate equation, and combining it with its complex conjugate counterpart, we obtain the discrete master equation: 
\begin{equation}
\frac{d}{dt} \langle P_j \rangle = \sum_{l \neq j} 2 Q(t) \left[ |t_{j-l}|^2 \langle P_l \rangle - |t_{l-j}|^2 \langle P_j \rangle \right],
\end{equation}
Because the noise is independent and identically distributed across different sites, this expectation value can be factored. By defining the site-independent phase coherence function $C_\phi(t-\tau) \equiv \langle e^{i[\phi_j(t) - \phi_j(\tau)]} \rangle$, the integral naturally reduces to a purely real quantity:
\begin{equation}
Q(t) = \int_{0}^{t} |C_{\phi}(t-\tau)|^2 d\tau.
\end{equation}
This confirms that $Q(t)$ is real and site-independent. Consequently, the final form of the discrete master equation simplifies to:
\begin{equation}\label{Discrete_Master_Equation2_en}
\frac{d}{dt} \langle P_j \rangle = 2Q(t)\sum_{l \neq j} \left( |t_{j-l}|^2 \langle P_l \rangle - |t_{l-j}|^2 \langle P_j \rangle \right).
\end{equation}

\subsection{Derivation of the Non-unitary Drift-Diffusion Equation}
To describe the transport properties in the long-wavelength limit, we introduce the lattice constant $a$ and define the continuum probability density $\rho(x, t) = \langle P_j(t) \rangle / a$, where $x = ja$. Performing a Taylor expansion of $\rho(x+na, t)$ around $x$ yields:
\begin{equation}
\rho(x+na, t) \approx \rho(x, t) + na \frac{\partial \rho}{\partial x} + \frac{(na)^2}{2} \frac{\partial^2 \rho}{\partial x^2}
\end{equation}Substituting this expansion into the discrete master equation~\eqref{Discrete_Master_Equation2_en} and collecting terms of the same order, we arrive at the non-unitary drift-diffusion equation:
\begin{equation}\label{Continuous_Master_Equation_en}
\frac{\partial \rho(x,t)}{\partial t} = S \rho(x,t) + v \frac{\partial \rho(x,t)}{\partial x} + D \frac{\partial^2 \rho(x,t)}{\partial x^2}
\end{equation} 

The coefficients governing the dynamics are identified as follows:
\begin{itemize}
    \item \textbf{Source term:} 
    \begin{equation}\label{S}
       S= 2Q(t) \sum_{n \neq 0} \left( |t_n|^2 - t_n t_{-n} \right)=2 Q(t) \sum_{n>0}\left(t_{-n}-t_{n}\right)^{2}
    \end{equation}
    \item \textbf{Drift velocity:} 
    \begin{equation}\label{V}
       v = - 2Q(t) \sum_{n \neq 0} na |t_n|^2= 2Q(t) a \sum_{n > 0} n \left( |t_{-n}|^2 - |t_n|^2 \right)
    \end{equation}
    \item \textbf{Diffusion coefficient:} 
    \begin{equation}\label{D}
        D = Q(t) \sum_{n \neq 0} (na)^2 |t_n|^2= Q(t) a^2 \sum_{n > 0} n^2 \left( |t_{-n}|^2 + |t_n|^2 \right)
    \end{equation}
\end{itemize}
\subsection{Solution of Non-unitary Drift-Diffusion Equation under OBC and Steady-State Lyapunov exponent}

To obtain the solution to Eq.~\eqref{Continuous_Master_Equation_en}, we need employ a transformation. We posit an ansatz of the form:
\begin{equation}\label{Ansatz}
\rho(x,t) = e^{\alpha x + \beta t} u(x,t)
\end{equation}
where $\alpha$ and $\beta$ are undetermined constants, and $u(x,t)$ is the transformed unknown function. Substituting Eq.~\eqref{Ansatz} into Eq.~\eqref{Continuous_Master_Equation_en} and rearranging the terms yields:
\begin{equation}
\frac{\partial u}{\partial t} = D \frac{\partial^2 u}{\partial x^2} + (v + 2D\alpha) \frac{\partial u}{\partial x} + (S + v\alpha + D\alpha^2 - \beta) u
\end{equation}
To reduce this equation to the standard diffusion equation, we must eliminate $\frac{\partial u}{\partial x}$ and $u$ term. This requires satisfying the following conditions:
\begin{equation}
\begin{split}
v + 2D\alpha &= 0, \\
S + v\alpha + D\alpha^2 - \beta &= 0.
\end{split}
\end{equation}
Solving this system of algebraic equations gives the parameters:
\begin{equation}
\begin{split}
 \alpha &= -\frac{v}{2D}, \\
 \beta &= S - \frac{v^2}{4D}.
 \end{split}
\end{equation}
Under this transformation, Eq.~\eqref{Continuous_Master_Equation_en} simplifies to the canonical diffusion equation:
 \begin{equation}
 \frac{\partial u}{\partial t} = D \frac{\partial^2 u}{\partial x^2}\label{diffusion}
 \end{equation}
The general solution to the diffusion equation under OBCs (i.e., $u(0,t)=u(L,t)=0$) can be expressed as a Fourier sine series:
 \begin{equation}
u(x,t) = \sum_{n=1}^{\infty} c_n \sin\left(\frac{n\pi}{L} x\right) e^{-D \frac{n^2 \pi^2}{L^2} t}.
 \end{equation}
The coefficients $c_{n}$ are determined by the initial condition $u(x,0) = f(x)$. Utilizing the orthogonality of the sine functions, $\int_0^L \sin\left(\frac{n\pi}{L} x\right) \sin\left(\frac{m\pi}{L} x\right) dx = \frac{L}{2} \delta_{nm}$, we obtain:
 \begin{equation}
 c_n = \frac{2}{L} \int_0^L f(x) e^{\frac{v}{2D}x} \sin\left(\frac{n\pi}{L} x\right) dx.
  \end{equation}
Consequently, the complete analytical solution under OBC is given by:
 \begin{equation}\label{rho_expression}
  \rho(x,t) = e^{-\frac{v}{2D}x} e^{\left(S - \frac{v^2}{4D}\right)t} \sum_{n=1}^{\infty} c_n \sin\left(\frac{n\pi}{L} x\right) e^{-D \frac{n^2 \pi^2}{L^2} t}.
  \end{equation}
We define the steady-state Lyapunov exponent of the wavefunction as $\lambda _{\infty } = \lim_{t \to \infty} \frac{1}{t} \ln \|\psi(t)\|$. In the long-time limit ($t \to \infty$), the dynamics are dominated by the lowest mode ($n=1$), as higher-order modes decay more rapidly. Therefore, independent of the initial state, the Lyapunov exponent is:
\begin{equation}\label{infty1}
\lambda_{\infty } = \frac{S}{2} - \frac{v^2}{8D} - \frac{D \pi^2}{2L^2}
\end{equation}
Substituting Eqs.~\eqref{S},~\eqref{V}, and~\eqref{D} into Eq.~\eqref{infty1}, and taking $L \to \infty$, we derive:
\begin{equation}\label{infty2}
\lambda_{\infty}=Q(\infty)\left[{\sum_{n>0}\left(t_{-n}-t_{n}\right)^{2}}-{\frac{1}{2} \frac{\left[\sum_{n>0} n\left(\left|t_{-n}\right|^{2}-\left|t_{n}\right|^{2}\right)\right]^{2}}{\sum_{n>0} n^{2}\left(\left|t_{-n}\right|^{2}+\left|t_{n}\right|^{2}\right)}}\right].
\end{equation}
Restricting our analysis to nearest-neighbor and next-nearest-neighbor hopping (consistent with the model discussed in the main text), the expression simplifies to:
\begin{equation}\label{infty_NN_NNN}
\lambda_{\infty} = Q(\infty) \left\{ \left[ (t_{-1} - t_1)^2 + (t_{-2} - t_2)^2 \right] - \frac{1}{2} \frac{\left[ \left( |t_{-1}|^2 - |t_1|^2 \right) + 2 \left( |t_{-2}|^2 - |t_2|^2 \right) \right]^2}{\left( |t_{-1}|^2 + |t_1|^2 \right) + 4 \left( |t_{-2}|^2 + |t_2|^2 \right)} \right\}.
\end{equation}

\subsection{Steady-State Lyapunov Exponent under PBC}
We now consider the case of PBCs. We define the time evolution of the total probability of the wave packet as $N(t) = \int_{0}^{L} \rho(x,t) dx$. Integrating Eq.~\eqref{Continuous_Master_Equation_en} over $[0, L]$, we obtain:
\begin{equation}\label{total_rate}
\frac{d N}{dt}=\frac{d}{dt} \int_{0}^{L} \rho dx = \int_{0}^{L} S \rho dx + v \int_{0}^{L} \frac{\partial \rho}{\partial x} dx + D \int_{0}^{L} \frac{\partial^2 \rho}{\partial x^2} dx
\end{equation}
Under PBC, both density and its derivatives are continuous at the boundaries, i.e., $\rho(0,t) = \rho(L,t)$ and $\partial_x \rho(0,t) = \partial_x \rho(L,t)$. Consequently, the boundary terms vanish, leading to:
\begin{equation}
\frac{d N(t)}{dt} = S N(t) \implies N(t) = N(0) e^{St}
\end{equation}
This implies that for PBC, regardless of the initial profile of the wave packet, the Lyapunov exponent is invariably $\frac{S}{2}$.

\subsection{Suppression of Lyapunov Exponents of Boundary Gradient Terms}

We begin by defining the instantaneous growth rate of the wave-function norm, where $N(t) \equiv \|\psi(t)\|^2$, as:
\begin{equation}
\zeta(t) \equiv \frac{1}{N(t)} \frac{d N(t)}{dt}.
\end{equation}
To establish the fundamental relationship between the FTLE $\lambda(t)$ and this instantaneous rate $\zeta(t)$, we directly integrate $\zeta(\tau)$ over time:
\begin{equation}
\int_0^t \zeta(\tau) d\tau = \int_0^t \frac{d \ln N(\tau)}{d\tau} d\tau = \ln N(t) - \ln N(0).
\end{equation}
Utilizing the definition of the FTLE, $\lambda(t) \approx \frac{1}{2t} \ln N(t)$, we can rearrange this integral to obtain the exact relation presented in the main text:
\begin{equation}
\lambda(t) = \frac{1}{2t} \int_0^t \zeta(\tau) d\tau + \frac{\ln N(0)}{2t}.
\end{equation}
For a general, unnormalized initial state (such as the deviation state $|\xi(t_1)\rangle$ immediately after scattering), the initial memory term $\frac{\ln N(0)}{2t}$ emerges. However, as argued in the main text, this term does not dictate the definitive dynamical evolution. First, in the long-time limit ($t \to \infty$), this $1/t$-dependent term naturally decays to zero. Second, even during the finite-time transient regime, because the initial norm of the spatially truncated deviation state is intrinsically smaller than or equal to that of the reference state ($N_\xi(t_1) \le N_\phi(t_1)$), this memory term merely provides a constant negative baseline shift in the logarithmic domain, which inherently favors the self-healing condition ($\lambda_\xi < \lambda_\phi$). Therefore, we can safely marginalize its impact and conclude that the non-trivial dynamical variations of the FTLE are overwhelmingly governed by the time-averaged accumulation of the instantaneous rate $\zeta(\tau)$.

To investigate the dynamical evolution of $\zeta(t)$, we spatially integrate the non-unitary drift-diffusion equation over the domain $[0, L]$. Imposing OBC (i.e., $\rho(0,t) = \rho(L,t) = 0$), we obtain:
\begin{equation}\label{zeta}
\zeta(t) = S + \frac{D}{N}\left(\left.\frac{\partial \rho}{\partial x}\right|_{L} - \left.\frac{\partial \rho}{\partial x}\right|_{0}\right).
\end{equation}
Compared to the exact analytical expression under periodic boundary conditions, an additional second term appears in this OBC configuration. Physically, this term originates directly from the diffusion process. Intrinsically, the diffusion term in the governing equation acts to spread the probability density as uniformly and diffusely as possible across the entire spatial domain. In stark contrast, the non-Hermitian skin effect continuously drives the wave function to strongly localize, effectively piling up the density against the boundary. To resolve this intense competition and mitigate the steep spatial density gradient enforced by the skin effect, the diffusion process actively suppresses the global norm of the wave function, thereby forcefully reducing the absolute magnitude of the gradient. Notably, this unique feedback mechanism—where boundary-induced spatial gradients dynamically suppress the overall macroscopic amplification rate—constitutes a fundamental characteristic of non-Hermitian skin dynamics.

\subsection{Discussion on the Convergence Behavior of the Lyapunov Exponent}
\begin{figure}
    \centering
    \includegraphics[width=1\linewidth]{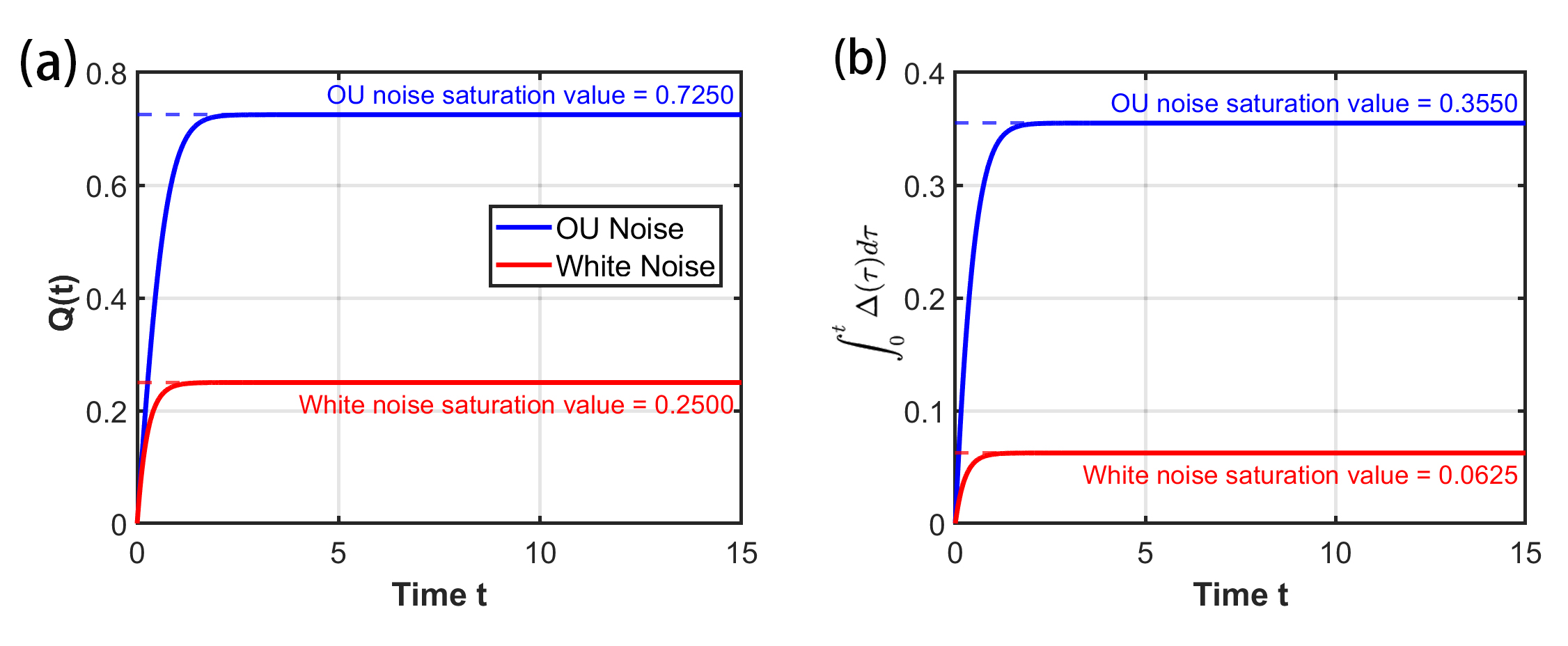}
    \caption{Time evolution of $Q(t)$ (a) and $\int_0^t \Delta(\tau) d\tau$ (b) for Ornstein-Uhlenbeck noise and white noise, where $\Delta(\tau) \equiv Q(\infty) - Q(\tau)$. The quantity $Q(t) = \int_{0}^{t} |C_{\phi}(t-\tau)|^2 d\tau$ is obtained by numerically integrating the analytically derived $|C_{\phi}(t)|^2$. For Gaussian noise, the  function is $|C_{\phi}(t)|^2 = \exp[ - \int_0^t dt_1 \int_0^t dt_2 \langle \xi(t_1)\xi(t_2) \rangle ]$. Substituting the Ornstein-Uhlenbeck noise correlation $\langle \xi(t_1)\xi(t_2) \rangle = \frac{\sigma^2}{2\theta}e^{-\theta|t_1-t_2|}$ yields $|C_{\phi}(t)|^2 = \exp\{ -\frac{\sigma^2}{\theta^2} [ t - \frac{1}{\theta}(1-e^{-\theta t}) ] \}$. For white noise, defined by $\langle \xi(t_1)\xi(t_2) \rangle = \Gamma \delta(t_1-t_2)$, this simplifies to $|C_{\phi}(t)|^2 = \exp(-\Gamma t)$. The parameters used in the simulations are $\Gamma = 4$, $\sigma = 2$, and $\theta = 1$. }
    \label{Different_noise}
\end{figure}

To analytically establish the convergence behavior of the FTLE in the long-time limit, we examine the asymptotic dynamics of the system. In general, the noise kernel integral $Q(t)$ becomes a slowly varying function of time. Consequently, the effective coefficients $S$, $v$, and $D$ in the non-unitary drift-diffusion equation \eqref{Continuous_Master_Equation_en} also become slowly varying time-dependent parameters. To evaluate $\zeta(t)$, we integrate Eq.~\eqref{Continuous_Master_Equation_en} spatially under OBC, yielding $\zeta(t) \equiv \dot{N}/N = S(t) + \frac{D(t)}{N(t)}(\partial_x\rho|_L - \partial_x\rho|_0)$. In the long-time limit, the spatial profile of the wave function asymptotically approaches the ansatz $\rho(x,t) \propto e^{\alpha(t) x}\sin(\frac{\pi}{L}x)$, where $\alpha(t)=-v(t)/2D(t)$. Substituting this asymptotic profile into the boundary gradient terms simplifies it to $-[\frac{v(t)^2}{4D(t)} + \frac{D(t)\pi^2}{L^2}]$. In this regime, the instantaneous growth rate $\zeta(t)$ can be approximated as:
\begin{equation}
\zeta(t) =  S(t) - \frac{v(t)^2}{4D(t)} - \frac{D(t)\pi^2}{L^2}.
\end{equation}
According to Eq.~\eqref{S}, Eq.~\eqref{V} and Eq.~\eqref{D}, the effective coefficients derived previously, $S$, $v$, and $D$ are all directly proportional to the noise kernel integral $Q(t)$. Therefore, it follows directly from the expression above that $\zeta(t)$ is also strictly proportional to $Q(t)$. For notational convenience, we express this relationship as $\zeta(t) = KQ(t)$, where $K$ summarizes the proportionality constants. Utilizing the fundamental relationship between the instantaneous rate $\zeta(t)$ and the FTLE $\lambda(t)$ established earlier, we obtain the exact expression:
\begin{equation}
\lambda(t) = \frac{1}{2t} \int_0^t \zeta(\tau) d\tau + \frac{\ln N(0)}{2t} = \frac{K}{2t} \int_0^t Q(\tau) d\tau + \frac{\ln N(0)}{2t}.
\end{equation}
For typical stochastic processes, the noise kernel integral $Q(t)$ evolves over time and eventually saturates to a steady-state constant $C$, as shown in Figs. \ref{Different_noise}(a)-(b). We define the transient deviation from this steady state as $\Delta(\tau) \equiv C - Q(\tau)$. Substituting $Q(\tau) = C - \Delta(\tau)$ into the integral yields:
\begin{equation}
\lambda(t) = \frac{K}{2t} \int_0^t [C - \Delta(\tau)] d\tau + \frac{\ln N(0)}{2t} = \frac{KC}{2} - \frac{K}{2t} \int_0^t \Delta(\tau) d\tau + \frac{\ln N(0)}{2t}.
\end{equation}
Clearly, as $t \to \infty$, the first term dictates the steady-state Lyapunov exponent: $\lambda_\infty = \frac{KC}{2}$. Furthermore, for typical exponentially decaying noise correlations, the integral of the deviation converges to a finite constant in the long-time limit, i.e., $\int_0^\infty \Delta(\tau) d\tau < \infty$. Consequently, the finite-time deviation from the asymptotic value can be rigorously expressed as:
\begin{equation}
\lambda(t) - \lambda_\infty \simeq \frac{1}{2t} \left( \ln N(0) - K \int_0^\infty \Delta(\tau) d\tau \right).
\end{equation}
This rigorous derivation analytically confirms that, in the strong-noise limit, the FTLE converges to its steady-state value strictly following a universal $1/t$ scaling law.

\section{Validity of the FTLE-Based Estimator and Statistical Evaluation of Self-Healing}
\begin{figure*}
    \centering
    \includegraphics[width=1\linewidth]{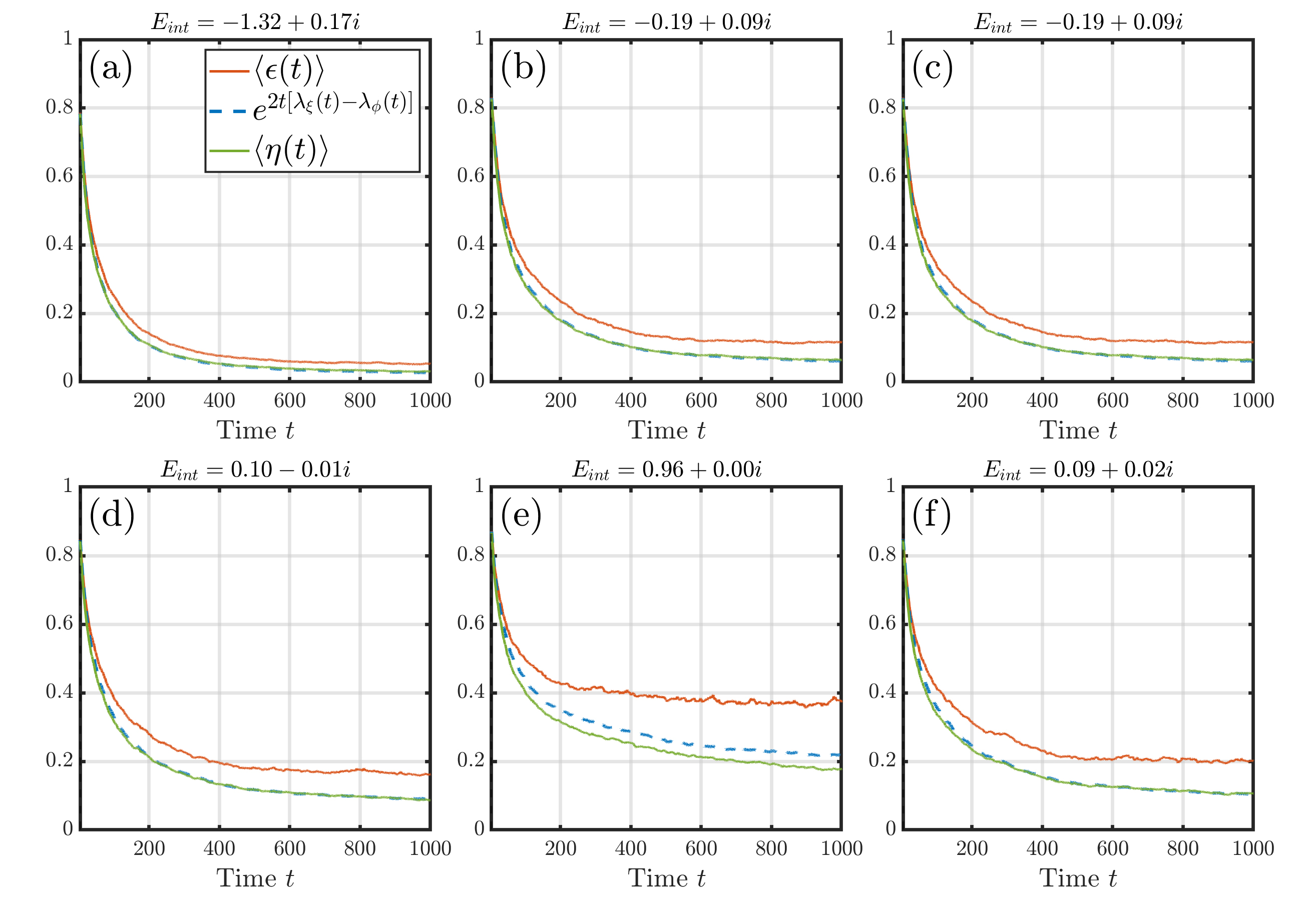}
    \caption{(a)–(f) display the time evolution of three metrics for six representative initial eigenstates with eigenenergies $E_{int}$. Simulations are performed on a 1D non-Hermitian tight-binding lattice ($L=100$) with non-reciprocal hopping parameters $(t_1, t_{-1}, t_2, t_{-2}) = (0.7, 1, 0.8, 1)$ (identical to those used in the main text). To perturb the left-boundary skin modes, a transient potential $\hat V_{\text{scat}}=-i\gamma\sum_{j=1}^{l} \hat c_j^\dagger \hat c_j$ ($\gamma=10$) is applied to the leftmost $l=10$ sites for $t \in [2, 4]$. The stochastic environment is modeled by on-site Ornstein-Uhlenbeck noise (relaxation rate $\theta = 5$, intensity $\sigma = 10$). The solid orange line represents the arithmetic mean of the relative deviation $\langle\epsilon(t)\rangle$, the dashed blue line depicts the FTLE-based estimator $e^{2t[\lambda_\xi(t)-\lambda_\phi(t)]}$, and the solid green line tracks the profile self-healing factor $\langle\eta(t)\rangle$. Results are averaged over 1000 noise realizations. }
    \label{eta_average}
\end{figure*}

In the main text, we approximate the typical ensemble behavior of the relative deviation using the FTLEs as $\langle \eta(t) \rangle \simeq \exp\{ 2t [ \lambda_{\xi}(t) - \lambda_{\phi}(t) ] \}$. To visually justify this approximation, Figs. \ref{eta_average} directly compares three distinct stochastic metrics quantifying the self-healing process across four representative initial eigenstates. The solid orange line represents the arithmetic mean of the relative deviation, $\langle \epsilon(t) \rangle$. The dashed blue line depicts the FTLE-based estimator $\exp\{ 2t [ \lambda_{\xi}(t) - \lambda_{\phi}(t) ] \}$, and the solid green line tracks the actual profile self-healing metric $\langle \eta(t) \rangle$.

A prominent feature universally observed across all panels (a)-(f) in Figs. \ref{eta_average} is that $\langle \epsilon(t) \rangle$ consistently maintains a notably higher value than the FTLE-based estimator $\langle \epsilon(t) \rangle \simeq \exp\{ 2t [ \lambda_{\xi}(t) - \lambda_{\phi}(t) ] \}$ throughout the entire time evolution. Mathematically, this persistent discrepancy is a direct consequence of Jensen's inequality, which guarantees that $\langle \epsilon(t) \rangle \ge \exp\{ \langle \ln \epsilon(t) \rangle \}$ always holds.

Most importantly, the numerical results in Figs. \ref{eta_average} reveal a striking quantitative agreement between the FTLE-based estimator $\exp\{ 2t [ \lambda_{\xi}(t) - \lambda_{\phi}(t) ] \}$ (dashed blue line) and the profile self-healing metric $\langle \eta(t) \rangle$ (solid green line). Regardless of the specific initial eigenenergy $E_{int}$, these two curves approximately coincide and decay toward the same stable, near-zero plateau. This effectively demonstrates that $\exp\{ 2t [ \lambda_{\xi}(t) - \lambda_{\phi}(t) ] \}$ serves as a reliable estimator for $\langle \eta(t) \rangle$.

\iffalse
Furthermore, it is worth noting that the temporal evolution of $\eta(t)$ presented in the main text is primarily based on single stochastic trajectories rather than ensemble averages. This distinction is crucial because genuine physical self-healing must manifest within the dynamical evolution of an individual wavefunction. In highly fluctuating stochastic systems, the decay of the ensemble average $\langle \eta(t) \rangle \to 0$ does not mathematically guarantee that every single trajectory will strictly approach zero over time. Nevertheless, $\langle \eta(t) \rangle$ remains a robust statistical metric for characterizing the self-healing phenomenon. In any single realization, while $\eta(t)$ inevitably exhibits random temporal fluctuations, its overall trend closely follows the decaying envelope of the ensemble average. Therefore, the asymptotic decay of $\langle \eta(t) \rangle$ statistically captures the dominant tendency of the system, serving as an effective macroscopic reference for evaluating the self-healing dynamics of individual wavefunctions.
\fi

\section{Relationship between Eigenstate Extension and Self-Healing Capability}

\subsection{Characterization of Eigenstate Localization: Skin Corner Weight}
To quantitatively characterize the localization properties of eigenstates across the complex energy spectrum, we employ the skin endpoint weight $w_{\text{CW}}^{(2)}(E_n)$\cite{PhysRevB.111.155121}. For a 1D chain with two endpoints, the corner weight for an eigenstate $|\psi_n\rangle$ is defined as:
\begin{equation}
w_{CW}^{(2)}(E_{n})\equiv\sum_{r}\sum_{i=1}^{2}(-1)^{i}|\psi_{n}(r)|^{4}e^{-|r-r_{i}|/\delta}, 
\end{equation}
where $r_i$ denotes the positions of the two ends ($r_1=1$ and $r_2=L$), and $\delta$ is a characteristic decay length chosen to be much smaller than the system size. The magnitude of $w_{\text{CW}}^{(2)}(E_n)$ quantifies the localization strength, while its sign indicates the direction: a negative value signifies localization at the left boundary ($r_1$), and a positive value signifies localization at the right boundary ($r_2$). As illustrated in Figs. \ref{Spectrum} (a), skin modes become increasingly extended as their eigenenergies approach the spectral edges.

\subsection{Connecting Eigenstate Extension to Dynamical Self-Healing via Spatial Truncation}
\begin{figure*}
    \centering
    \includegraphics[width=1.1\linewidth]{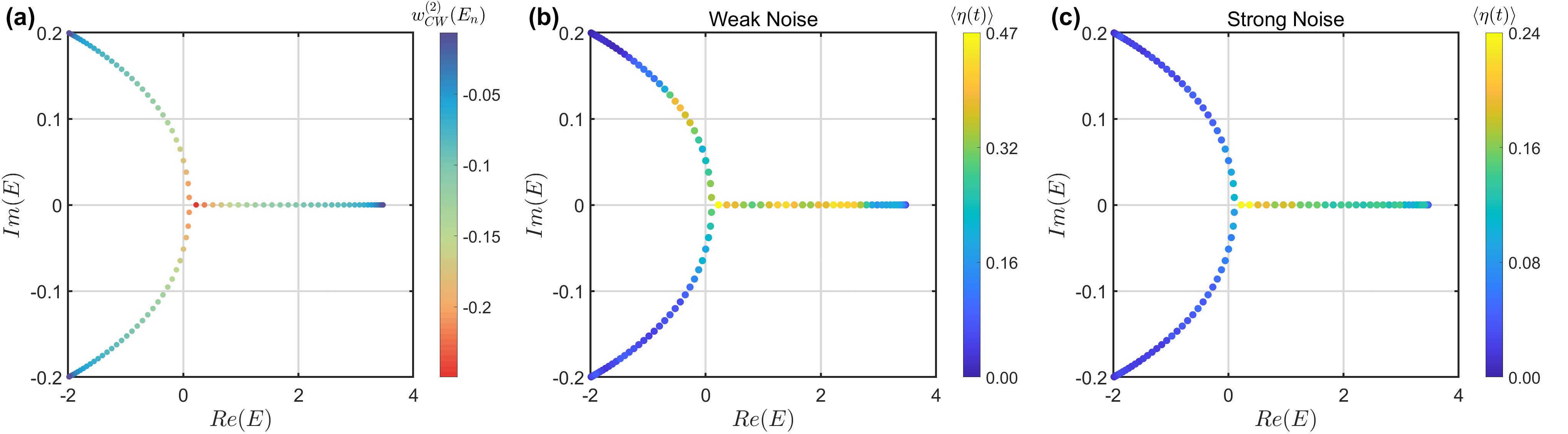}
    \caption{ (a) Distribution of the skin endpoint weight $w_{\text{CW}}^{(2)}(E_n)$ for all eigenstates on the complex energy plane under OBC with $\delta=10$. The color scale quantifies the degree of localization: more negative values indicate stronger localization at the left boundary, while values near zero correspond to more extended modes. Notably, eigenstates closer to the spectral edges exhibit significantly larger spatial extensions (less negative $w_{\text{CW}}^{(2)}$). (b)-(c) Snapshots of the ensemble-averaged profile self-healing metric $\langle \eta(t) \rangle$ across the entire spectrum under (b) weak noise $[(\theta, \sigma) = (1, 0.1)]$ at $t = 60$ and (c) strong noise $[(\theta, \sigma) = (5, 10)]$ at $t = 1000$. Simulations are performed on a 1D non-Hermitian tight-binding lattice of size $L=100$ with non-reciprocal hopping parameters $(t_1, t_{-1}, t_2, t_{-2}) = (0.7, 1, 0.8, 1)$. A transient scattering potential $\hat V_{\text{scat}}=-i\gamma\sum_{j=1}^{l} \hat c_j^\dagger \hat c_j$ is applied to the leftmost $l = 10$ sites during $t \in [2, 4]$. Results in (b) and (c) are averaged over 100 independent realizations of on-site Ornstein-Uhlenbeck noise.}
    \label{Spectrum}
\end{figure*}
The spectral non-uniformity of the degree of eigenstate localization provides a direct and profound physical explanation for the initial-state dependence of the dynamical self-healing capability. Based on our proposed spatial truncation picture, for a fixed scattering range $l$, an extended mode undergoes more severe spatial truncation during the scattering process compared to a highly localized one. This truncation effect directly modulates the instantaneous growth rate $\zeta(t) = \frac{1}{N}\frac{dN}{dt}$; as the steeper boundary density gradients induced in the truncated deviation state $|\xi\rangle$ result in a larger negative contribution to $\zeta_{\xi}$ relative to the reference state $\zeta_{\phi}$ according to Eq.~\eqref{zeta}.

The resulting suppression of $\zeta_{\xi}$ further induces a markedly diminished initial $\lambda_{\xi}(t_1)$. This discrepancy creates a steeper negative initial slope for the evolution of the cumulative difference $t[\lambda_{\xi}(t) - \lambda_{\phi}(t)]$ and drives its long-time saturation toward a more negative regime. According to the proxy relation $\langle\eta(t)\rangle \simeq e^{2t[\lambda_{\xi}(t)-\lambda_{\phi}(t)]}$, such a deeper saturation ensures a more potent exponential suppression of the profile mismatch $\eta(t)$ for extended skin modes.

As illustrated in Figs. \ref{Spectrum} (a), eigenstates closer to the spectral edges exhibit enhanced spatial extension, which elucidates the numerical observations in Figs. \ref{Spectrum} (b) and (c). Specifically, eigenstates residing near the spectral center yield larger ensemble-averaged $\eta$ values, signifying inherently weaker self-healing capabilities. These numerical results strongly validate the truncation theory mentioned above, revealing how non-Hermitian skin effects determine the robustness of dynamic processes through boundary gradient feedback mechanisms.

\section{Universality of Noise-Enhanced Self-Healing and the Lattice Discreteness Effect}
\begin{figure*}
    \centering
    \includegraphics[width=1.1\linewidth]{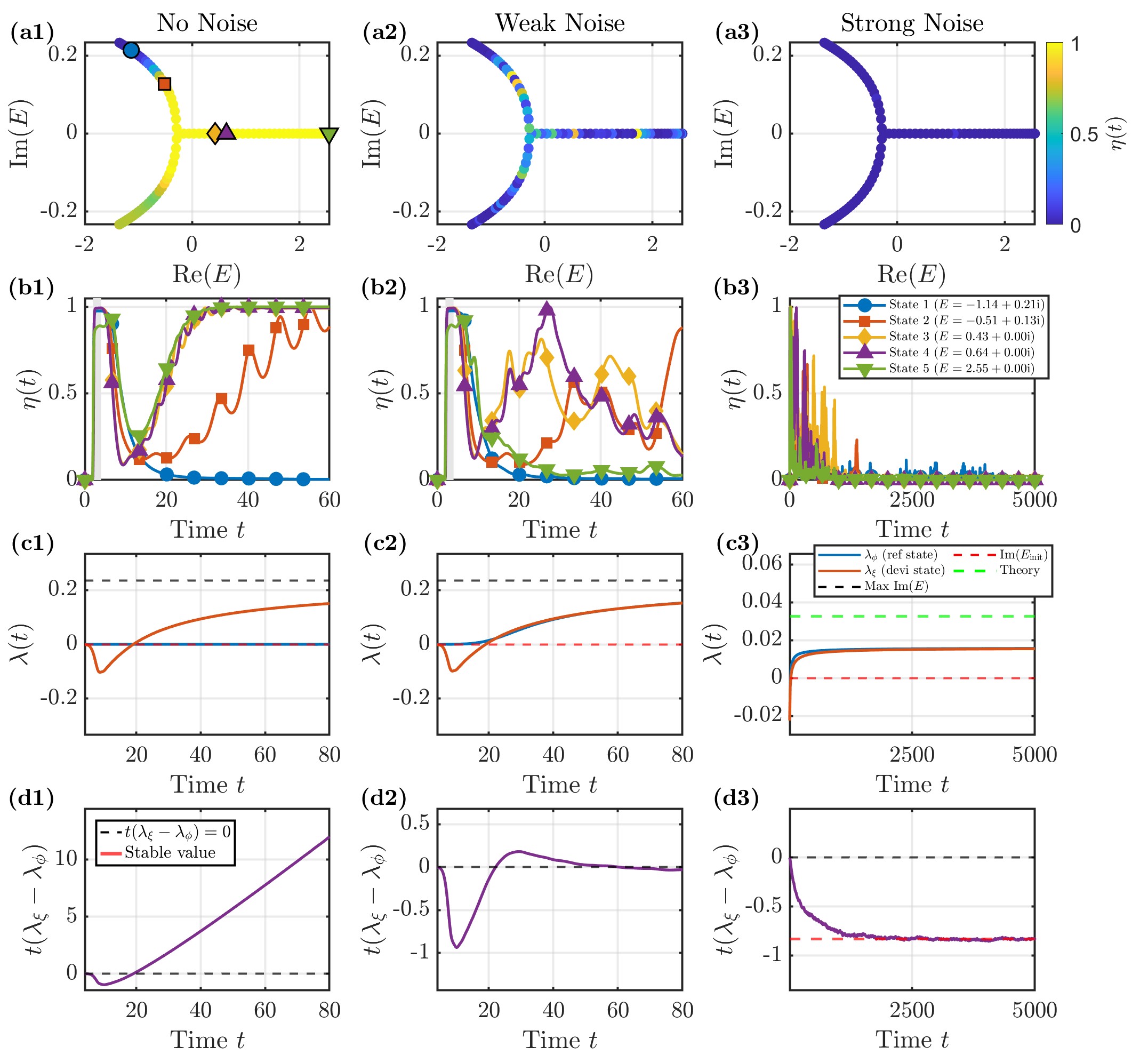}
    \caption{Numerical simulation of noise-enhanced self-healing dynamics and FTLE evolution for a 1D non-Hermitian lattice. The figure presents numerical results simulated on a lattice of size $L=100$ with non-reciprocal hoppings $(t_1,t_{-1},t_2,t_{-2},t_3,t_{-3})=(0.5,1,0.4,0.8,0,0)$. To simulate the scattering process, we apply a diagonal scattering potential $\hat V_{\text{scat}}=-i\gamma\sum_{j=1}^{l} \hat c_j^\dagger \hat c_j$ ($\gamma=10$) localized at the left boundary with a spatial extent of $l=10$ sites, which is turned on only during the time interval $t \in [2, 4]$. We consider three scenarios defined by the Ornstein-Uhlenbeck noise parameters $(\theta, \sigma)$: No Noise in the left column, Weak Noise $(1, 0.1)$ in the middle column, and Strong Noise $(5, 10)$ in the right column, where noise is independently generated and added to the diagonal elements of the Hamiltonian at each time step $dt=0.1$.  (a1)-(a3) display snapshots of the profile self-healing metric $\eta(t)$ plotted on the unperturbed energy spectrum, taken at $t=60$ for No/Weak noise and $t=5000$ for Strong noise, with 5 randomly selected eigenstates marked by distinct symbols. (b1)-(b3) track the real-time evolution of $\eta(t)$ specifically for these 5 selected states. (c1)-(c3) and (d1)-(d3) illustrate the FTLEs ($\lambda_\phi$, $\lambda_\xi$) and the cumulative difference $t(\lambda_\xi - \lambda_\phi)$, respectively. These quantities are calculated specifically for the 3rd sorted state among the 5 randomly selected ones, and the results are averaged over an ensemble of $1000$ independent noise realizations. In (c3), the green dashed line represents the theoretically calculated asymptotic limit $\lambda_{\text{theory}}$, while in (d3), the red dashed line indicates the numerically extracted stable value of the difference averaged over the final $20\%$ of the long-time evolution. }
    \label{S4}
\end{figure*}

\begin{figure*}
    \centering
    \includegraphics[width=1.1\linewidth]{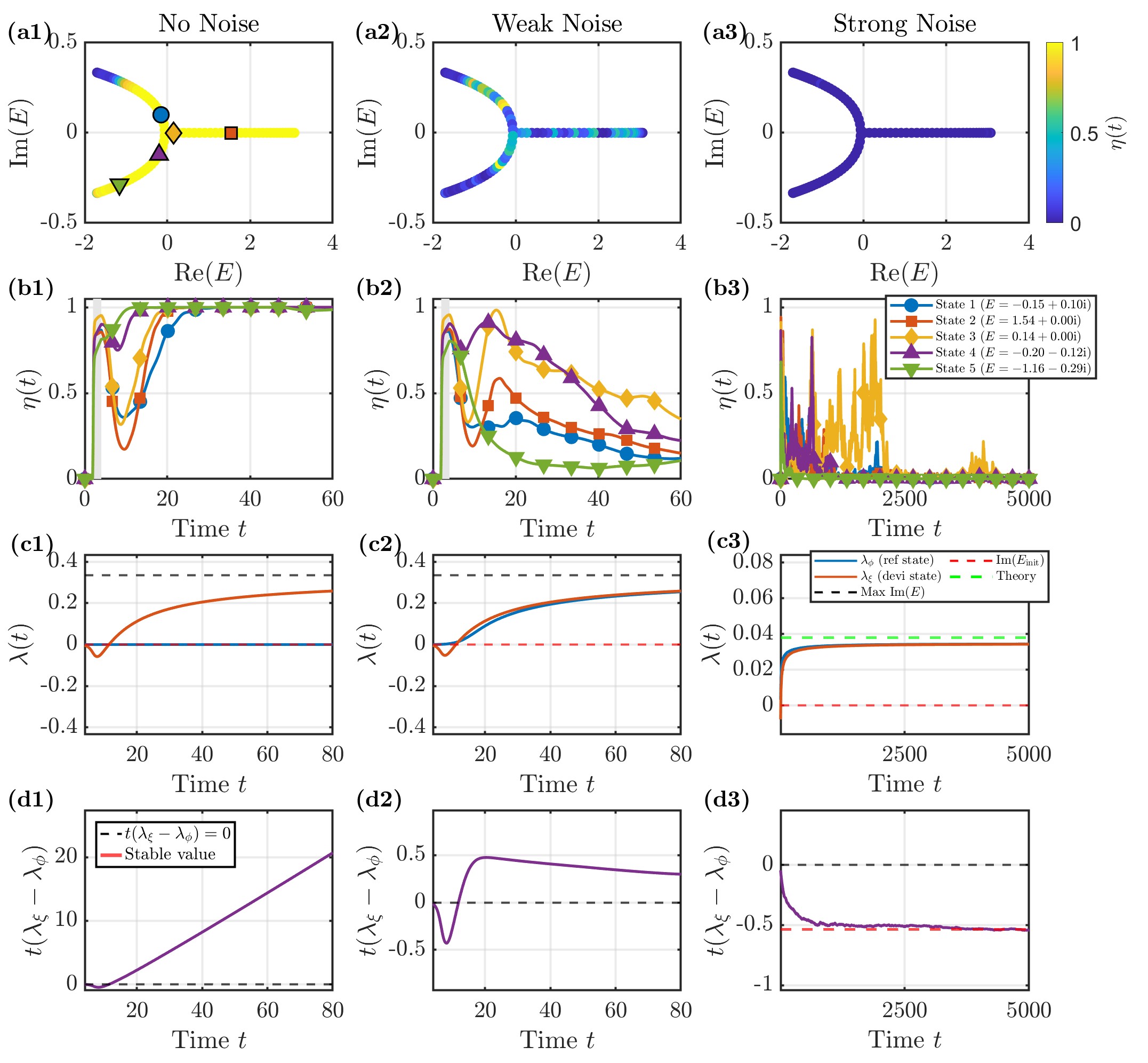}
    \caption{Numerical simulation of noise-enhanced self-healing dynamics and FTLE evolution for a 1D non-Hermitian lattice. The figure presents numerical results simulated on a lattice of size $L=100$ with non-reciprocal hoppings $(t_1,t_{-1},t_2,t_{-2},t_3,t_{-3})=(0.6,1,0.7,0.8,0,0)$. To simulate the scattering process, we apply a diagonal scattering potential $\hat V_{\text{scat}}=-i\gamma\sum_{j=1}^{l} \hat c_j^\dagger \hat c_j$ ($\gamma=10$) localized at the left boundary with a spatial extent of $l=10$ sites, which is turned on only during the time interval $t \in [2, 4]$. We consider three scenarios defined by the Ornstein-Uhlenbeck noise parameters $(\theta, \sigma)$: No Noise in the left column, Weak Noise $(1, 0.1)$ in the middle column, and Strong Noise $(5, 10)$ in the right column, where noise is independently generated and added to the diagonal elements of the Hamiltonian at each time step $dt=0.1$.  (a1)-(a3) display snapshots of the profile self-healing metric $\eta(t)$ plotted on the unperturbed energy spectrum, taken at $t=60$ for No/Weak noise and $t=5000$ for Strong noise, with 5 randomly selected eigenstates marked by distinct symbols. (b1)-(b3) track the real-time evolution of $\eta(t)$ specifically for these 5 selected states. (c1)-(c3) and (d1)-(d3) illustrate the FTLEs ($\lambda_\phi$, $\lambda_\xi$) and the cumulative difference $t(\lambda_\xi - \lambda_\phi)$, respectively. These quantities are calculated specifically for the 3rd sorted state among the 5 randomly selected ones, and the results are averaged over an ensemble of $1000$ independent noise realizations. In (c3), the green dashed line represents the theoretically calculated asymptotic limit $\lambda_{\text{theory}}$, while in (d3), the red dashed line indicates the numerically extracted stable value of the difference averaged over the final $20\%$ of the long-time evolution.}
    \label{S5}
\end{figure*}

\begin{figure*}
    \centering
    \includegraphics[width=1.1\linewidth]{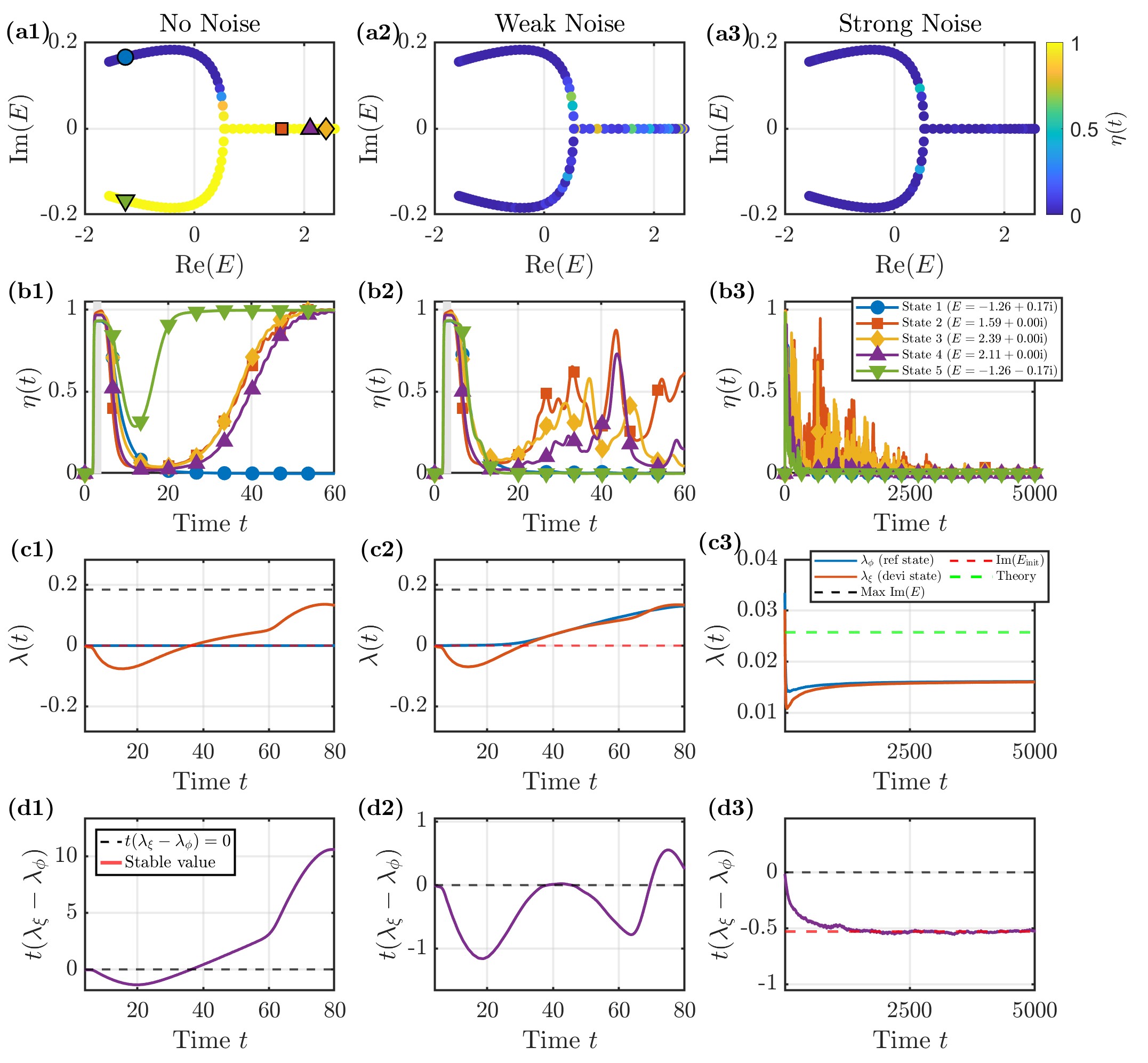}
    \caption{Numerical simulation of noise-enhanced self-healing dynamics and FTLE evolution for a 1D non-Hermitian lattice. The figure presents numerical results simulated on a lattice of size $L=100$ with non-reciprocal hoppings $(t_1,t_{-1},t_2,t_{-2},t_3,t_{-3})=(0.2,0.5,0.6,1,0.1,0.3)$. To simulate the scattering process, we apply a diagonal scattering potential $\hat V_{\text{scat}}=-i\gamma\sum_{j=1}^{l} \hat c_j^\dagger \hat c_j$ ($\gamma=10$) localized at the left boundary with a spatial extent of $l=10$ sites, which is turned on only during the time interval $t \in [2, 4]$. We consider three scenarios defined by the Ornstein-Uhlenbeck noise parameters $(\theta, \sigma)$: No Noise in the left column, Weak Noise $(1, 0.1)$ in the middle column, and Strong Noise $(5, 10)$ in the right column, where noise is independently generated and added to the diagonal elements of the Hamiltonian at each time step $dt=0.1$.  (a1)-(a3) display snapshots of the profile self-healing metric $\eta(t)$ plotted on the unperturbed energy spectrum, taken at $t=60$ for No/Weak noise and $t=5000$ for Strong noise, with 5 randomly selected eigenstates marked by distinct symbols. (b1)-(b3) track the real-time evolution of $\eta(t)$ specifically for these 5 selected states. (c1)-(c3) and (d1)-(d3) illustrate the FTLEs ($\lambda_\phi$, $\lambda_\xi$) and the cumulative difference $t(\lambda_\xi - \lambda_\phi)$, respectively. These quantities are calculated specifically for the 3rd sorted state among the 5 randomly selected ones, and the results are averaged over an ensemble of $1000$ independent noise realizations. In (c3), the green dashed line represents the theoretically calculated asymptotic limit $\lambda_{\text{theory}}$, while in (d3), the red dashed line indicates the numerically extracted stable value of the difference averaged over the final $20\%$ of the long-time evolution.}
    \label{S6}
\end{figure*}

To demonstrate that the noise-enhanced edge self-healing mechanism does not originate from fine-tuned parameters but is an intrinsic dynamical property of non-Hermitian systems exhibiting the skin effect, we further investigated three alternative lattice models with distinct parameter configurations, as shown in Figs. \ref{S4}, \ref{S5} and \ref{S6}.

As can be seen, despite the differences in their underlying hopping configurations, the introduction of environmental noise consistently induces highly similar dynamical evolution across these models: weak noise significantly prolongs the time window for wave-packet self-healing, while strong noise ensures universal self-healing for all eigenstates in the long-time limit.

It is worth noting that a careful observation of the long-time FTLEs in the strong-noise regime [e.g., Figs. \ref{S4}(c3) and  \ref{S6}(c3)] reveals a slight discrepancy between the theoretically predicted asymptotic limit (green dashed line) and the exact numerical steady-state values. This minor deviation is not a breakdown of the physical mechanism but stems from the continuum approximation utilized in our strong-noise perturbation theory.

To quantitatively explain this phenomenon, we define a characteristic correlation length of the steady-state envelope as $\xi = D_0 / v_0$, which characterizes its spatial decay profile $\rho(x) \sim e^{-x/\xi}$, where, $v_0 = \sum n(|t_{-n}|^2 - |t_n|^2)$ and $D_0 = \frac{1}{2} \sum n^2(|t_{-n}|^2 + |t_n|^2)$. Intrinsically, mapping the discrete lattice master equation to a continuous non-unitary drift-diffusion equation requires the spatial variation of the wave function to be relatively smooth over the lattice constant $a=1$. In practice, a correlation length greater than 3 (i.e., $\xi > 3$) is already sufficient to provide a highly accurate continuum approximation.

By substituting the specific hopping parameters, we calculate the correlation length for the main-text model to be $\xi_{\text{main}} \approx 3.27$, where it can be seen that the theoretical prediction and numerical results show excellent agreement. Similarly, the model in Figs. \ref{S5}, with a correlation length of $\xi_{\text{S5}} \approx 3.13$, also satisfies this condition and exhibits high analytical accuracy. However, for other configurations in the Supplementary Material, this correlation length is significantly reduced (e.g., $\xi_{\text{S6}} \approx 1.92$ for the model in  \ref{S6}, and further down to $\xi_{\text{S4}} \approx 1.30$ for \ref{S4}). As the characteristic length approaches the lattice constant, the aforementioned continuum approximation becomes less accurate. This leads to the observable gap between the strictly continuous analytical prediction and the exact discrete numerical dynamics in models with smaller correlation lengths, such as Figs. \ref{S4} and \ref{S6}.

In conclusion, the consistent behaviors observed in Figs. \ref{S4}, \ref{S5} and \ref{S6} compellingly verify that the noise-enhanced self-healing phenomenon is not a coincidence under specific parameters, but a highly universal characteristic of non-Hermitian skin dynamics in noisy environments.

\section{Fragility of Coherent Non-Hermitian Dynamics under Stochastic Noise}
In the absence of noise, real-time wave-packet dynamics in non-Hermitian systems are fundamentally governed by delicate phase coherence and precise interference among participating eigenmodes. Building upon this underlying physical picture and employing saddle-point approximations, recent theoretical works have predicted two notable phenomena: the bulk self-healing effect \cite{llbb-pcgk} and the suppressed short-time plateau of FTLE \cite{xue2025nonblochedgedynamicsnonhermitian}. However, our numerical results reveal that these coherent dynamical phenomena are extraordinarily fragile, rapidly collapsing even under extremely weak stochastic noise.

Physically, continuous noise introduces random phase accumulations that act as a dephasing mechanism, directly leading to the breakdown of both bulk self-healing and the short-time Lyapunov exponent plateau. Theoretically, the saddle-point approximation fundamentally relies on a global, coherent momentum integration over the generalized Brillouin zone, which intrinsically assumes strict translational invariance and deterministic phase evolution. Once stochastic fluctuations break this spatial symmetry and scramble phase coherence, the mathematical foundation of this integration framework becomes invalid.

\begin{figure*}[htbp]
    \centering

 % 第二张图 (Figure S8)
    \includegraphics[width=0.7\linewidth]{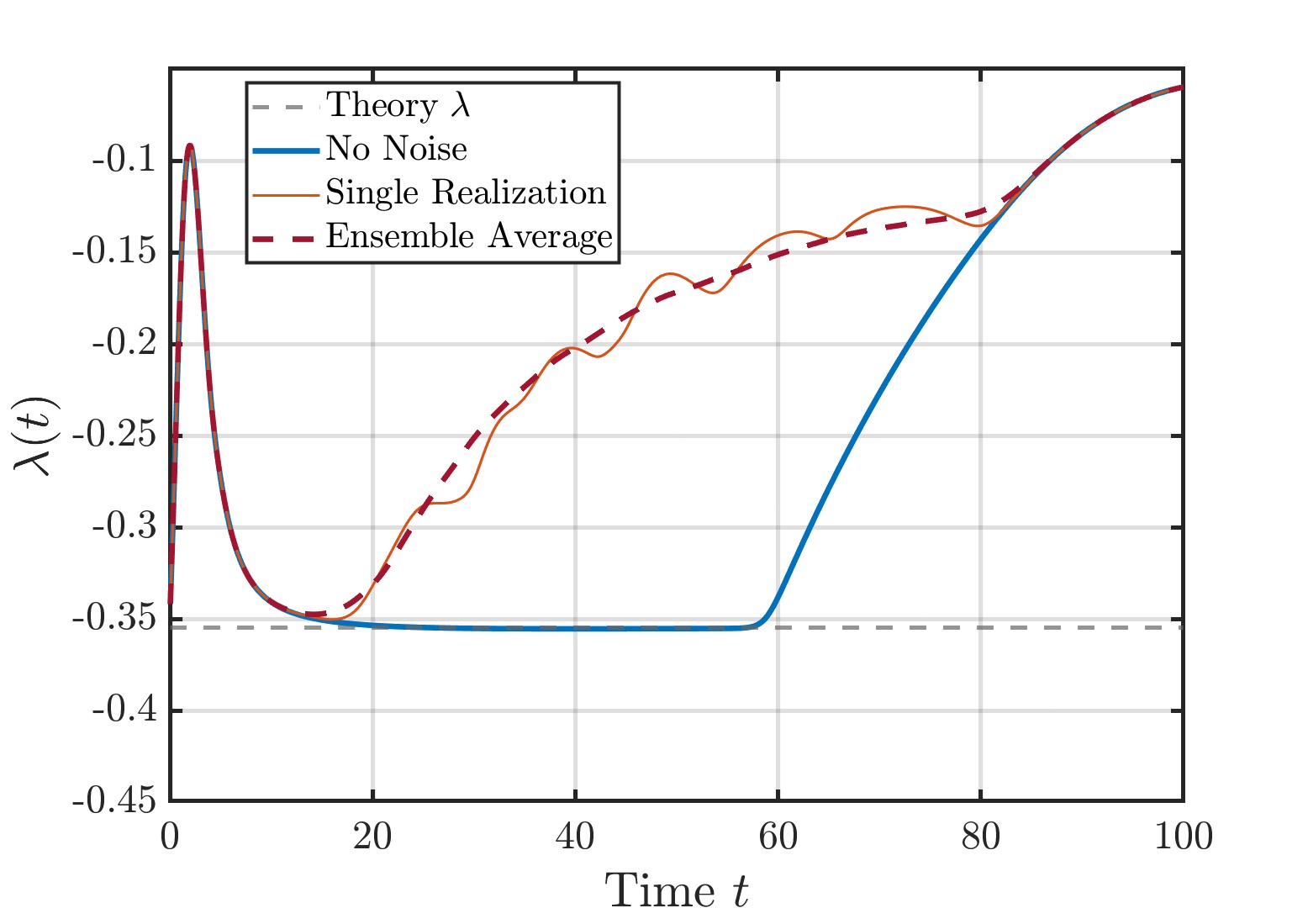}
    \caption{Numerical simulation of FTLE under clean and weakly noisy conditions. The time evolution of the FTLE, $\lambda(t) = \frac{1}{t} \ln ||\psi(t)||$ (consistent with the definition in \cite{xue2025nonblochedgedynamicsnonhermitian}), is simulated for a 1D non-Hermitian lattice with 100 sites. The unperturbed Hamiltonian parameters are $t_{-1} = 1.2i$, $t_{1} = -0.8i$, $t_{-2} = 0.35i$, and $t_{2} = 0.05i$, with a uniform onsite loss $\kappa = 0.35$. The system is initialized with a strictly localized wave packet $|\psi(0)\rangle = \delta_{x,5}$. The blue solid line shows the deterministic evolution under the noiseless Hamiltonian. The light orange solid line depicts a single stochastic realization of the dynamics perturbed by an onsite Ornstein-Uhlenbeck noise, with a relaxation rate $\theta = 1.0$ and a weak intensity $\sigma = 0.01$. The dark red dashed line represents the ensemble average over 100 independent noise realizations. The numerical integration is performed using a time step $dt = 0.05$. The black dashed horizontal line marks the theoretical short-time reference value ($\lambda_{theo} \approx -0.3549$), analytically predicted by the saddle-point approximation (Lefschetz thimble method) detailed in recent literature \cite{xue2025nonblochedgedynamicsnonhermitian}.}
    \label{S8}

    % 第一张图 (Figure S7)
    \includegraphics[width=0.95\linewidth]{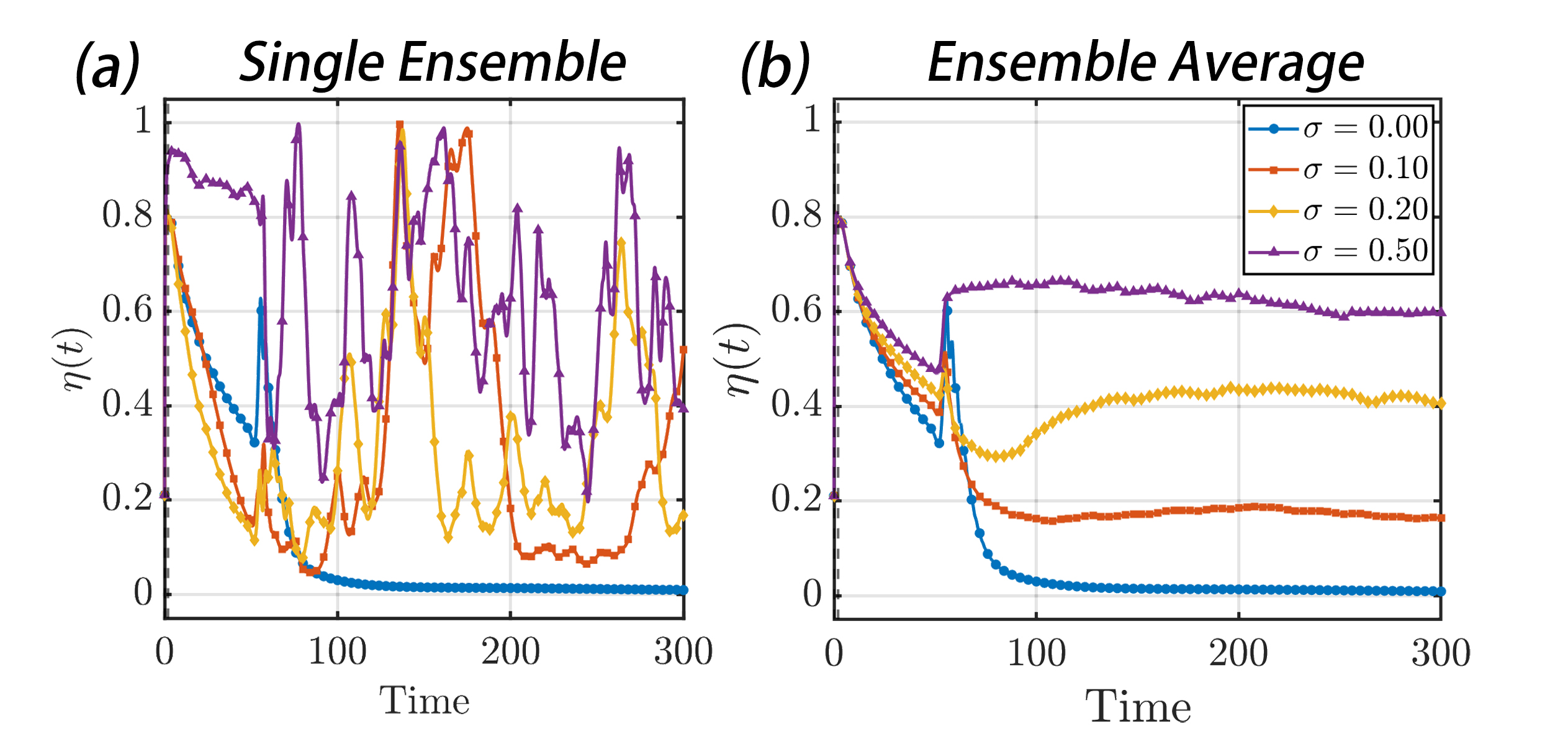}
    \caption{Evolution of the profile self-healing metric $\eta(t)$ in the bulk regime under various noise intensities $\sigma$. (a) Results for a single ensemble realization. (b) Ensemble-averaged results over 1000 realizations. In both cases, the initial state is a Gaussian wave packet localized at the system center. A transient scattering potential $V_{\text{scat}} = -i\gamma$ ($\gamma = 10$) is applied at the center over a region of length $l = 5$ for $t \in [0, 2]$. Other model parameters are $(t_1, t_{-1}, t_2, t_{-2}, L) = (0.7, 1, 0.8, 1, 500)$.}
    \label{bulk_selfHealing_theta1}
    
    \vspace{0.8cm} % 在两张图之间加一点垂直间距，避免太挤。如果排版放不下，可以把 0.8cm 改小比如 0.4cm

\end{figure*}

To explicitly corroborate the breakdown of this integration framework, we examine the short-time FTLE dynamics in Fig. \ref{S8}. Here, we initialize a strictly localized wave packet ($|\psi(0)\rangle = \delta_{x,5}$) within a 1D non-Hermitian lattice. Under purely deterministic evolution, the FTLE closely tracks the theoretical short-time plateau ($\lambda_{theo} \approx -0.3549$) analytically predicted by the Lefschetz thimble (saddle-point) method. However, once extremely weak Ornstein-Uhlenbeck noise ($\sigma = 0.01$) is introduced, the protective interference effect that momentarily restrains the overall amplitude is abruptly stripped away. As evidenced by the evolutionary curves in Fig. \ref{S8}, whether for a single stochastic realization or the ensemble average, the real-time FTLE no longer exhibits the short-time Lyapunov plateau predicted by the saddle-point approximation.

Beyond FTLE dynamics, the fragility of these coherent systems is equally evident in the loss of bulk self-healing capabilities, as illustrated in Fig. \ref{bulk_selfHealing_theta1}. In this simulation, the system is initialized with a Gaussian wave packet localized at the center and is subjected to a transient scattering potential $\hat{V}_{scat}$. We observe that while the long-time convergence value of $\eta(t)$ in the ensemble-averaged sense gradually increases with noise , introducing a noise intensity of merely $\sigma=0.1$ is sufficient to completely destroy the bulk self-healing property for a single evolutionary trajectory. This occurs because the violent oscillations of $\eta(t)$ at the single-trajectory level are random across different realizations; performing a statistical average over an ensemble (e.g., 1000 realizations) artificially smooths them out, yielding a deceptively stable decay in Fig. \ref{bulk_selfHealing_theta1}(b). 

Therefore, converging to a small $\langle \eta(t) \rangle$ at long times is strictly a necessary, but not sufficient, condition for true self-healing. This susceptibility to deceptive statistical artifacts directly dictates our presentation strategy in the main text: we exclusively display the temporal evolution of $\eta(t)$ for single stochastic trajectories rather than relying on ensemble averages. Genuine physical self-healing must manifest as a stable profile recovery within the dynamical evolution of an individual wavefunction. Nevertheless, it must be emphasized that $\langle \eta(t) \rangle$ remains an indispensable statistical metric. While individual trajectories inevitably exhibit random temporal fluctuations, their overall decaying trends stochastically oscillate around and align with the statistical baseline established by the ensemble average. Thus, the asymptotic decay of $\langle \eta(t) \rangle$ captures the dominant relaxation tendency of the system, serving as a macroscopic statistical indicator rather than as definitive evidence of strict individual profile recovery.

\bibliography{reference}% Produces the bibliography via BibTeX.